\documentclass{jfm}
\usepackage{graphicx}
\usepackage{subcaption}
\usepackage{amsmath,bm}
\usepackage{natbib}
\usepackage{multirow}
\usepackage{hhline}
\usepackage{booktabs}
\usepackage{makecell}
\usepackage{tikz}
\usepackage{wasysym}
\usepackage{enumerate,enumitem}

\ifCUPmtlplainloaded \else
  \checkfont{eurm10}
  \iffontfound
    \IfFileExists{upmath.sty}
      {\typeout{^^JFound AMS Euler Roman fonts on the system,
                   using the 'upmath' package.^^J}%
       \usepackage{upmath}}
      {\typeout{^^JFound AMS Euler Roman fonts on the system, but you
                   dont seem to have the}%
       \typeout{'upmath' package installed. JFM.cls can take advantage
                 of these fonts,^^Jif you use 'upmath' package.^^J}%
      }
  \else
  \fi
\fi

\ifCUPmtlplainloaded \else
  \checkfont{msam10}
  \iffontfound
    \IfFileExists{amssymb.sty}
      {\typeout{^^JFound AMS Symbol fonts on the system, using the
                'amssymb' package.^^J}%
       \usepackage{amssymb}%
       \let\le=\leqslant  \let\leq=\leqslant
       \let\ge=\geqslant  
      }{}
  \fi
\fi

\ifCUPmtlplainloaded \else
  \IfFileExists{amsbsy.sty}
    {\typeout{^^JFound the 'amsbsy' package on the system, using it.^^J}%
     \usepackage{amsbsy}}
    {}
\fi

\newsavebox{\astrutbox}
\sbox{\astrutbox}{\rule[-5pt]{0pt}{20pt}}

\title[]{Invariant states in inclined layer convection. Part 2. Bifurcations and connections between branches of invariant states}

\author[F. Reetz \textit{et al.}]%
{FLORIAN REETZ$^1$ \ns
PRIYA SUBRAMANIAN$^{2,3}$ \ns \& \ns TOBIAS M.\ns SCHNEIDER$^1$\thanks{Email address for correspondence: tobias.schneider@epfl.ch}}
\affiliation{$^1$ Emergent Complexity in Physical Systems Laboratory (ECPS), Ecole Polytechnique F{\'e}d{\'e}rale de Lausanne, CH-1015, Switzerland\\
$^2$ School of Mathematics, University of Leeds, Leeds LS2 9JT, UK\\
$^3$ Mathematical Institute, University of Oxford, Woodstock Road, Oxford OX2 6GG, UK}

\date{?? and in revised form ??}

\begin{document}

\maketitle
%\tableofcontents
%

\begin{abstract}
Convection in a layer inclined against gravity is a thermally driven non-equilibrium system, in which both buoyancy and shear forces drive spatio-temporally complex flow. As a function of the strength of thermal driving and the angle of inclination, a multitude of convection patterns is observed in experiments and numerical simulations. Several observed patterns have been linked to exact invariant states of the fully nonlinear 3D Oberbeck-Boussinesq equations. These exact equilibria, traveling waves and periodic orbits reside in state space and, depending on their stability properties, are transiently visited by the dynamics or act as attractors. To explain the dependence of observed convection patterns on control parameters, we study the parameter dependence of the state space structure. Specifically, we identify the bifurcations that modify the existence, stability and connectivity of invariant states. We numerically continue exact invariant states underlying spatially periodic convection patterns at $\mathrm{Pr}=1.07$ under changing control parameters for temperature difference between the walls and inclination angle. The resulting state branches cover various inclinations from horizontal layer convection to vertical layer convection and beyond. The collection of all computed branches represents an extensive bifurcation network connecting 16 different invariant states across control parameter values. Individual bifurcation structures are discussed in detail and related to the observed complex dynamics of individual convection patterns. Together, the bifurcations and associated state branches indicate at what control parameter values which invariant states coexist. This provides a nonlinear framework to explain the multitude of complex flow dynamics arising in inclined layer convection. 
\end{abstract}

\section{Introduction}
%%%%%%%%%%%%

%
Thermal convection in a gap between two parallel infinite walls maintained at different fixed temperatures, a system known as Rayleigh-B\'enard convection, is a thermally driven nonequilibrium system that exhibits many different complex convection patterns \citep[e.g.][]{Cross2009}. When inclining the walls against gravity, hot and cold fluid flows up and down the incline, respectively, creating a cubic laminar flow that breaks the isotropy of a horizontal layer and produces shear forces. This system is known as inclined layer convection (ILC). ILC has three control parameters: the temperature difference between the walls, the Prandtl number $\mathrm{Pr}$ parametrising the diffusive properties of the fluid, and the angle of inclination against gravity. 

Recent experiments of ILC using compressed $CO_2$ at a pressure of $41.4$ bars and a mean temperature of $27^{\circ}C$ yielding a Prandtl number of $\mathrm{Pr}=1.07$ have systematically varied the temperature difference and the inclination angle over a wide range, and report ten different spatio-temporal convection patterns \citep{Daniels2000}. In these experiments, the flow domain has a lateral extent much larger than the gap height and thereby allows large-scale patterns to form. The observed convection patterns show spatio-temporally complex dynamics. This includes intermittent temporal bursting of spatially localized convection structures, observed both at small angles of inclination \citep{Busse2000,Daniels2000} as well as at large angles of inclination \citep{Daniels2003}. Other examples include transient oblique patterns forming unsteady interfaces between spatial domains of differently oriented wavy roll patterns \citep{Daniels2002a,Daniels2008}, bimodal patterns, turbulent patterns like crawling rolls at intermediate inclinations \citep{Daniels2008} and chaotically switching diamond panes. These convection patterns have also been reproduced in direct numerical simulations of ILC \citep{Subramanian2016}. How the large variety of patterns at different values of the control parameters emerges from the nonlinear equations describing the flow is however not completely understood.

Theoretical approaches towards explaining spatio-temporal convection patterns in ILC can be described as either an approach `close to thresholds' or an approach `far above thresholds'. Approaches `close to thresholds' include linear stability analysis and the construction of weakly nonlinear amplitude equations. At critical stability thresholds, flow states become unstable and give rise to new pattern motifs. Linear stability analysis of laminar ILC \citep{Gershuni1969,Vest1969,Hart1971b,Ruth1980,Chen1989,Fujimura1993} identified two different types of primary instabilities. A buoyancy driven instability gives rise to straight convection rolls oriented along the base flow at small inclinations. A shear driven instability gives rise to straight convection rolls oriented transverse to the base flow at large inclinations. Secondary instabilities of finite amplitude straight convection rolls and subsequent tertiary instabilities at increased temperature difference and certain angles of inclination have been investigated using Floquet analysis of two- and three-dimensional states \citep{Clever1977a,Busse1992,Clever1995,Busse1996,Subramanian2016}.  Such stability analysis can only explain the onset of convection patterns at, or very close, to the critical stability thresholds in control parameters.

Theoretical approaches to convection patterns `far above thresholds' include the construction of finite amplitude states within a nonlinear analysis at control parameter values far above the critical stability thresholds. Finite amplitude states can be constructed by choosing a Galerkin projection for the governing equations of ILC, often motivated by pattern motifs and their symmetries as identified in a stability analysis at critical stability thresholds \citep{Busse1996,Golubitsky2002}. Galerkin approximations can then be evolved in time under the fully nonlinear governing equations until their amplitudes saturate at finite values with either steady or periodic time evolution \citep{Subramanian2016}. Alternative to forward time integration, finite amplitudes of a Galerkin projection may also be calculated using a Newton-Raphson iteration giving access also to dynamically unstable finite amplitude states \citep{Busse1992,Fujimura1993,Subramanian2016}. If Galerkin projections invoke a complete basis and fully resolve all spatial scales and modal interactions in the three-dimensional flow, exact finite-amplitude states with steady or periodic time evolution can be found. These so-called invariant states are time-invariant exact solutions of the full nonlinear partial differential equations governing the flow. Depending on their temporal dynamics, invariant states are steady equilibrium states, traveling waves or periodic orbits, all of which capture particular structures in the flow. Invariant states can either be dynamically stable or dynamically unstable. In subcritical shear flows like pipe or Couette flow, the construction and analysis of unstable invariant states has lead to significant progress in understanding the complex dynamics of weakly turbulent flow by describing chaotic state space trajectories relative to invariant states \citep[][and references therein]{Kerswell2005,Eckhardt2007,Kawahara2012}.

In ILC, only few highly resolved three-dimensional invariant states had been constructed \citep{Busse1992,Clever1995} before \citet[][referred to as RS20 in the following]{Reetz2020a} identified stable and unstable invariant states underlying various convection patterns at $\mathrm{Pr}=1.07$ observed in experiments \citep{Daniels2000} and simulations \citep{Subramanian2016}. These invariant states are found to transiently attract and repel the dynamics of ILC that is numerically simulated in minimal periodic domains. Minimal periodic domains accommodate only a single spatial period of a periodic convection pattern. Any invariant state computed in minimal periodic domains is also an invariant state in larger extended domains where the pattern of the state periodically repeats in space. To capture a specific pattern with an invariant state in a minimal periodic domain, the size of the domain must be chosen appropriately to match the wavelengths of the pattern. A suitable domain size for a specific pattern can be suggested by Floquet analysis which determines the most unstable pattern wavelength of an instability. At the critical thresholds of instabilities, invariant states emerge in bifurcations and may continue as state branches far above critical thresholds. Thus, bifurcations provide a connection between instabilities `at thresholds' and invariant states `far above thresholds'.

In general, bifurcations are structural changes in a system's state space across which the dynamics of the system changes qualitatively \citep{GuckenheimerHolmes}. Emerging stable invariant states that may continue `far above thresholds' correspond to a supercrtical, forward bifurcation leading to continuous changes in the dynamics. Subcritical bifurcations however, create discontinuous changes in the dynamics allowing for sudden transitions from one state of the system to a very different state. Prominent and potentially harmful examples of such bifurcations, also called tipping points, have been identified in the earth's climate system \citep{Lenton2008} or in combustion chambers \citep{Juniper2018}. In low-dimensional nonlinear model systems, like the three-dimensional Lorenz model for thermal convection \citep{Lorenz1963}, various types of bifurcations have been found and related to different routes to chaos \citep[see][for a review]{Argyris1993}. Thus, different types of bifurcations change the dynamics in different ways. Complex temporal dynamics may be observed where invariant states coexist at equal control parameters (RS20). Complex spatial dynamics, like spatial coexistence of different states in a non-conservative system as ILC, suggests that the spatially coexisting states also coexist as individual states at equal control parameters \citep{Knobloch2015}. Coexistence of invariant states at equal control parameters is a consequence of bifurcations creating these invariant states. Thus, bifurcations creating invariant states that underlie observed convection patterns in ILC not only provide a parametric connection between invariant states and instabilities, but may also explain the state space structure underlying the spatio-temporally complex dynamics observed in spatially extended domains.

Computing bifurcation diagrams in nonlinear dynamical systems requires in practice to numerically continue branches of stable and unstable invariant states under changes of control parameters \citep[see][for a review]{Dijkstra2014}. Numerically fully resolved invariant states in minimal periodic domains of ILC have between $\sim 10^4$ and $\sim 10^6$ degrees of freedom (RS20), fewer than the earth's climate system but much more than the Lorenz equations. Due to the numerically demanding size of the state space, not many prior studies have computed bifurcation diagrams in ILC. Using $4$ degrees of freedom, \citet{Fujimura1993} traced states of mixed longitudinal and transverse modes in almost vertical fluid layers. Using $\sim 10^3$ degrees of freedom, \citet{Busse1992} continued invariant states underlying three-dimensional wavy rolls at selected $\mathrm{Pr}$ and angles of inclinations, and \citet{Clever1995} followed a sequence of supercritical bifurcations in vertical fluid layers. Bifurcation diagrams of two-dimensional invariant states have been computed in vertical convection \citep{Mizushima2002b,Mizushima2002a} and horizontal convection \citep{Waleffe2015}, not addressing three-dimensional dynamics. Recent advances in matrix-free algorithms and computer hardware allow to efficiently construct and continue fully resolved three-dimensional invariant states in double-periodic domains with channel geometry \citep{Viswanath2007,Gibson2008}. We use an extension to the existing numerical framework of the MPI-parallel code \emph{Channelflow 2.0} \citep{Gibson2019} that also handles ILC (RS20).

%\textbf{Objective:}\\
The aim of this paper is to systematically compute and describe bifurcations in ILC. These bifurcations explain the spatio-temporal complexity observed both experimentally and numerically. Using numerical continuation, we trace invariant states that have been constructed in RS20 and that underlie the observed basic convection patterns at specific values of the control parameters. While RS20 analyses dynamical connections between invariant states at those fixed specific values of the control parameters, the present article discusses bifurcations and connections between state branches when control parameters are varied. The analysis covers the same range of control parameters as recent experimental \citep{Daniels2000} and theoretical work \citep{Subramanian2016} at $\mathrm{Pr}=1.07$ and leads to an extensive network of bifurcating branches across values of the control parameters. To understand how temporal and spatio-temporal complexity arises in ILC, we specifically address the following three questions:\\
\begin{itemize}[leftmargin=0.9cm,itemindent=0.0cm]
\item[\emph{Q1} ] \emph{Bifurcation types}: Complex temporal dynamics between coexisting invariant states is a result of bifurcations creating the associated invariant states. Different bifurcation types change the dynamics in different ways. What types of bifurcations create invariant states underlying the observed convection patterns in ILC? \\

\item[\emph{Q2} ] \emph{Connection to instabilities}: Floquet analysis characterises instabilities at critical control parameter values. Results from such an analysis are valid close to the critical thresholds for small amplitude solutions. Do the fully nonlinear invariant states, found in RS20 to underlie the observed convection patterns far from critical thresholds in ILC, bifurcate at the corresponding secondary instabilities reported from a Floquet analysis in \citet{Subramanian2016}?\\

\item[\emph{Q3} ] \emph{Range of existence}: Spatio-temporally complex dynamics suggests existence of invariant states at the associated control parameter values. How do the bifurcation branches of invariant states in ILC continue across control parameter values and what are the limits of their existence? \\
\end{itemize}

%\textbf{Outline}: \\
The present article is structured in the following way. Section \ref{sec:method} describes the numerical methods and outlines the systematic bifurcation analysis. The results of the bifurcation analysis are stated in Section \ref{sec:results}. In five subsections, we report in detail on selected bifurcation diagrams explaining individual convection patterns. The results are discussed in response to \emph{Q1}-\emph{Q3} in Section \ref{sec:discussion}.

\section{Bifurcation analysis of invariant states}
\label{sec:method}
%%%%%%%%%%%%
%Description of ILC
Before introducing the approach of the bifurcation analysis in Section \ref{sec:conti}, we summarize the basic numerical concepts underlying direct numerical simulations of ILC (Section \ref{sec:ilc}), and describe the invariant states that capture relevant convection patterns (Section \ref{sec:invstates}). More details on the direct numerical simulations and identified invariant states are described elsewhere (RS20).

%\\
\subsection{Direct numerical simulation of inclined layer convection}
\label{sec:ilc}
%%%%%%%%%%%%
ILC is studied by numerically solving the nondimensionalised Oberbeck-Boussinesq equations for the velocity $\bm{U}$, temperature $\mathcal{T}$ and pressure $p$ relative to the hydrostatic pressure
\begin{align}
\frac{\partial \bm{U}}{\partial t} + \left(\bm{U}\cdot \nabla \right)\bm{U} &=- \nabla p + \tilde{\nu}\nabla^2 \bm{U} -\hat{\bm{g}}\,\mathcal{T} \ ,\label{eq:obe1}\\
\frac{\partial \mathcal{T}}{\partial t} + \left(\bm{U}\cdot\nabla \right) \mathcal{T} &= \tilde{\kappa}  \nabla^2  \mathcal{T} \ , \label{eq:obe2}\\
\nabla \cdot \bm{U} &=0 \ ,\label{eq:obe3} 
\end{align}
in numerical domains with $x$, $y$ and $z$ indicating the streamwise, the spanwise and the wall-normal dimension. The domains are bounded in $z$ by two parallel walls at $z=\pm 0.5$. In the streamwise dimension $x$ and the spanwise dimension $y$ periodic boundary conditions are imposed at $x=[0,L_x]$ and $y=[0,L_y]$, respectively. The walls are stationary with $\bm{U}(z=\pm 0.5)=0$, have prescribed temperatures $\mathcal{T}(z=\pm 0.5)=\mp 0.5$, and are inclined against the gravitational unit vector $\hat{\bm{g}}=-\sin(\gamma) \bm{e}_x - \cos(\gamma)\bm{e}_z $ by inclination angle $\gamma$. With these boundary conditions, Equations (\ref{eq:obe1}-\ref{eq:obe3}) admit the laminar solution
\begin{align}
\bm{U}_0(z)&= \frac{\sin(\gamma)}{6\,\tilde{\nu}} \left(z^3  - \frac{1}{4}z\right) \bm{e}_x\ , \label{eq:baseU}\\
\mathcal{T}_0(z)&= -z      \  , \label{eq:baseT} \\
p_0(z)&=   \Pi -\cos(\gamma)z^2/2    \  , \label{eq:baseP}
\end{align}
with arbitrary pressure constant $\Pi$. Equations (\ref{eq:obe1}-\ref{eq:obe3}) are nondimensionalised by three characteristic scales of the system. We have chosen the temperature difference $\Delta \mathcal{T}$ between the walls, the gap height $H$, and the free-fall velocity $U_f=(g\,\alpha\,\Delta \mathcal{T}H)^{1/2}$ as characteristic scales. This nondimensionalisation defines the parameters $\tilde{\nu}=(\mathrm{Pr}/\mathrm{Ra})^{1/2} $ and $\tilde{\kappa}=(\mathrm{Pr}\,\mathrm{Ra})^{-1/2} $ in terms of the Rayleigh number $\mathrm{Ra}=g\,\alpha\,\Delta T\,H^3/(\nu \kappa)$ and the Prandtl number $\mathrm{Pr}=\nu/\kappa$. Here, $\alpha$ is the thermal expansion coefficient, $\nu$ is the kinematic viscosity, and $\kappa$ is thermal diffusivity. Thus, ILC has three control parameters, $\gamma$, $\mathrm{Ra}$, and $\mathrm{Pr}$, of which we fix $\mathrm{Pr}=1.07$, in line with previous studies using compressed $CO_2$ as a working fluid \citep{Daniels2000,Subramanian2016}.

Time is measured in free-fall units $H/U_f$ but will also be compared with other relevant time scales of ILC, like the heat diffusion time $H^2/\kappa$, and the laminar mean advection time $L_x/\bar{U}_0$. The latter follows from the laminar velocity profile (\ref{eq:baseU}) integrated over the lower half of the domain where $-0.5\le z\le 0$, that is $\bar{U}_0=\sin(\gamma)/384\tilde{\nu}$. 

The pseudo-spectral code \emph{Channelflow 2.0} \citep{Gibson2019} has been extended to solve (\ref{eq:obe1}-\ref{eq:obe3}) using Fourier-Chebychev-Fourier expansions with $N$ spectral modes in space and a 3rd order implicit-explicit multistep algorithm to march forward in time (see RS20; and references therein). Any time evolution computed with \emph{Channelflow-ILC} represents a unique state vector trajectory $\bm{x}(t)=[\bm{u},\theta](x,y,z,t)$ in a state space with $N$ dimensions. This state space contains all solenoidal velocity fluctuations $\bm{u}=\bm{U}-\bm{U}_0$ and temperature fluctuations $\theta=\mathcal{T}-\mathcal{T}_0$.

\subsection{Computing invariant states}
\label{sec:invstates}
%%%%%%%%%%%%

Invariant states are particular state vectors $\bm{x}^*(t)$ representing roots of a recurrent map 
\begin{equation}
\mathcal{G}(\bm{x}^*,\mu)=\sigma \mathcal{F}^T(\bm{x}^*,\mu) - \bm{x}^* = 0\ . \label{eq:map}
\end{equation}
Here, $\mathcal{F}^T(\bm{x},\mu)$ is the dynamical map integrating (\ref{eq:obe1}-\ref{eq:obe3}) from state $\bm{x}$ over time period $T$ at control parameter $\mu\in [\gamma,\mathrm{Ra},\mathrm{Pr}]$. The invariant state is either an equilibrium state if $T$ is a free parameter, or a periodic orbit if $T$ must match a specific period. Definition (\ref{eq:map}) includes a symmetry transformation $\sigma\in S_{ilc}$. The symmetry group $S_{ilc}=O(2) \times O(2)$, where $\times$ is the direct product, is an equivariance of Equations (\ref{eq:obe1}-\ref{eq:obe3}) in $x$-$y$-periodic domains. $S_{ilc}$ is generated by spanwise $y$-reflection $\pi_y$, streamwise $x$-$z$-reflection $\pi_{xz}$, and $x$- and $y$-translations $\tau(a_x,a_y)$ such that 
\begin{align}
\pi_y[u,v,w,\theta](x,y,z)&=[u,-v,w,\theta](x,-y,z)\label{eq:sympiy} \ , \\
\pi_{xz}[u,v,w,\theta](x,y,z)&=[-u,v,-w,-\theta](-x,y,-z) \label{eq:sympixz}\ ,\\
\tau(a_x,a_y)[u,v,w,\theta](x,y,z)&=[u,v,w,\theta](x+a_x L_x,y+a_y  L_y,z) \label{eq:symtau} \ ,
\end{align}
with shift factors $a_x,a_y\in[0,1)$ scaling the spatial periods $L_x$ and $L_y$ of the periodic domain. All invariant states discussed here are invariant under transformations within subgroups of $S_{ilc}=\langle \pi_y, \pi_{xz}, \tau(a_x,a_y) \rangle$, where angle brackets $\langle \rangle $ imply all products of elements given in the brackets. The specific coordinate transformations for reflection symmetries depend on the spatial phase of the flow structure relative to the origin. We choose the spatial phase such that three-dimensional inversion $\pi_{xyz}=\pi_y \pi_{xz}$, where applicable to invariant states, applies with respect to the domain origin at $(x,y,z)=(0,0,0)$.

If $\sigma\neq 1$ in (\ref{eq:map}), the invariant state is a \emph{relative} invariant state. A traveling wave state, where $\sigma=\tau(a_x,a_y)$ with specific shift factors $a_x$ and $a_y$, is a relative equilibrium state. A relative periodic orbit is either traveling, where $\sigma=\tau(a_x,a_y)$ must be applied after period $T$, or is `pre-periodic' if $\sigma= 1$ after a full period $T$ but $\sigma\neq 1$ after time interval $T'=T/n$ with $n\in \mathbb{N}$.  

Invariant states are computed by solving (\ref{eq:map}) with a Newton-Raphson iteration using matrix-free Krylov methods \citep{Gibson2019}. Practically, we stop iterations if $||\mathcal{G}(\bm{x},\mu)||_2<10^{-13}$ where
\begin{equation}
 ||\bm{x}||_2= \left(\frac{1}{L_x L_y}\int\limits_0^{L_x}\int\limits_0^{L_y}\int\limits_{-0.5}^{0.5} u^2+v^2+w^2+\theta^2 dx\, dy\, dz\right)^{1/2} \ . \label{eq:statenorm}
\end{equation}
A residual of $<10^{-13}$ is sufficiently close to double machine precision to consider the iteration as fully converged. 

Invariant states may be dynamically stable or unstable. The dynamical stability is characterised by the eigenvalues and eigenmodes of the linearised equations computed using Arnoldi iteration \citep{Gibson2019} and depends on the specific symmetry subspace defined by size $[L_x,L_y]$ of the periodic domain and potentially imposed discrete symmetries $\sigma\in S_{ilc}$. We impose $\sigma$ on a state vector $\bm{x}(t)$ using a projection $(\bm{x}(t)+ \sigma \bm{x}(t))/2$ which requires $\sigma^2=1$. We will specify the considered symmetry subspace for each computation of the eigenvalue spectrum.

\begin{figure}
\center
   	\includegraphics[width=\linewidth,trim={0.0cm 0.0cm 0.0cm 0.0cm},clip]{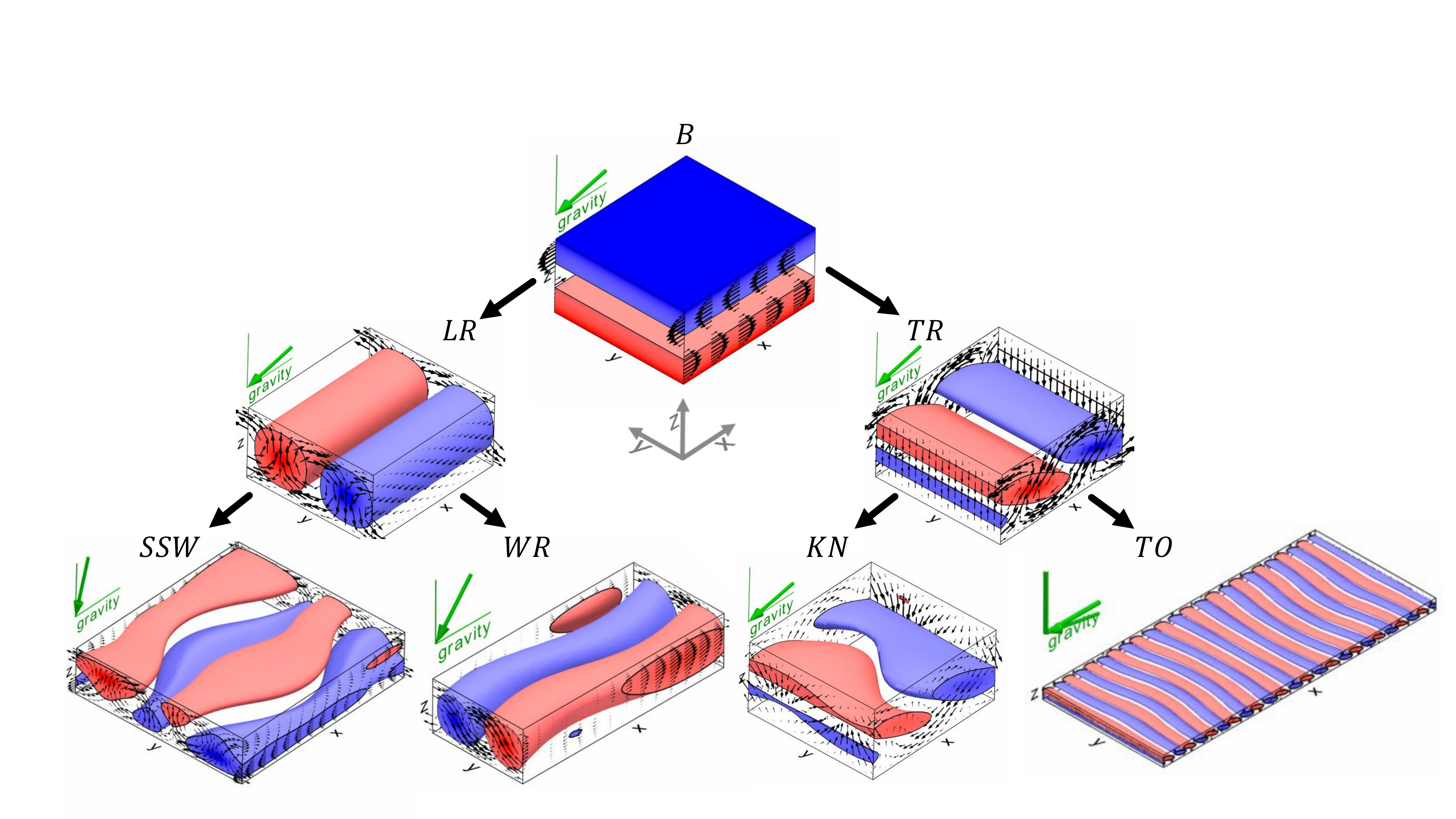}
      	\caption{\label{fig:transitions} Invariant states found as sequentially visited states in temporal transitions at selected control parameter values (RS20). Black arrows indicate the direction of temporal transitions, starting from laminar base flow $B$, via longitudinal and transverse rolls (equilibria $LR$ and $TR$), to four different tertiary invariant states representing the convection patterns of subharmonic standing waves (periodic orbit $SSW$), wavy rolls (equilibrium $WR$), knots (equilibrium $KN$) and transverse oscillations (periodic orbit $TO$). While $B$ is shown as the vector fields of total velocity $\bm{U}$ at the box sides and 3D contours of total temperature $\mathcal{T}$, all other states are shown in terms of velocity and temperature fluctuations, $\bm{u}$ and $\theta$, around $B$. Iso-contours of temperature are at $1/3[\min(\theta),\max(\theta)]$. The states are computed at different inclinations of the domain (green gravity vectors): $B$, $LR$ and $TR$ at $\gamma_{c2}$, and the four tertiary states at the control parameters marked in Figure \ref{fig:overview}. }
\end{figure}

%
%\textbf{Summarize the collection of ECS from part 1}\\
Previously, invariant states underlying observed convection patterns in ILC have been identified by combining direct numerical simulations in small periodic domains with Newton-Raphson iteration (RS20). There, simulations from unstable laminar flow perturbed by small-amplitude noise lead to temporal transitions between seven invariant states. All of these seven invariant states are either linearly stable or posses few unstable eigendirections, depending on the symmetry subspace corresponding to the chosen periodic domain, control parameter values and potentially imposed discrete symmetries. As a consequence, the temporal dynamics is either asymptotically or transiently attracted to these invariant states. Moreover, the temporal dynamics is found to visit these invariant states in a specific sequential order. Figure \ref{fig:transitions} summarizes the observed transition sequences and visualises the flow structures of all seven invariant states. Which invariant state is visited by the dynamics depends on the values of the control parameters. Following \citet{Daniels2000} and \citet{Subramanian2016}, we explore the two-dimensional parameter space of inclination angle $\gamma$ and Rayleigh number $\mathrm{Ra}$ at fixed $\mathrm{Pr}=1.07$. When traversing this two-dimensional parameter space, the laminar base flow ($B$) may lose stability in one of two different primary instabilities corresponding to two different transitions. At any Prandtl number, sufficiently increasing $\mathrm{Ra}$ yields a transition from $B$ to either longitudinal rolls ($LR$), for angles $\gamma<\gamma_{c2}$, or from $B$ to transverse rolls ($TR$), for angles $\gamma>\gamma_{c2}$. Only at a single point in this two-dimensional parameter space, $[\gamma_{c2},\mathrm{Ra}_{c2}]$, the two instabillities occur simultaneously. Such points are commonly referred to as codimension-2 points. At $\mathrm{Pr}=1.07$, we determine the codimension-2 point accurately at $[\gamma_{c2},\mathrm{Ra}_{c2}]=[77.7567^{\circ},8053.1]$ when computing $LR$ and $TR$ in a domain with periodicity $[L_x,L_y]=[\lambda_x,\lambda_y]$ where $[\lambda_x,\lambda_y]=[2.2211,2.0162]$ and grid size $[n_x,n_y,n_z]=[32,32,25]$. Wavelengths $\lambda_x$ and $\lambda_y$ are suggested by linear stability analysis \citep{Subramanian2016}. %In contrast to stability analysis with Floquet vectors \citep{Subramanian2016}, invariant solution can only be computed at specific pattern wavelengths defined by the choice of the periodic domain. 
As in RS20, we fix the domain periodicity $[\lambda_x,\lambda_y]$ and the grid resolution $[n_x,n_y,n_z]$ throughout this study and choose all computational domains as multiples of this minimal periodic box. Subharmonic standing waves ($SSW$) are computed in a domain with periodicity $[L_x,L_y]=[2\lambda_x,2\lambda_y]$, wavy rolls ($WR$) with $[L_x,L_y]=[2\lambda_x,\lambda_y]$, knots ($KN$) with $[L_x,L_y]=[\lambda_x,\lambda_y]$, and transverse oscillations ($TO$) with $[L_x,L_y]=[12\lambda_x,6\lambda_y]$. The grid size is scaled accordingly. Choosing all domains as integer multiples of the same minimal box ensures commensurable wavelengths and thus allows for potential bifurcations between invariant states.

%The SV case
The approach of combining direct numerical simulations from unstable laminar flow with Newton-Raphson iteration allows to determine all of the above invariant states. However, this approach fails in the case of the pattern emerging from the skewed varicose instability at $\gamma=0^{\circ}$ (RS20). There, the dynamics does not asymptotically approach or transiently visit an invariant state underlying the pattern, suggesting that no associated invariant state exists above thresholds. Therefore, we search for the bifurcating branch below critical parameters of the skewed varicose instability by taking the following steps. The bifurcating eigenmode that destabilizes $x$-aligned straight convection rolls at wavelength $\lambda_y$ in a domain of periodicity $[L_x,L_y]=[4\lambda_x,4\lambda_y]$ is computed using Arnoldi iteration. Since ILC for $\gamma=0^{\circ}$, corresponding to the Rayleigh-B\'enard system, has isotropic symmetry, there is no distinction between `longitudinal' and `transverse' rolls. Consequently, we indicate straight convection rolls by $R_{\lambda 2}$ where the subscript indicates their approximate wavelength. Different finite amplitude perturbations of $R_{\lambda 2}$ with the bifurcating eigenmode are integrated forward in time to generate a large set of initial states for brute-force Newton-Raphson iterations below critical threshold parameters of the instability. Using this approach we identified an unstable equilibrium state that underlies the skewed varicose pattern and is described in Section \ref{sec:sv}. Consequently, invariant states in thermal convection cannot be assumed to generically exist above critical control parameter values, but may also be found below thresholds suggesting a backward bifurcation. Whether bifurcations are forward or backward in control parameters, is studied in the present bifurcation analysis.

\subsection{Bifurcation analysis}
\label{sec:conti}
%%%%%%%%%%%%
%1. Explain predictor-corrector
%2. Explain arc-length continuation
%3. Explain periodic orbit specific: fixing phase, plotting min/max, multishooting
%4. Systematic bifurcation analysis
%\textbf{Continuation methods}\\
The general approach of our bifurcation analysis is to compute bifurcation branches of invariant states in ILC and to characterize the resulting bifurcation diagrams. Branches of invariant states are computed using continuation methods to solve (\ref{eq:map}) under a changing value of a control parameter $\mu$ \citep{Dijkstra2014}. There are two iterative schemes for numerical continuation implemented in \emph{Channelflow-ILC}. The control parameter continuation uses quadratic extrapolation to predict a state vector $\bm{x}$ for a specific value of $\mu$ which is fixed in the following Newton-Raphson iteration. The pseudo-arclength continuation does not fix $\mu$ but solves for $\mu$ as additional unknown entry in state vector $\bm{x}$ under an additional arclength constraint. Depending on the shape of the continued state branch, one continuation scheme might outperform the other \citep{Gibson2019}. Continuation of periodic orbits with long periods may require a multi-shooting method to converge \citep{Gibson2019}. Where invariant states have discrete reflection symmetries $\pi_y$ or $\pi_{xz}$ (\ref{eq:sympiy}-\ref{eq:sympixz}) we impose reflections during numerical continuation because they fix the spatial phase of the flow relative to the $x$- or $y$-coordinates. If the spatial phase is free, states may translate under numerical continuation reducing the computational efficiency. Since both continuation schemes solve (\ref{eq:map}) and the algorithmic details do not change the resulting bifurcation diagrams, we use the better performing scheme for each branch.

\begin{figure}
\center
       \includegraphics[width=0.7\linewidth,trim={0.1cm 2.7cm 0.0cm 0.5cm},clip]{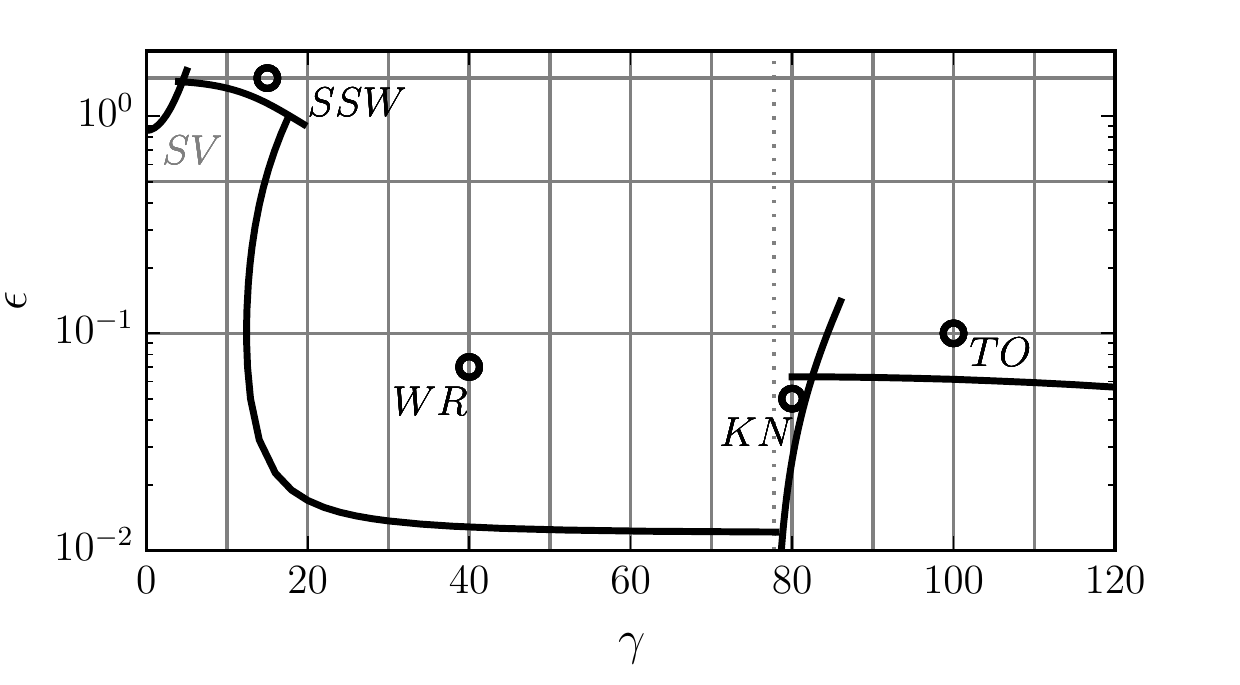}
      	\caption{\label{fig:overview} The considered parameter space is spanned by inclination angle $\gamma$ and $\epsilon=(\mathrm{Ra}-\mathrm{Ra}_c)/\mathrm{Ra}_c$ at $\mathrm{Pr}=1.07$. The parameters at which invariant states have been identified in RS20 (Figure \ref{fig:transitions}) are marked by circles. The invariant state underlying skewed varicose pattern ($SV$) is not described in RS20 but identified in this work. Thick solid lines indicate critical thresholds in control parameters of the five secondary instabilities determined by \citet{Subramanian2016}. Bifurcations between invariant states are computed along 15 sections across the parameter space (grey solid lines). The thin grey dotted line marks the inclination angle of the codimension-2 point at $\gamma_{c2}=77.7567^{\circ}$.}
\end{figure}

%\textbf{Systematic bifurcation analysis:}\\
Continuations of the invariant states cover \emph{a priori} chosen sections across the considered parameter space at $\mathrm{Pr}=1.07$ covering $0^{\circ}\le \gamma< 120^{\circ}$ and $0\le \epsilon \le 2$, as illustrated by thin grey lines in Figure \ref{fig:overview}. The control parameter $\epsilon=(\mathrm{Ra}-\mathrm{Ra}_c(\gamma))/\mathrm{Ra}_c(\gamma)$ indicates $\mathrm{Ra}$ normalised by a critical threshold function $\mathrm{Ra}_c(\gamma)$ which here, approximates the true critical control parameters $\mathrm{Ra}'_c(\gamma)$ of the primary instability in ILC (see Figure 2a in RS20). Thus, the primary instability defining the onset of convection is always at $\epsilon\approx0$, independent of the inclination angle. Critical thresholds of bifurcation points are denoted as $\epsilon_c$. To continue invariant states in $\gamma$ at $\epsilon=\mathrm{const.}$, also $\mathrm{Ra}$ needs to be adjusted accounting for variations in $\mathrm{Ra}_c(\gamma)$. Since the true critical $\mathrm{Ra}'_c(\gamma)$ cannot be expressed in closed-form, we define the function
\begin{align}
\mathrm{Ra}_c(\gamma \leq \gamma_{c2}) &= \frac{\mathrm{Ra}'_c(\gamma=0^{\circ})}{\cos(\gamma)} \label{eq:critRa1} \\
\mathrm{Ra}_c(\gamma > \gamma_{c2}) &= \frac{1}{41}(\gamma-\gamma_{c2})^3+\frac{5}{14}(\gamma-\gamma_{c2})^2+29(\gamma-\gamma_{c2}) + \mathrm{Ra}'_{c2}  \label{eq:critRa2}
\end{align}
to keep $\epsilon= (\mathrm{Ra}-\mathrm{Ra}_c(\gamma))/\mathrm{Ra}_c(\gamma)=\mathrm{const.}$ under $\gamma$-continuations. The definition of function $\mathrm{Ra}_c(\gamma)$ has three precise coefficients, namely the critical parameters for horizontal convection $\mathrm{Ra}'_c(\gamma=0^{\circ})=1707.76$ \citep{Busse1978} and the codimension-2 point $[\gamma_{c2},\mathrm{Ra}'_{c2}]=[77.7567^{\circ},8053.1]$. Relation (\ref{eq:critRa1}), already found by \citet{Gershuni1969}, is a geometric consequence of the linear laminar temperature profile. Polynomial (\ref{eq:critRa2}) is a least-square-fit of the empirical critical thresholds for $\gamma_{c2}<\gamma<120^{\circ}$ reported in \citet{Subramanian2016}, and is an approximation of the true $\mathrm{Ra}'_c(\gamma)\approx \mathrm{Ra}_c(\gamma)$. The purpose of defining $\mathrm{Ra}_c(\gamma)$ in (\ref{eq:critRa1}-\ref{eq:critRa2}) is not to most accurately capture the true $\mathrm{Ra}'_c(\gamma)$ but to provide a closed-form function for converting values between $\mathrm{Ra}$ and $\epsilon$. The conversion allows $\gamma$-continuations at $\epsilon=\mathrm{const.}$ and a comparison of the present results with other work reported in terms of a similar $\epsilon'$, based on the empirically determined primary instabilities.

Linear stability analysis of invariant states is performed at selected points along continued branches. 
%Automatic stability analysis as part of a continuation scheme, would not be very meaningful in many cases because the result of Arnoldi iteration depends on the considered symmetry subspace, which differs if invariant states are computed at different periodicity.  - This is already said at end of paragraph after equation 2.9!
Under continuation, we consider invariant states in their minimal periodic domain capturing only one spatial period of the pattern. In order to compare the dynamical stability between different invariant states, Arnoldi iterations must be performed in identical symmetry subspaces. This implies using the same periodic domains and imposing the same discrete symmetries for all considered states. Wherever we compute the dynamical stability along selected bifurcation branches, we specifically choose and report the symmetry subspace for the full branch.

Many bifurcation types of vector fields are known \citep[e.g.][]{GuckenheimerHolmes}. The most common bifurcations we encounter in ILC are pitchfork bifurcations, Hopf bifurcations, saddle-node bifurcation and mutual annihilation of two periodic orbits, all of which are also well-known bifurcations in ordinary differential equations \citep[e.g.][]{Schaeffer2016}. The two latter types we simply refer to as `folds'. If bifurcations are not one of these four common types, we provide explicit references that discuss the bifurcation type in detail, as such discussions would be beyond the scope of the present work. When discussing symmetry-breaking bifurcations, the classification into supercritical/subcritical bifurcations refers to a `more stable'/'less stable' bifurcating branch in comparison to the stability of the coexisting parent branch \citep{Tuckerman1990}. The orientation of symmetry-breaking bifurcations along a control parameter $\mu$ is given specifically as $\mu$-forward or $\mu$-backward.

\section{Results}
\label{sec:results}
%%%%%%%%%%%%

\begin{figure}
\includegraphics[width=0.99\linewidth,trim={1.2cm 1.4cm 2.2cm 2cm},clip]{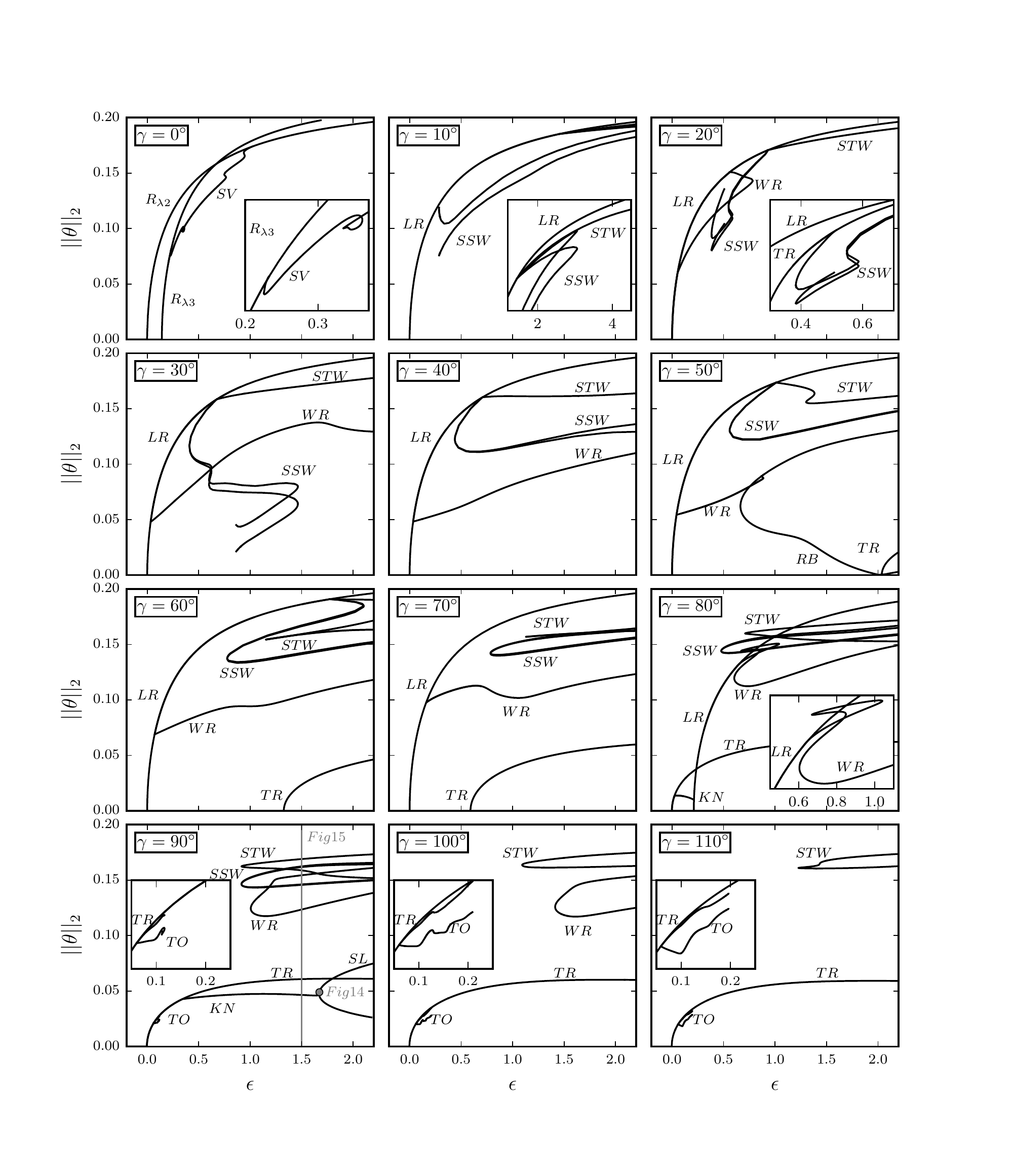}
\caption{\label{fig:bifeps3} Summary of all bifurcation branches of invariant states continued in $\epsilon$ at constant $\gamma \in \{0^{\circ},10^{\circ},20^{\circ},...,110^{\circ}\}$. Each branch is plotted in terms of $||\theta||_2$ (Equation \ref{eq:tempnorm}) and is labeled by the name of the invariant state to which the branch belongs. Insets enlarge or isolate features of the bifurcations diagrams. All panels are labeled by the angle of inclination and share the same axes. $TR$ is left out in panels $\gamma \in \{0^{\circ},10^{\circ},20^{\circ}\}$ to avoid clutter. In panel $\gamma=90^{\circ}$, the grey vertical line crosses the bifurcation branches where invariant states are shown in Figure 15 and discussed in Appendix \ref{sec:app:vertical}. $KN$ at $\gamma=90^{\circ}$ connect to equilibrium state $SL$ showing a subharmonic lambda pattern (Figure 14), discussed in Appendix \ref{sec:app:additional}. }
\end{figure}

\begin{figure}
\includegraphics[width=0.99\linewidth,trim={0.5cm 1.5cm 4.3cm 2cm},clip]{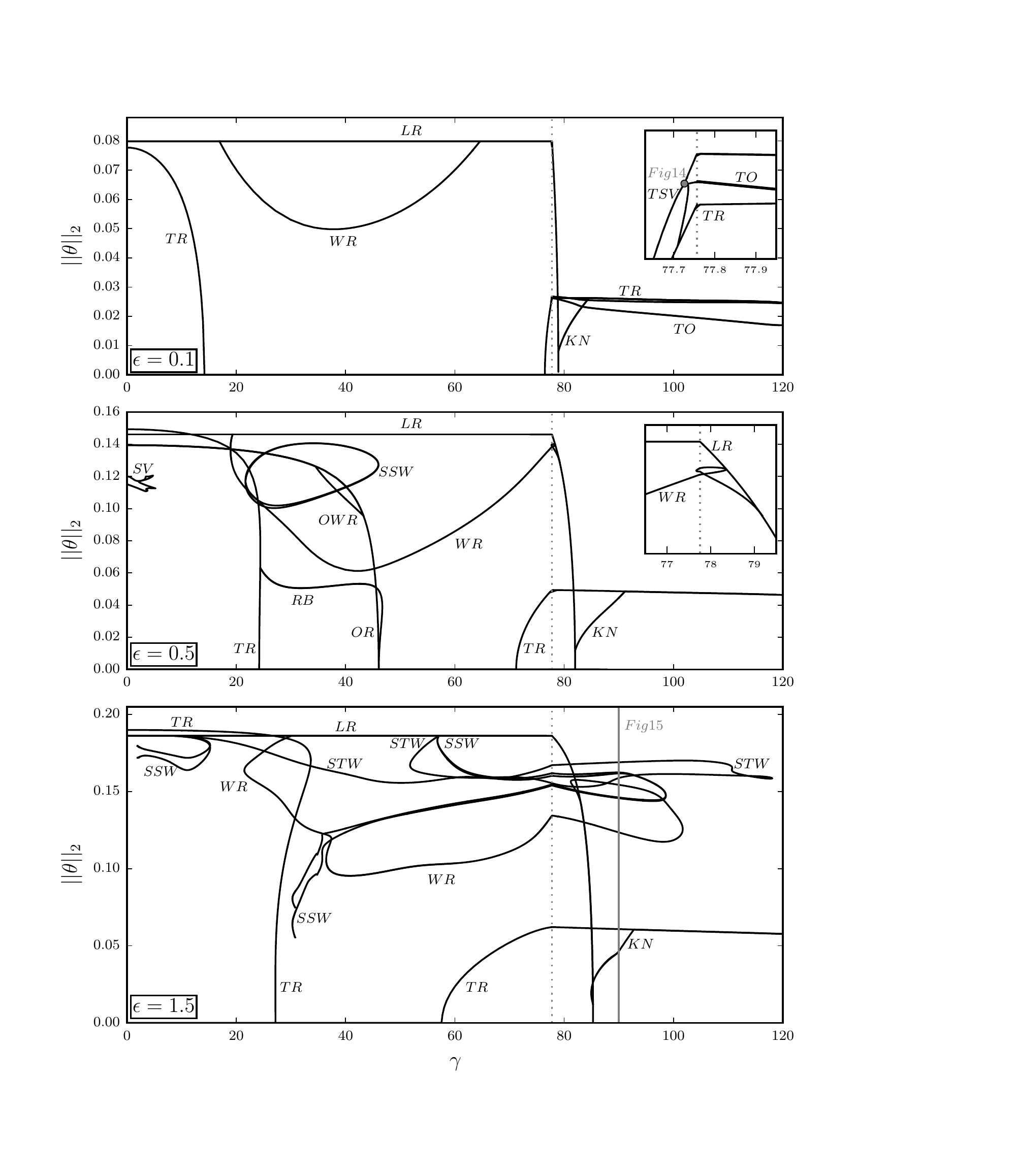}
\caption{\label{fig:bifgamma3} Summary of all bifurcation branches of invariant states continued in $\gamma$ at constant $\epsilon \in \{0.1,0.5,1.5\}$. Each branch is plotted in terms of $||\theta||_2$ (Equation \ref{eq:tempnorm}) and is labeled by the name of the invariant state to which the branch belongs. All panels are labeled by the $\epsilon$ value and share the same axes. Grey dotted lines mark the inclination angle of the codimension-2 point. Insets enlarge features of the bifurcations diagrams around the inclination angle of the codimension-2 point. In panel $\epsilon=1.5$, the grey vertical line crosses the bifurcation branches where invariant states are shown in Figure 15 and discussed in Appendix \ref{sec:app:vertical}. $TO$ at $\epsilon=0.1$ (inset) connect to equilibrium state $TSV$ showing a transverse subharmonic varicose pattern (Figure 14), discussed in Appendix \ref{sec:app:additional}. }
\end{figure}

%\textbf{Outline:}\\
We first provide an overview of the results from the bifurcation analysis. Considering twelve sections at constant $\gamma$ and three sections at $\epsilon=\mathrm{const.}$ (Figure \ref{fig:overview}), we present 15 bifurcation diagrams here in Figures \ref{fig:bifeps3} and \ref{fig:bifgamma3}. In addition to branches of the seven invariant states shown in Figure \ref{fig:transitions}, the bifurcation diagrams also contain branches of nine additional invariant states. We refer to these states as subharmonic traveling wave ($STW$), skewed varicose state ($SV$), ribbons ($RB$), oblique rolls ($OR$), oblique wavy rolls ($OWR$), disconnected wavy rolls ($DWR$), longitudinal subharmonic varicose state ($LSV$), transverse subharmonic varicose state ($TSV$), and subharmonic lambda plumes ($SL$). For three out of those nine additional states their existence was suggested previously. The existence of $SV$ and $OR$ has been suggested by stability analysis \citep[e.g.][]{Subramanian2016} and $TSV$ is similar, yet not identical, to an equilibrium discussed by \citet{Clever1995}. Table \ref{tab:overviewtable} provides an index of all invariant states considered here. A complete systematic analysis of all branches in these diagrams is beyond the scope of this paper. Instead in this section, we first summarise the bifurcation diagrams and then focus on selected state branches covering the control parameter values where spatio-temporally complex convection patterns are observed and temporal dynamics between invariant states has been studied (RS20). 

\begin{table}
 \begin{center}
   \caption{Invariant states considered in the bifurcation analysis. The states are of one of three types: steady equilibrium states (EQ), traveling wave states (TW), pre-periodic orbits (PPO). Their spatial periodicity along the $x$- and $y$-dimension is given in units of $\lambda_x=2.2211$ and $\lambda_y=2.0162$. }
   \label{tab:overviewtable}
   \begin{tabular}{| l | l | l | r | c | l |} 
     \toprule
     state & full name & type & periodicity & bifurcates off & branches in figure  \\ \midrule
     $R_{\lambda}$ & rolls at $\gamma=0^{\circ}$ & EQ & various & $B$ & 3($0^{\circ}$),5d \\
     $LR$ & longitudinal rolls & EQ & $[\infty, 1\lambda_y]$ & $B$ & 3($10^{\circ}...80^{\circ}$),4,6c,9e,11 \\     
     $TR$ & transverse rolls & EQ & $[1\lambda_x,\infty]$ & $B$ & 3($50^{\circ}...110^{\circ}$),4,8a,11,12c \\    
     $OR$ & oblique rolls & EQ & $[2\lambda_x,1\lambda_y]$ & $B$ & 4($0.5$),9,10 \\
     $RB$ & ribbons & EQ & $[2\lambda_x,1\lambda_y]$ & $B$ & 3($50^{\circ}$),4($0.5$),9e,10 \\
     $SV$ & skewed varicose state & EQ & $[2\lambda_x,1\lambda_y]$ & $R_{\lambda}$ & 3($0^{\circ}$),4($0.5$),5d \\
     $SSW$ & subh. standing wave & PPO & $[2\lambda_x,2\lambda_y]$ & $LR$, $TR$, $LSV$ &  3($10^{\circ}...90^{\circ}$),4($0.5,1.5$),6,8 \\
     $STW$ & subh. traveling wave & TW & $[2\lambda_x,2\lambda_y]$ & $LR$ & 3($10^{\circ}...110^{\circ}$),4($1.5$),6 \\
     $WR$ & wavy rolls & EQ & $[2\lambda_x,1\lambda_y]$ & $LR$ & 3($20^{\circ}...100^{\circ}$),4,9\\
     $OWR$ & oblique wavy rolls & EQ & $[2\lambda_x,1\lambda_y]$ & $OR$, $RB$ & 4($0.5$),9 \\
     $DWR$ & disconnected wavy rolls & EQ & $[2\lambda_x,1\lambda_y]$ & $OR$, $WR$ & 9 \\
     $KN$ & knots & EQ & $[1\lambda_x,1\lambda_y]$ & $LR$, $TR$, $SL$& 3($80^{\circ},90^{\circ}$),4,11 \\
     $TO$ & transverse oscillations & PPO & $[12\lambda_x,6\lambda_y]$ & $TR$, $TSV$ & 3($90^{\circ}...110^{\circ}$),4($0.1$),12 \\
     $LSV$ & long. subh. varicose state & EQ & $[2\lambda_x,2\lambda_y]$ & n.a. & 6c \\
     $TSV$ & trans. subh. varicose state & EQ & $[3\lambda_x,3\lambda_y]$ & $TR$ & 4($0.1$) \\
     $SL$ & subh. lambda plumes & EQ & $[1\lambda_x,1\lambda_y]$ & n.a. & 3($90^{\circ}$)\\
   \end{tabular}
 \end{center}
\end{table}

We specifically discuss the branches that bifurcate from straight convection rolls via the five secondary instabilities that were identified by \citet{Subramanian2016}. These are, skewed-varicose instability, longitudinal subharmonic oscillatory instability, wavy roll instability, knot instability and transverse oscillatory instability. Branches of equilibrium and travelling wave states are plotted in terms of the norm of the temperature fluctuations,
\begin{equation}
 ||\theta||_2= \left(\frac{1}{L_x L_y}\int\limits_0^{L_x}\int\limits_0^{L_y}\int\limits_{-0.5}^{0.5} \theta^2(x,y,z,t_{\pm}) dx\, dy\, dz\right)^{1/2} \ , \label{eq:tempnorm}
\end{equation}
as a function of the bifurcation parameter. Periodic orbits are illustrated by a pair of branches indicating the minimum and maximum of $||\theta||_2$ over one orbit period, at instances $t_{\pm}$. Bifurcation branches are labeled inside the diagram with the name of the invariant state. We recommend reading each diagram panel by first identifying the branches of $LR$ and/or $TR$. In most cases, $LR$ or $TR$ have the largest $||\theta||_2$ and tertiary branches bifurcate to lower $||\theta||_2$. See Figure \ref{fig:bifeps3} for bifurcations while varying $\epsilon$ and Figure \ref{fig:bifgamma3} for bifurcations while varying $\gamma$. 


The $\epsilon$-bifurcations at fixed $\gamma$, confirm the common observation that $LR$ and $TR$ always bifurcate in supercritical, $\epsilon$-forward pitchfork bifurcations from the laminar base state. At $\gamma=0$, longitudinal and transverse rolls are related via symmetries, and we refer to both of them as $R_{\lambda}$, where the subscript indicates the wavelength of the roll pattern. At $0<\gamma\le 20^{\circ}$, branches of $LR$ and $TR$ still are very close to each other. Only the $LR$-branches, defining the onset of convection, are plotted to avoid clutter. At $30^{\circ}\le \gamma \le 40^{\circ}$, $TR$ bifurcates outside the considered interval \hbox{of~$0\le \epsilon\le 2$}. At $\gamma \ge 50^{\circ}$, the branches of $TR$ and $LR$ bifurcate again closer to each other. The branches however differ significantly in amplitude and functional form. $TR$-branches show non-monotonic behaviour in $||\theta||_2$, e.g., a local maximum at $\epsilon=1.8$ and $\gamma=90^{\circ}$. Non-monotonic branches of $TR$ were also computed in vertical convection at $\mathrm{Pr}=0.71$ \citep{Mizushima2002b,Mizushima2002a}. 

$LR$-branches monotonically increase in $||\theta||_2$ with $\epsilon$. For further increasing $\epsilon$, $LR$ appears to eventually approach the same $\epsilon$-scaling law as reported for invariant states underlying straight convection rolls in horizontal convection at $\gamma=0^{\circ}$ \citep{Waleffe2015}. The large-$\epsilon$ behaviour is observed for all $\gamma$. The observation that a scaling law of straight convection rolls at $\gamma=0^{\circ}$ also describes $LR$-branches at $\gamma\neq 0^{\circ}$ suggests a particular scaling invariance of the nonlinear Oberbeck-Boussinesq equations under changes of inclination angle $\gamma$. This scaling invariance is discussed in the following paragraph.

%\textbf{Scaling invariance:}\\
At fixed $\epsilon$, all $\gamma$-continuations of $LR$ result in horizontal lines in $||\theta||_2$ for $0^{\circ}\le \gamma\le \gamma_{c2}$, e.g. the branch with $||\theta||_2=0.08$ at $\epsilon=0.1$ in Figure \ref{fig:bifgamma3}. These horizontal lines are a remarkable feature of the bifurcation diagrams and can be explained as a consequence of a scaling invariance of the nonlinear Oberbeck-Boussinesq equations, that holds for patterns or states that are steady in time and uniform in $x$, like $LR$:\\
For $\gamma<\gamma_{c2}$, keeping $\epsilon=\mathrm{const.}$ implies $\mathrm{Ra}\sim 1/\cos(\gamma)$ due to (\ref{eq:critRa1}). The laminar solution (\ref{eq:baseU}-\ref{eq:baseP}) thus scales with $\gamma$ as $U_0(z) \sim \sin(\gamma)/\sqrt{\cos(\gamma)}$, $\mathcal{T}_0(z)\sim 1$ and $p_0(z)\sim -\cos(\gamma)$. Here, $\mathrm{Pr}$ is constant. Inserting the base-fluctuation decomposition $\bm{U}=\bm{U}_0+\bm{u}$ and $\mathcal{T}=\mathcal{T}_0+\theta$ into (\ref{eq:obe1}-\ref{eq:obe3}) and assuming $\partial_t[\bm{u},\theta]=0$ and $\partial_x[\bm{u},\theta]=0$ for steady stripe/roll states, we observe: If temperature and velocity fluctuations are scaled as $u(y,z) \sim \sin(\gamma)/\sqrt{\cos(\gamma)}$, $v(y,z)\sim w(y,z)\sim \sqrt{\cos(\gamma)}$, and $\theta(y,z) \sim 1$, for all components $[u,v,w,\theta,c]$, each term in the governing equations for the respective component is proportional to a common $\gamma$-dependent factor. This factor differs between components of the governing equations but is common for all terms within a single equation:
\begin{align}
u:\quad&	v\partial_y u + w\partial_z (u + U_0(z)) &=& -\partial_x p + (\mathrm{Pr}/\mathrm{Ra})^{1/2}\nabla^2 u + \sin(\gamma)\theta & \sim \sin(\gamma) \label{eq:scinvu}\\
v:\quad& v\partial_y v + w\partial_z v &=& -\partial_y p + (\mathrm{Pr}/\mathrm{Ra})^{1/2}\nabla^2 v & \sim \cos(\gamma) \label{eq:scinvv}\\
w:\quad&	v\partial_y w + w\partial_z w &=& -\partial_z p + (\mathrm{Pr}/\mathrm{Ra})^{1/2}\nabla^2 w + \cos(\gamma)\theta & \sim \cos(\gamma) \label{eq:scinvw}\\
\theta:\quad&	v\partial_y \theta + w \partial_z (\theta+\mathcal{T}_0(z)) &=& \qquad\qquad (\mathrm{Pr} \mathrm{Ra})^{-1/2}\nabla^2 \theta & \sim \sqrt{\cos(\gamma)}\\
c:\quad& \qquad\qquad\quad \partial_y v + \partial_z w &=& \qquad 0 & \sim \sqrt{\cos(\gamma)} \label{eq:scinvc}
\end{align}
The common scaling factors $\sin(\gamma)$, $\cos(\gamma)$ and $\sqrt{\cos(\gamma)}$, as indicated above on the right-hand side, can be absorbed. The resulting scaling implies that any equilibrium at one value of $\gamma$ corresponds to a whole family of equilibria for $0^{\circ}\le \gamma \le 90^{\circ}$. The temperature scaling $\theta(y,z)\sim1$ directly implies that $||\theta||_2$ of $LR$ remains invariant under changes in $\gamma$ with $\epsilon=\mathrm{const.}$ This leads to self-similar $LR$-branches under $\epsilon$-continuation at fixed $\gamma$ (Figure \ref{fig:bifeps3}) and horizontal $LR$-branches under $\gamma$-continuation at fixed $\epsilon$ (Figure \ref{fig:bifgamma3}). Moreover, any $x$-uniform and steady invariant state for $0<\gamma\le 90^{\circ}$ corresponds to a specific invariant state in the horizontal Rayleigh-B\'enard case at $\gamma=0^{\circ}$. A similar relation has previously been reported for the infinite $\mathrm{Pr}$ limit only \citep{Clever1973}. The scaling relation provided here is valid for all $\mathrm{Pr}$ and a property of the full nonlinear Oberbeck-Boussinesq equations.

In the limit of a vertical gap ($\gamma\rightarrow 90^{\circ}$), the $\cos^{-1}(\gamma)$-scaling implies diverging $\mathrm{Ra}$. In this limit, the amplitude of the fixed $u(y,z)$-profile diverges and the cross-flow components vanish, $v,w\rightarrow 0$. The temperature field $\theta(y,z)$ remains fixed. Consequently in a vertical gap, hot and cold streamwise jets without cross flow and diverging streamwise velocity amplitude are invariant states in the $\mathrm{Ra}\rightarrow\infty$ limit. Any temperature field of $LR$ found at $\gamma<90^{\circ}$ is a valid temperature field for these jets at infinite $\mathrm{Ra}$. 

The subsequent sections discuss selected bifurcation diagrams covering the parameters where temporal dynamics between invariant states has been studied (RS20). We do not systematically explain the bifurcations at all covered values of the control parameters but rather highlight important features of the bifurcations at selected control parameter values. In each section we summarise key features of the bifurcation structure and relate those to observed spatio-temporally complex dynamics of the flow. The sections are ordered by increasing values of the angle of inclination.

\subsection{Skewed varicose state - subcritical connector of bistable rolls}
\label{sec:sv}
%%%%%%%%%%%%
%\textbf{Contrast observation and key result:}\\
%\textbf{Subsection summary:}\\ 
%\subsubsection{Subsection summary}
The skewed varicose instability of Rayleigh-B\'enard convection, first found as spatially periodic instability at $\mathrm{Pr}=7$ \citep{Busse1979}, is experimentally observed to trigger a spatially localized transient pattern at $\mathrm{Pr}=1.07$ with very subtle varicose features \citep[][Figure 7]{Bodenschatz2000}. This section reports on a bifurcation from straight convection rolls to an equilibrium state capturing the observed skewed varicose pattern in a periodic domain. The bifurcating branch is subcritical, exists only below $\epsilon_c$ of the skewed varicose instability, and connects two bistable straight convection rolls at different wavelengths and orientations. The subcritical coexistence of the skewed varicose equilibrium with bistable straight convection rolls may explain the spatial localization of the transiently observed pattern.

\subsubsection{Bifurcation branch of skewed varicose states}
%\subsubsection{Observation and key result}
When convection patterns in experiments or numerical simulations exhibit complex dynamics, we expect the existence of invariant states underlying the pattern dynamics. For the pattern dynamics emerging from the skewed varicose instability of straight convection rolls $R_{\lambda}$ at $\gamma=0^{\circ}$ we however do not find invariant states at the control parameter values where the dynamics is observed. Direct numerical simulations in a minimal periodic domain can reproduce the transient dynamics of the skewed varicose pattern, but previous analysis of the temporal dynamics did not yield an underlying invariant state (RS20).

An equilibrium state resembling the observed skewed varicose pattern ($SV$) is identified below $\epsilon_c$ of the skewed varicose instability, as described in Section \ref{sec:invstates}. Numerical continuation of $SV$ reveals a subcritical $\epsilon$-backward pitchfork bifurcation from $R_{\lambda 2}$ at $\epsilon_c=1.019$. The bifurcation breaks the continuous translation symmetry $\tau(a_x,0)$ of straight convection rolls $R_{\lambda 2}$. Here we consider the rolls to be $x$-aligned and periodic with wavelength $\lambda_y$. The bifurcating eigenmode shows a skewed three-dimensional flow structure. The bifurcating equilibrium $SV$ is $[4\lambda_x,4\lambda_y]$-periodic and invariant under transformations of the symmetry group $S_{sv}=\langle \pi_{xyz},\tau(0.25,0.25)\rangle$. From the bifurcation point, the $SV$-branch continues down in $\epsilon$, undergoes a sequence of folds, and terminates at $\epsilon=0.206$ in a bifurcation from straight convection rolls $R_{\lambda 3}$ with wavelength $\lambda=2.8$ (panel $\gamma=0^{\circ}$ in Figure \ref{fig:bifeps3}). Thus, the equilibrium state $SV$ connects two equilibrium states representing straight convection rolls at different wavelengths. $SV$ exists only below the critical threshold parameter $\epsilon_c$. The pure subcritical existence of $SV$ explains why no temporal transition to an underlying invariant state at $\epsilon>\epsilon_c$ has been found in RS20.

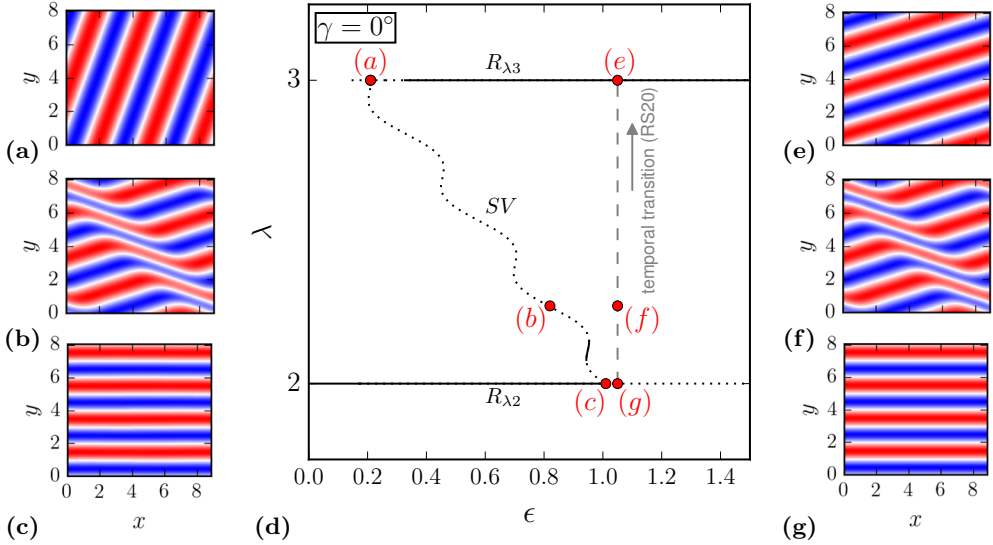
\begin{figure}
\center
       \begin{subfigure}[b]{0.21\textwidth}
		\begin{tikzpicture}
   	\draw (0, 0) node[inner sep=0] {\includegraphics[width=\linewidth,trim={0.0cm 0.0cm -0.3cm 0.0cm},clip]{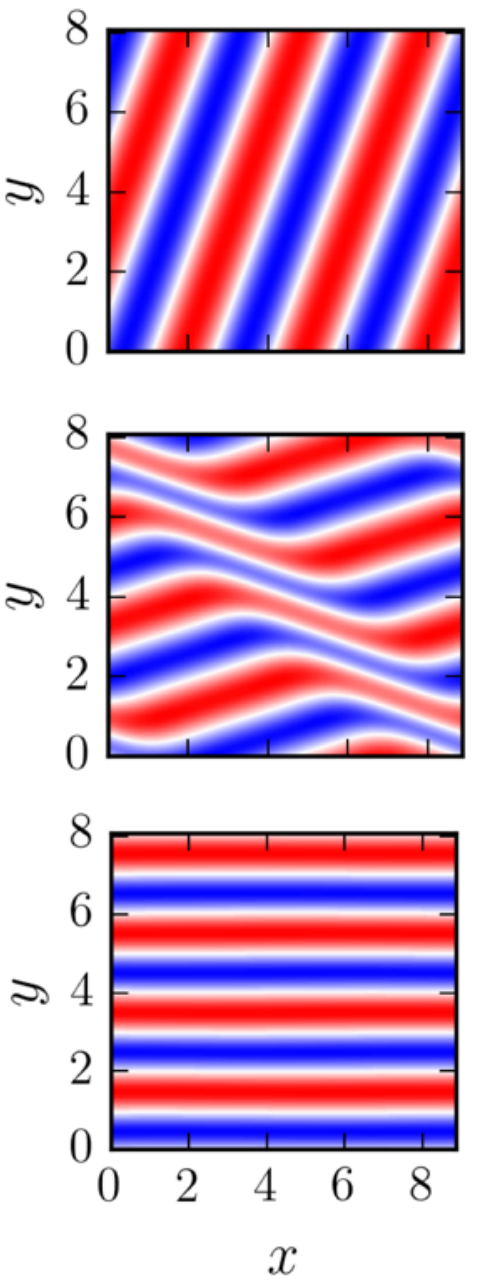}};
   	\draw (-1.4,1.5) node {\textbf{(a)}};
   	\draw (-1.4,-1) node {\textbf{(b)}};
   	\draw (-1.4,-3.5) node {\textbf{(c)}};
		\end{tikzpicture}
       \end{subfigure}
       \begin{subfigure}[b]{0.54\textwidth}
		\begin{tikzpicture}
   	\draw (0, 0) node[inner sep=0] {\includegraphics[width=\linewidth,trim={2.0cm 2.0cm 1.0cm 0cm},clip]{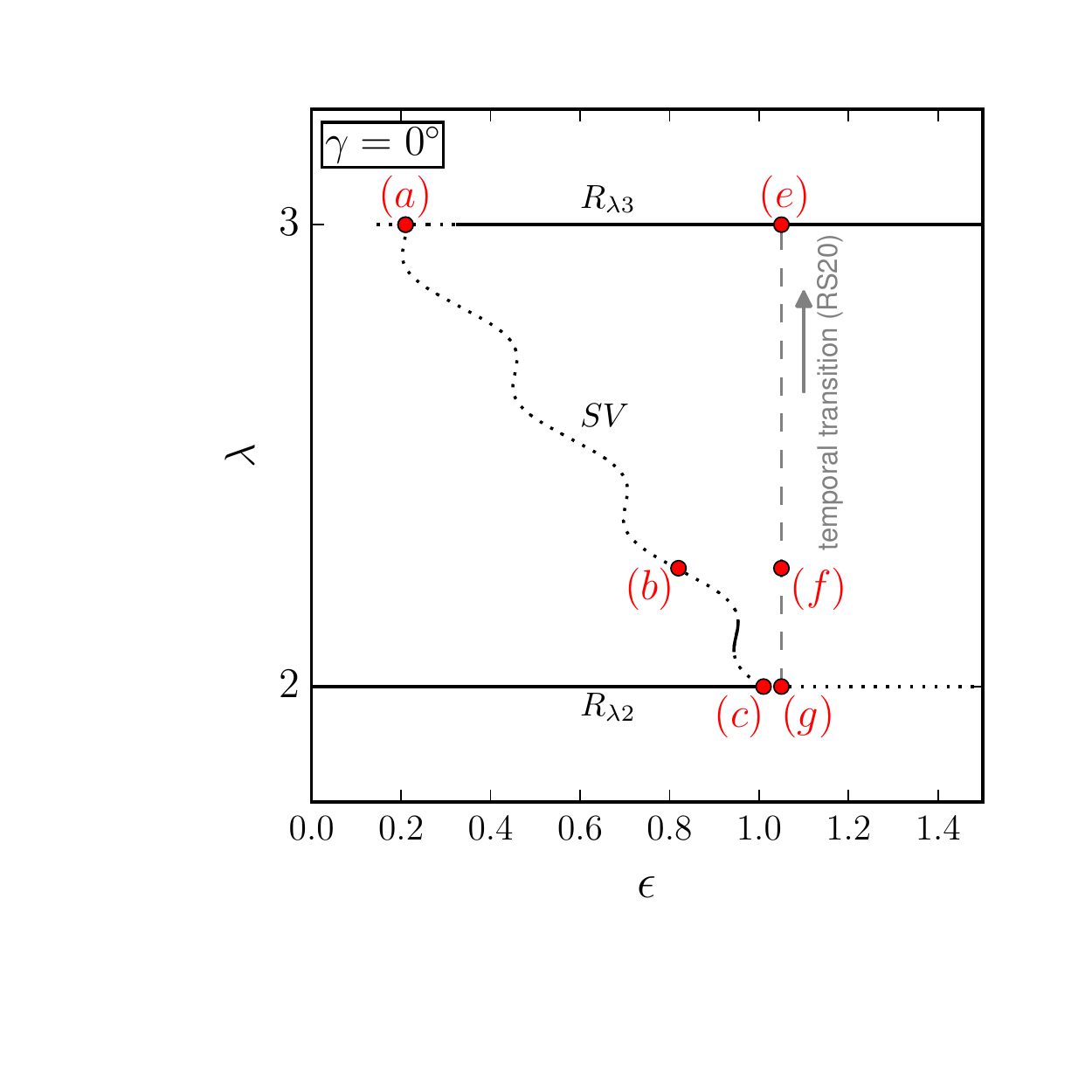}};
   	\draw (-2.9,-3.9) node {\textbf{(d)}};
		\end{tikzpicture}
       \end{subfigure}
       \begin{subfigure}[b]{0.21\textwidth}
		\begin{tikzpicture}
   	\draw (0, 0) node[inner sep=0] {\includegraphics[width=\linewidth,trim={0.0cm 0.0cm -0.3cm 0.0cm},clip]{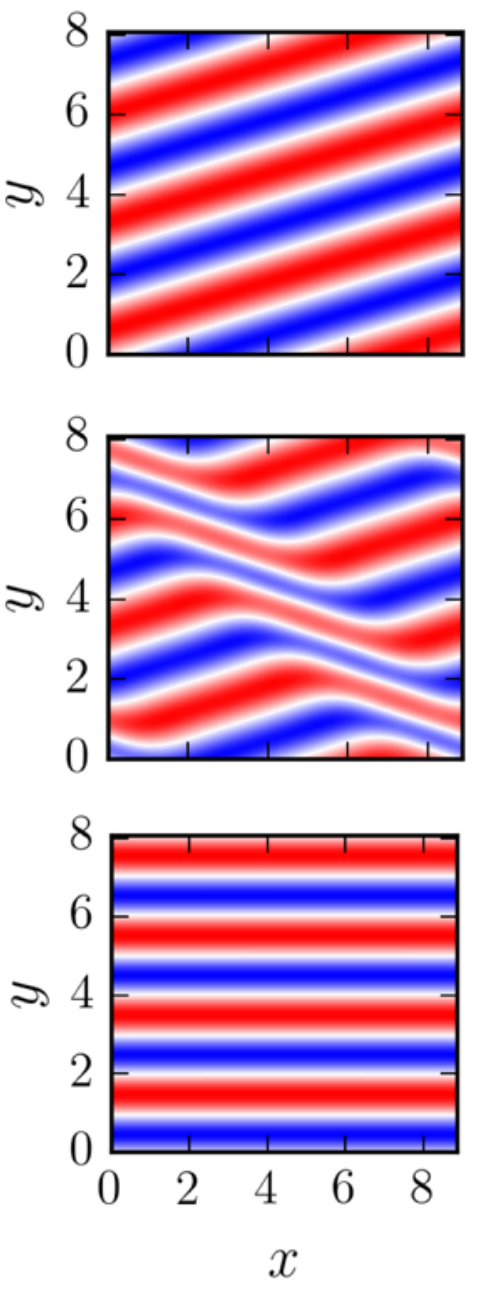}};
   	\draw (-1.4,1.5) node {\textbf{(e)}};
   	\draw (-1.4,-1) node {\textbf{(f)}};
   	\draw (-1.4,-3.5) node {\textbf{(g)}};
		\end{tikzpicture}
       \end{subfigure}
      	\caption{\label{fig:sv} For $\gamma=0^{\circ}$ (Rayleigh-B\'enard), the subcritical bifurcation of the skewed varicose state ($SV$) connects two equilibrium states of straight convection rolls at different wavelength and orientation. Midplane temperature fields of equilibrium states \textbf{(a-c)} are chosen along the $SV$-branch, plotted here in a schematic bifurcation diagram in terms of an approximate dominating pattern wavelength $\lambda$ over $\epsilon$ in \textbf{(d)}. A simulated temporal transition from unstable $R_{\lambda 2}$ to stable $R_{\lambda 3}$ at supercritical $\epsilon>\epsilon_c$ (RS20) is indicated by the dashed line. Snapshots from the temporal transition \textbf{(g,f,e)} show matching skewed varicose patterns between the equilibrium state \textbf{(b)} for $\epsilon<\epsilon_c$ and the transient state \textbf{(f)} at $\epsilon>\epsilon_c$. The orientation of $R_{\lambda 3}$ differs between the terminating bifurcation branch at $\epsilon=0.2$ \textbf{(a)} and the attracted temporal dynamics at $\epsilon=1.05$ \textbf{(e)}.}
\end{figure}

Since the bifurcation branches are very cluttered at $\gamma=0^{\circ}$ in Figure \ref{fig:bifeps3}, we reproduce the bifurcation diagram schematically. In Figure \ref{fig:sv}, the bifurcation branches are plotted in terms of their approximate dominating pattern wavelength $\lambda$ as a function of $\epsilon$. Along the $SV$-branch, convection rolls develop skewed relative orientations (Figure \ref{fig:sv}b) until the rolls pinch-off and reconnect at an oblique orientation (Figure \ref{fig:sv}c). At the bifurcation point, $R_{\lambda 3}$ is rotated by $74.6^{\circ}$ against the orientation of $R_{\lambda 2}$ (Figure \ref{fig:sv}a,c). To link these two different roll orientations, the continuous deformations in the skewed varicose pattern skip two instances for potential reconnection to straight rolls with orientations in between. Each of the potential reconnection points corresponds to a pair of folds along the $SV$-branch. Here, three pairs of folds are observed but this number is specific to the chosen domain size. In between the first two folds at $0.94<\epsilon<0.95$, the $SV$-branch is bistable with $R_{\lambda 2}$ and $R_{\lambda 3}$ in a symmetry subspace of $S_{sv}$. The stability of all branches is indicated by the linestyle. Overall, the bifurcation diagram indicates coexistence of stable (or weakly unstable) $SV$ with stable $R_{\lambda 2}$ and $R_{\lambda 3}$ over a range of $\epsilon$. The coexistence of invariant states suggests spatial coexistence of straight convection rolls and the skewed varicose patterns.

The convection pattern along the $SV$-branch at $\epsilon<\epsilon_c$ may be compared to the convection pattern observed transiently in time along a simulated transition at $\epsilon=1.05>\epsilon_c$ (dashed line in Figure \ref{fig:sv}d and Section 4.3 in RS20). The midplane temperature contours of $SV$ along its subcritical bifurcation branch partly match the transient patterns along the supercritical temporal transition. We find matching patterns at initial instances in time when straight convection rolls are observed (Figure \ref{fig:sv}c,g), and at intermediate time when the transient pattern of skewed varicose pattern emerges (Figure \ref{fig:sv}b,f). Thus, $SV$ indeed captures the observed transient pattern triggered by the skewed varicose instability, but the comparison is for different $\epsilon$. This observation raises the question how the transient temporal dynamics observed above critical thresholds can be related to an equilibrium state existing only below critical thresholds.

\subsection{Subharmonic oscillations - standing and traveling waves}
\label{sec:so}
%%%%%%%%%%%%
%\textbf{Contrast observation and key result:}\\
%\textbf{Summary}. 
%\subsubsection{Subsection summary}
Subharmonic oscillations are observed as standing wave patterns emerging in spatially localized patches that may travel across extended domains \citep{Daniels2000,Subramanian2016}. Here, the periodic orbit $SSW$, underlying the standing wave, is found to coexist with a traveling wave state. Standing and traveling wave states always bifurcate together in equivariant Hopf bifurcations. Both, standing and traveling waves capture observed patterns of spatially subharmonic oscillations. The existence of a traveling wave state explains the observed traveling dynamics of the pattern.

%
%SSW pattern
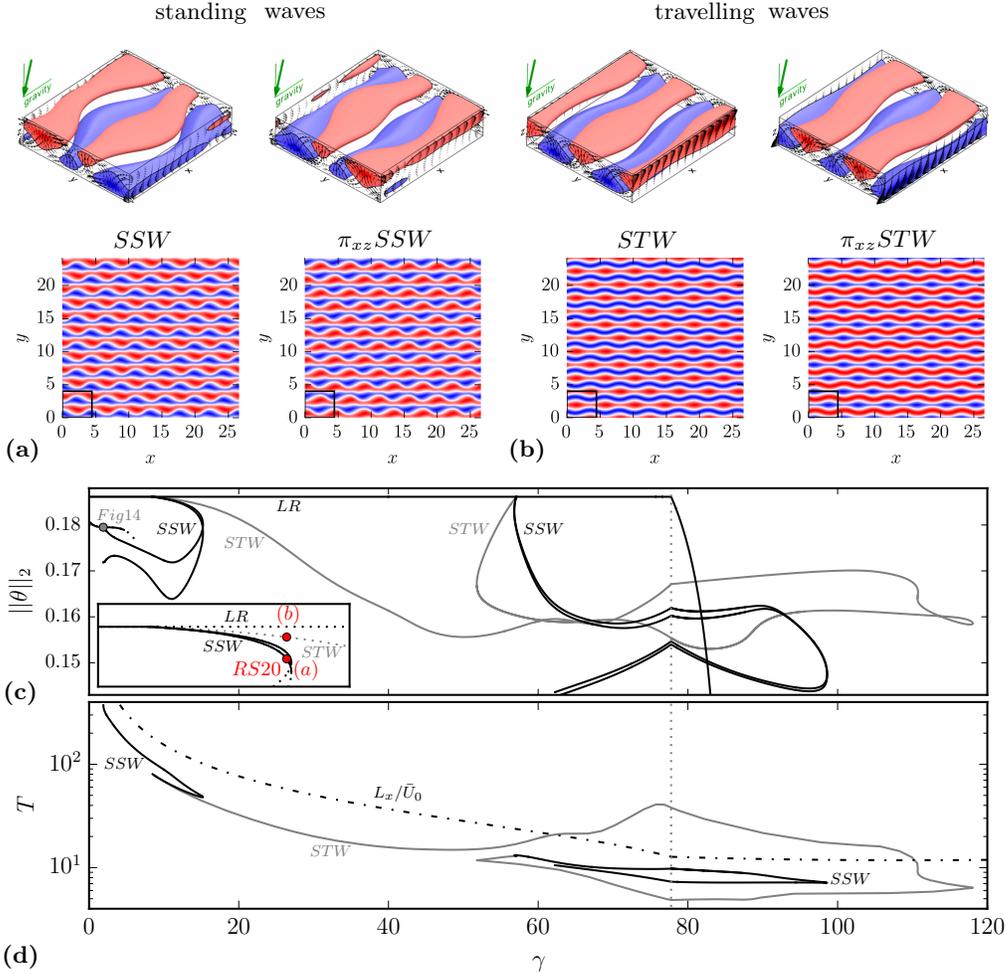
\begin{figure}
\centering
       \begin{subfigure}[b]{0.24\textwidth}
		\begin{tikzpicture}
   	\draw (0, 0) node[inner sep=0] {\includegraphics[width=\linewidth,trim={-4.1cm 1.1cm 0.1cm 1cm},clip]{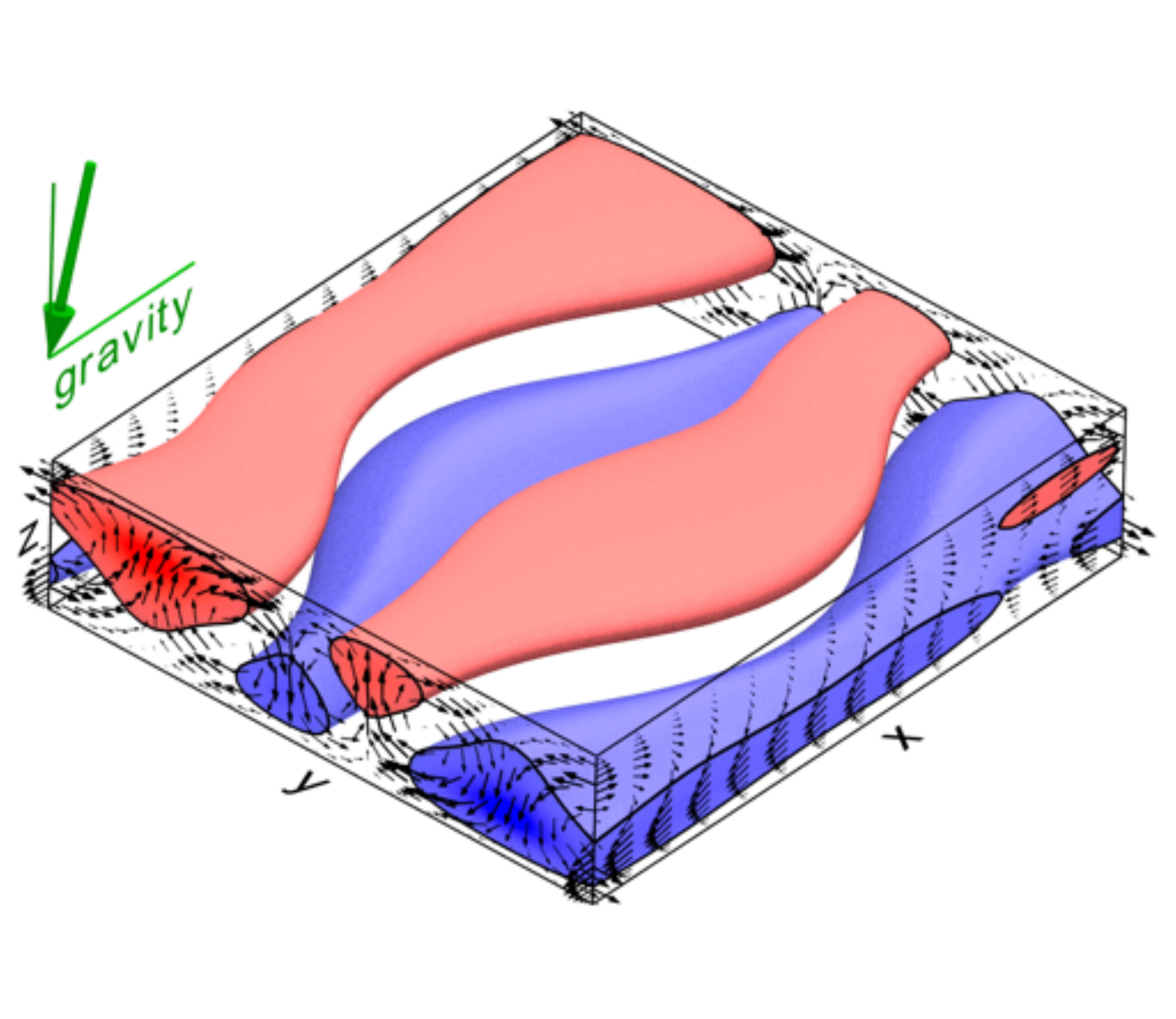}};
   	\draw (1.1,1.5) node {standing};
		\end{tikzpicture}
       \end{subfigure}
       \begin{subfigure}[b]{0.24\textwidth}
		\begin{tikzpicture}
   	\draw (0, 0) node[inner sep=0] {\includegraphics[width=\linewidth,trim={-4.1cm 1.1cm 0.1cm 1cm},clip]{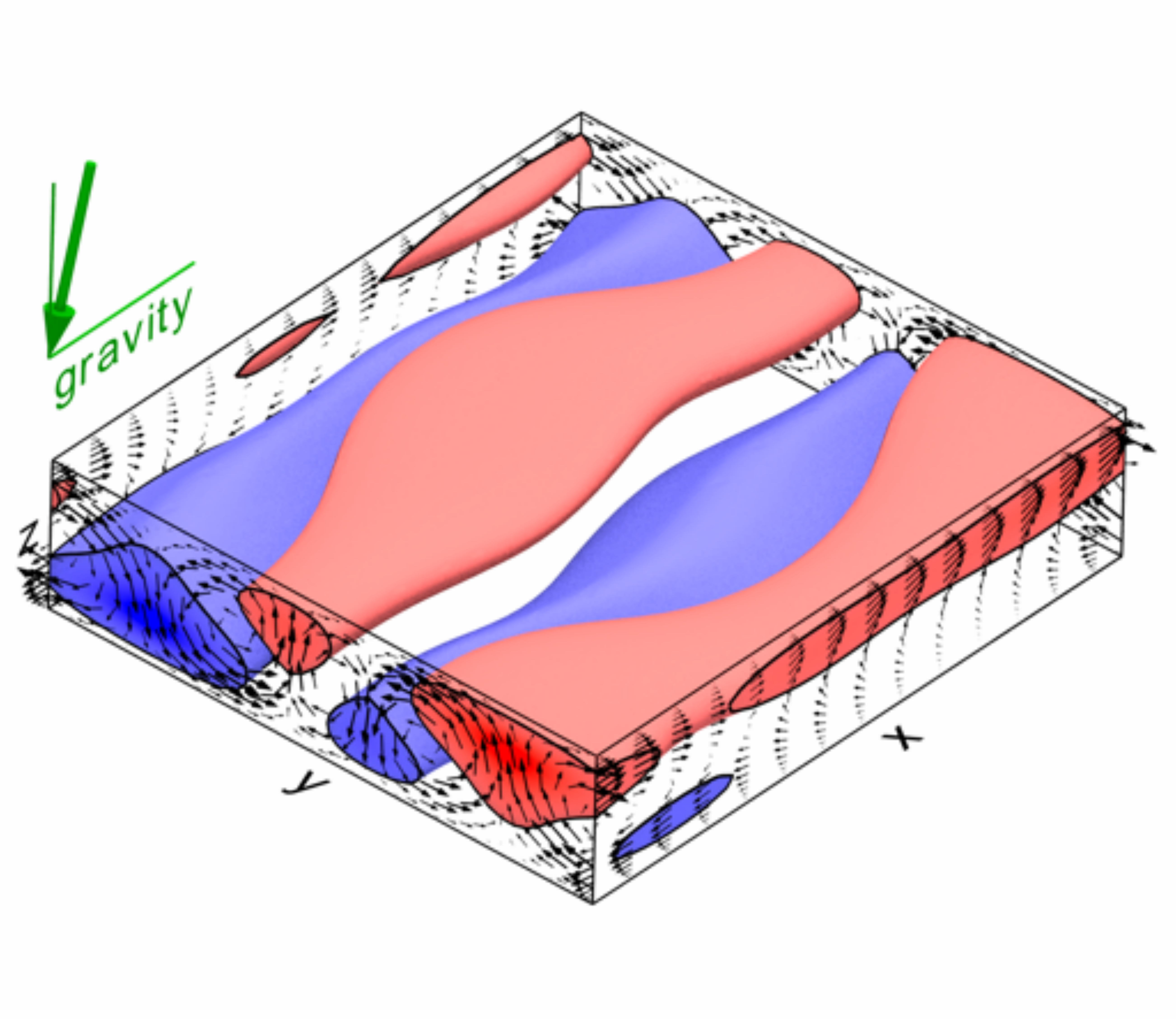}};
   	\draw (-1,1.5) node {waves};
		\end{tikzpicture}
       \end{subfigure}
       \begin{subfigure}[b]{0.24\textwidth}
		\begin{tikzpicture}
   	\draw (0, 0) node[inner sep=0] {\includegraphics[width=\linewidth,trim={-4.1cm 1.1cm 0.1cm 1cm},clip]{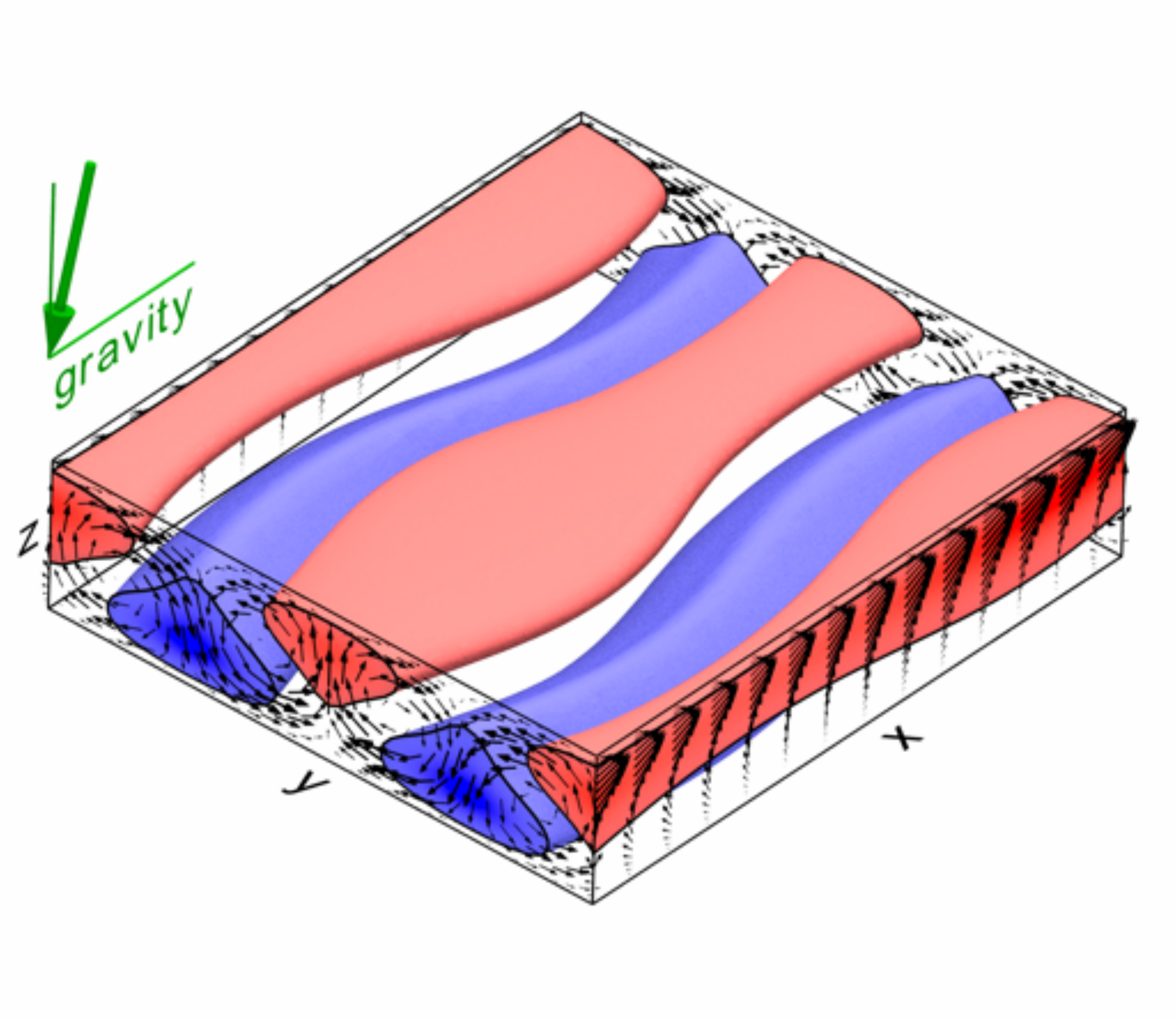}};
   	\draw (1.1,1.5) node {travelling};
		\end{tikzpicture}
       \end{subfigure}
       \begin{subfigure}[b]{0.24\textwidth}
		\begin{tikzpicture}
   	\draw (0, 0) node[inner sep=0] {\includegraphics[width=\linewidth,trim={-4.1cm 1.1cm 0.1cm 1cm},clip]{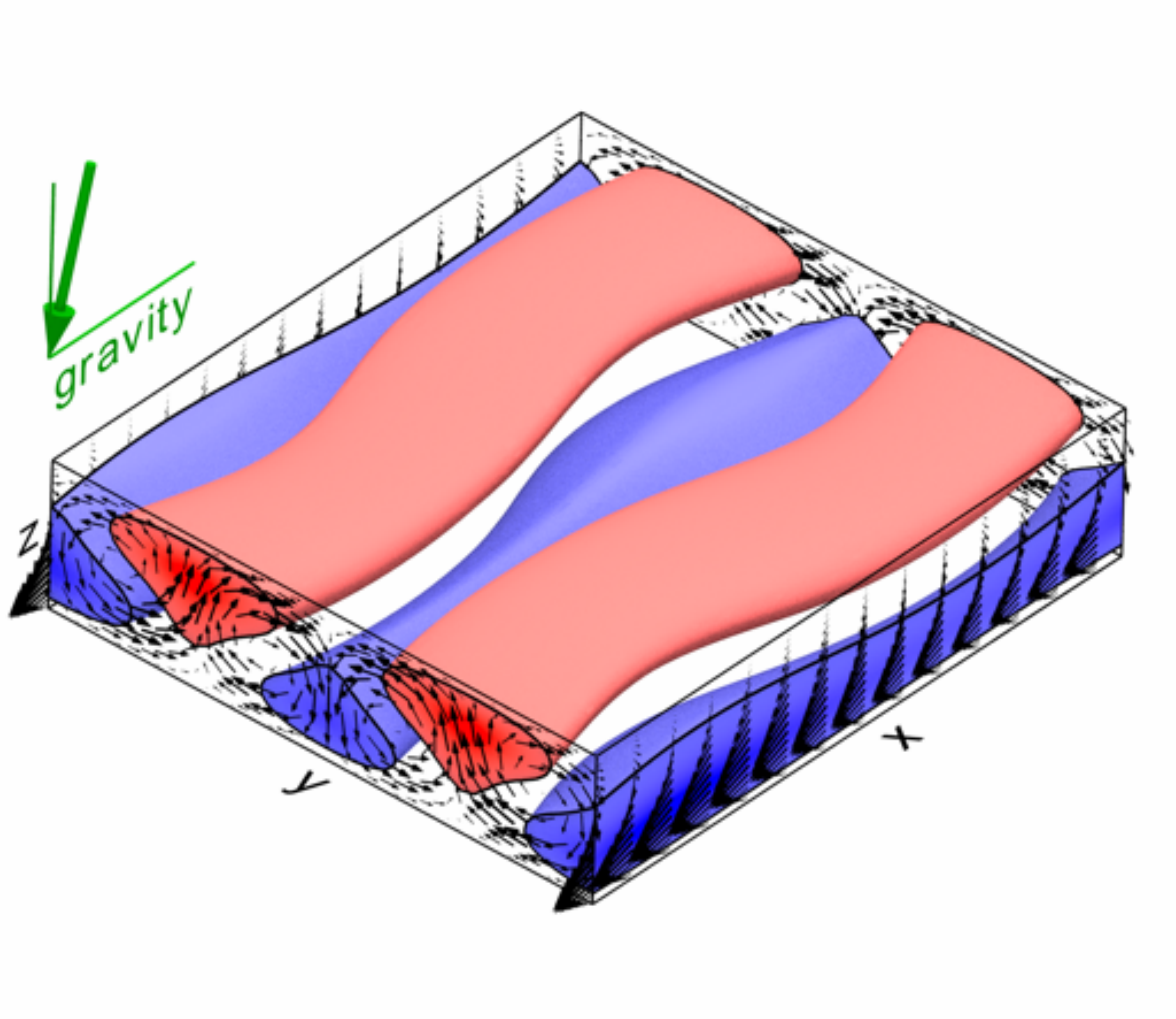}};
   	\draw (-1,1.5) node {waves};
		\end{tikzpicture}
       \end{subfigure}
       \begin{subfigure}[b]{0.24\textwidth}
		\begin{tikzpicture}
   	\draw (0, 0) node[inner sep=0] {\includegraphics[width=\linewidth,trim={0.1cm 0.1cm 0cm 0cm},clip]{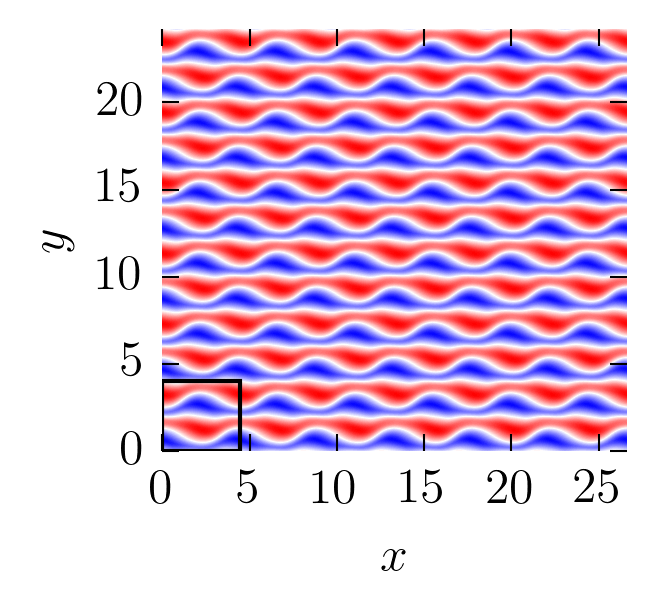}};
   	\draw (-1.4,-1.2) node {\textbf{(a)}};
   	\draw (0.2,1.6) node {$SSW$};
		\end{tikzpicture}
       \end{subfigure}
       \begin{subfigure}[b]{0.24\textwidth}
		\begin{tikzpicture}
   	\draw (0, 0) node[inner sep=0] {\includegraphics[width=\linewidth,trim={0.1cm 0.1cm 0cm 0cm},clip]{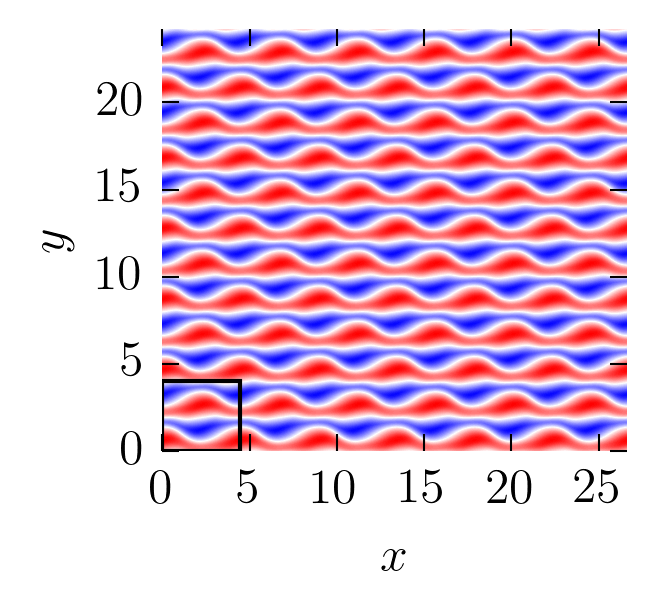}};
   	\draw (0.2,1.6) node {$\pi_{xz}SSW$};
   	%\draw (1.8,-0.9) -- (1.8,1.9);
		\end{tikzpicture}
       \end{subfigure}
       \begin{subfigure}[b]{0.24\textwidth}
		\begin{tikzpicture}
   	\draw (0, 0) node[inner sep=0] {\includegraphics[width=\linewidth,trim={0.1cm 0.1cm 0cm 0cm},clip]{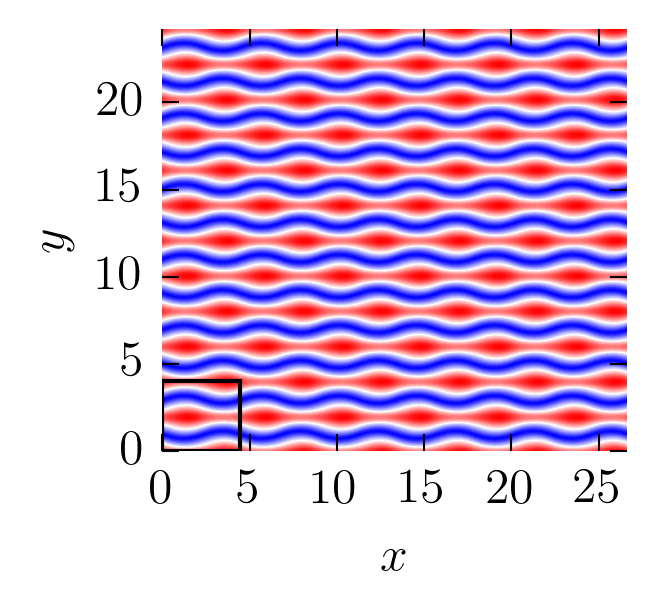}};
   	\draw (-1.4,-1.2) node {\textbf{(b)}};
   	\draw (0.2,1.6) node {$STW$};
		\end{tikzpicture}
       \end{subfigure}
       \begin{subfigure}[b]{0.24\textwidth}
		\begin{tikzpicture}
   	\draw (0, 0) node[inner sep=0] {\includegraphics[width=\linewidth,trim={0.1cm 0.1cm 0cm 0cm},clip]{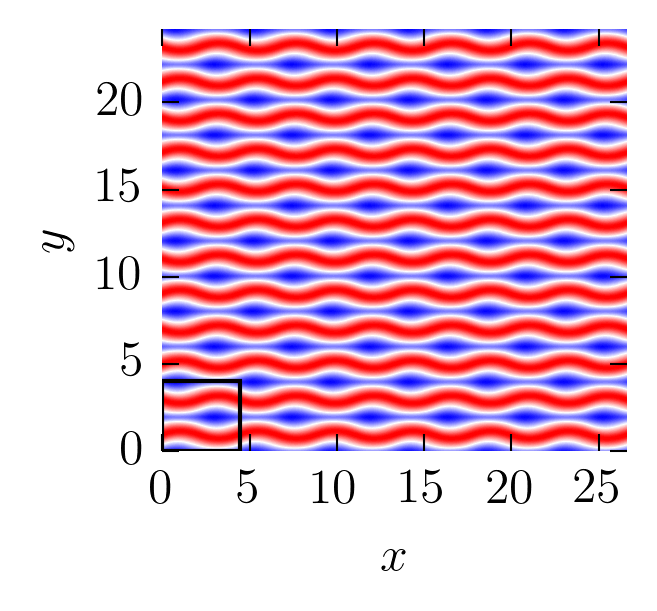}};
   	\draw (0.2,1.6) node {$\pi_{xz}STW$};
		\end{tikzpicture}
       \end{subfigure}
       \begin{subfigure}[b]{0.99\textwidth}
		\begin{tikzpicture}
   	\draw (0, 0) node[inner sep=0] {\includegraphics[width=\linewidth,trim={1.1cm 0.4cm 1.6cm 0.8cm},clip]{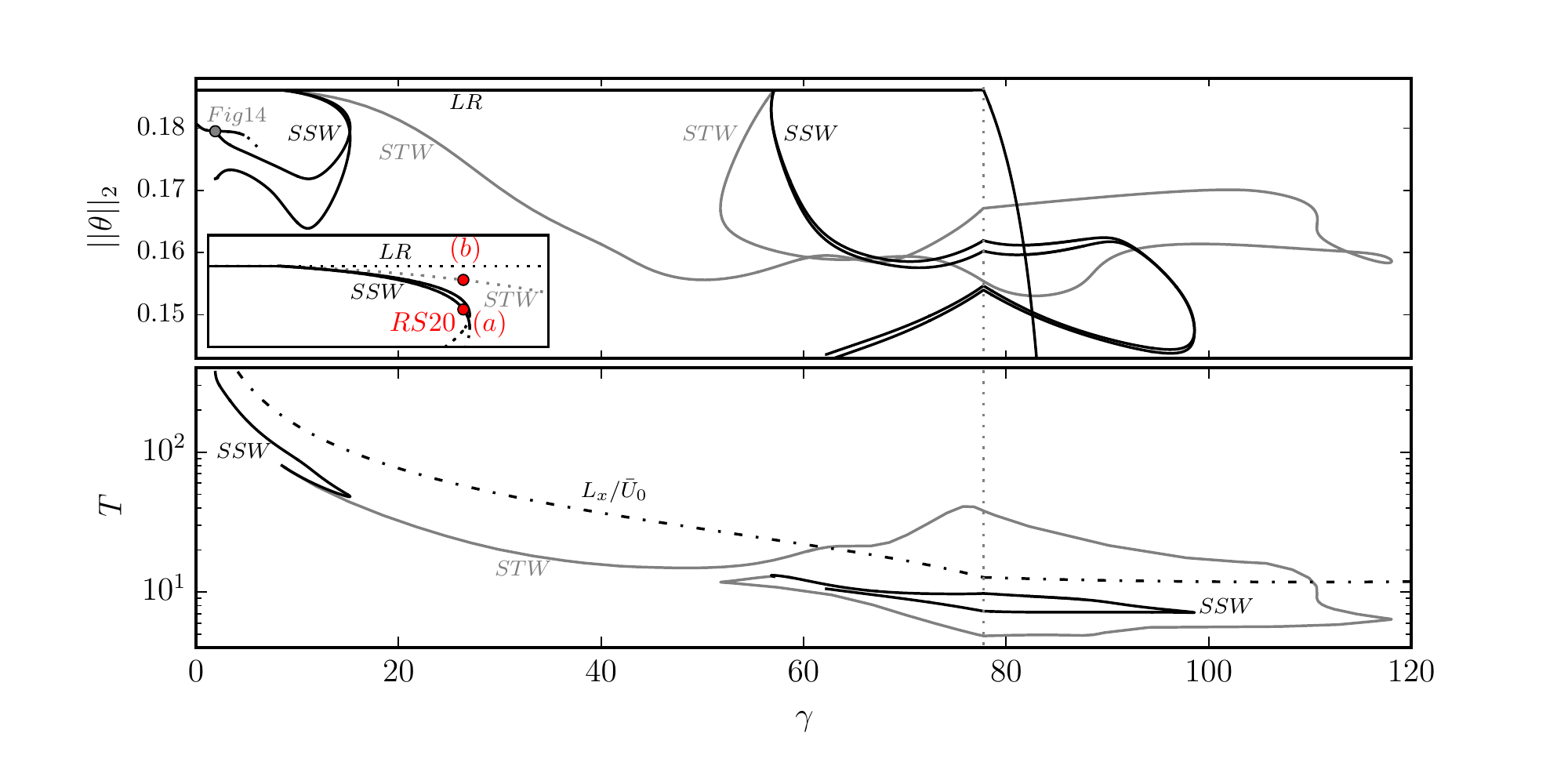}};
   	\draw (-6.5,0.5) node {\textbf{(c)}};
   	\draw (-6.5,-3.0) node {\textbf{(d)}};
		\end{tikzpicture}
       \end{subfigure}
      	\caption{\label{fig:lsw} Symmetry-related pairs of standing wave states $SSW$ \textbf{(a)} and traveling wave states $STW$ \textbf{(b)} bifurcate together in equivariant Hopf-bifurcations from $LR$. The pattern of $SSW$ and $STW$ is visualised by temperature contours in three and two dimensions (midplane). The bifurcation diagram in \textbf{(c)} shows the branches of $SSW$ (black) and $STW$ (grey) at $\epsilon=1.5$. Other branches shown in Figure \ref{fig:bifgamma3} for $\epsilon=1.5$ are suppressed for clarity. The inset indicates stable (solid) and unstable (dotted) branches close to the bifurcation at $\gamma=8.5^{\circ}$ giving rise to the states in \textbf{(a)} and \textbf{(b)} at $\gamma=15^{\circ}$. `RS20' labels the control parameter value where the temporal evolution has been studied in RS20. Panel \textbf{(d)} illustrates how the orbit period $T_{\mathrm{ssw}}$ of $SSW$ and the travel time $T_{\mathrm{lsw}}$ of $STW$ across $L_x$ approximately follow the mean advection time $L_x/\bar{U}_0$ of the laminar base flow $B$ (dashed-dotted line). Note the `hydrolic jumps' where phase and advection velocity match at $\gamma=65^{\circ}$ and at $\gamma=110^{\circ}$ of the $STW$-branch.}
\end{figure}

%Hopf bifurcations to SSW and STW
%\textbf{Detailed discussion of $\epsilon=1.5$:}\\
%required to introduce STW
\subsubsection{Equivariant Hopf bifurcation from longitudinal rolls}
Starting from the periodic orbit $SSW$, found at $[\gamma,\epsilon]=[15^{\circ},1.5]$ in RS20, we continue the branch down in $\gamma$. The $SSW$-branch bifurcates from $LR$ at $[\gamma_c,\epsilon_c]=[8.5^{\circ},1.5]$. This bifurcation is a $\gamma$-forward, supercritical Hopf bifurcation with a critical orbit period of $T_{\mathrm{ssw}}=80.2$. The bifurcation corresponds to four complex eigenvalues crossing the imaginary axis at $\omega_i=\pm 0.078$ controlling the period $T_{\mathrm{ssw}}=2\pi/|\omega_i|$. The Hopf bifurcation to $SSW$ accounts only for one pair of complex neutral eigenmodes. The other pair gives rise to a traveling wave state that we term subharmonic traveling wave ($STW$) with a critical phase speed of $c=L_x/T_{\mathrm{lsw}}=0.055$ where $T_{\mathrm{lsw}}=T_{\mathrm{ssw}}$. Both invariant states, the $\pi_{y}$-symmetric $STW$ and the $\pi_{xyz}$-symmetric $SSW$ (shown in left panels of Fig. \ref{fig:lsw}($a$) and ($b$)), each have a counter-propagating sibling state obtained via $\pi_{xz}$-transformation. Both invariant states capture a subharmonic oscillatory convection pattern invariant under $\tau(0.5,0.5)$. An equilibrium pattern very close to $STW$ can be observed in spatially forced horizontal convection \citep{Weiss2012}. $STW$ also resembles the subharmonic ``sinucose'' state arising from an instability of longitudinal streaks in pure shear flow \citep{Waleffe1997}.

%Explain the equivariant Hopf bifurcation
%\textbf{General discussion of bifurcations:}\\
At parameters where $SSW$ and $STW$ bifurcate locally from $LR$, they share the same bifurcation point (see panels for $10^{\circ}\leq \gamma \leq 60^{\circ}$ in Figure \ref{fig:bifeps3}. This robust feature in the bifurcation diagram is a consequence of equivariant Hopf bifurcations. It is known that Hopf bifurcations that break the $O(2)$-symmetry of a flow must result in two branches originating from the bifurcation: A standing wave branch and a traveling wave branch \citep{Knobloch1986}. At most one of the two branches is stable. In the present case, the Hopf bifurcation breaks the $O(2)$-symmetry of the $x$-uniform state $LR$. The bifurcation at $[\gamma_c,\epsilon_c]=[8.5^{\circ},1.5]$ (inset panel in Figure \ref{fig:lsw}c) has both the branches bifurcating supercritically. $SSW$ is initially stable and $STW$ is unstable. This corresponds to one specific of six discussed cases in \citet{Knobloch1986}. However, here the bifurcation is secondary. While in \citet{Knobloch1986}, bifurcations from a non-patterned two-dimensional primary state with a spatial $O(2)$-symmetry are discussed, here, the bifurcating secondary equilibrium state $LR$ is a three-dimensional state that is symmetric under transformations of $O(2) \times Z_2$. The additional third dimension and the additional reflection symmetry do not affect the conditions necessary for equivariant Hopf bifurcations \citep{Knobloch1986}.

%
%describe continuations along the branches
Continuations of $SSW$ and $STW$ in $\gamma$ reveal their existence over a large range of inclination angles $\gamma$, including $\gamma>90^{\circ}$ where their parent state $LR$ does not exist anymore (Figure \ref{fig:lsw}c). Over this range in $\gamma$, the orbit period of $SSW$ and the propagation time $T_{\mathrm{lsw}}=L_x/c$ of $STW$ with phase velocity $c$ and $L_x=2\lambda_x$ follow approximately the mean laminar advection time $L_x/\bar{U}_0$ (Figure \ref{fig:lsw}d). While the state branch of $STW$ at $\epsilon=1.5$, shown as light grey line in Figure \ref{fig:lsw}c, connects two Hopf bifurcations, one at small and one at large $\gamma$, the $SSW$ branches at $\epsilon=1.5$, shown as black lines in Figure \ref{fig:lsw}c, originating from these two bifurcations remain disconnected. The branch bifurcating forward at $\gamma_c=8.5^{\circ}$ undergoes a fold at $\gamma=15.2^{\circ}$ (Figure \ref{fig:lsw}c) explaining why no temporal state transition from $LR$ to $SSW$ was found at $\gamma=17^{\circ}$ in RS20. The fold destabilizes $SSW$ and connects to a state branch reaching to subcritical parameters where $LR$ is linearly stable. Similar folds occur under $\epsilon$-continuations at $\gamma=10^{\circ}$ and $\gamma=20^{\circ}$. Beyond these folds, $SSW$ state branches terminate and show `loose ends' in the bifurcation diagrams. These terminations correspond to global bifurcations which we explain in the next subsection.
%

%\textbf{Detailed discussion of $\gamma=10$:}\\
\subsubsection{Global bifurcation to subharmonic standing waves}
\begin{figure}
       \begin{subfigure}[b]{0.99\textwidth}
       \begin{tikzpicture}
   	\draw (0, 0) node[inner sep=0] {\includegraphics[width=0.99\linewidth,trim={0cm 0cm 0cm 0cm},clip]{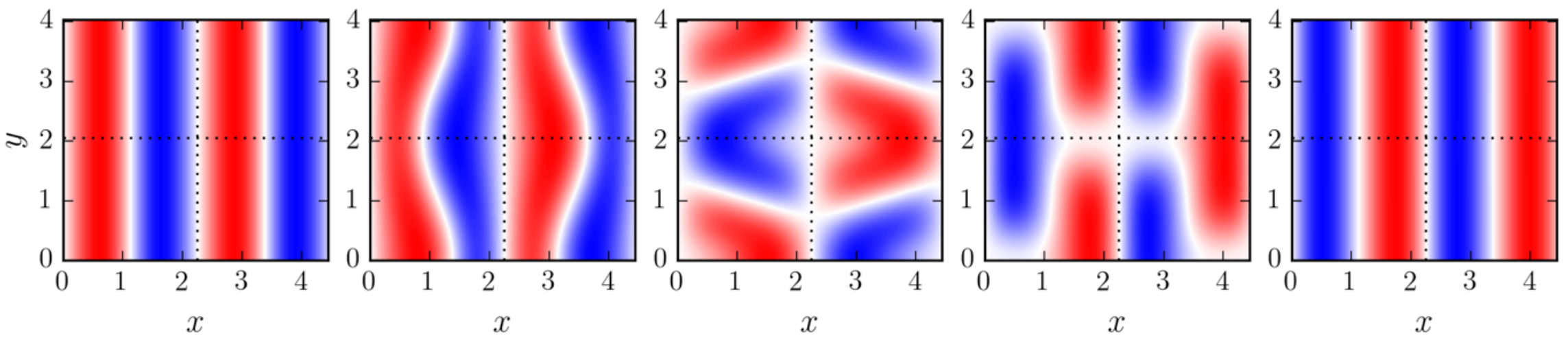}};
   	\draw (-6.4,-1.3) node {\textbf{(a)}};
   	\draw (-3.6,-1.3) node {\textbf{(b)}};
   	\draw (-1.0,-1.3) node {\textbf{(c)}};
   	\draw (1.6,-1.3) node {\textbf{(d)}};
   	\draw (4.2,-1.3) node {\textbf{(e)}};
   	\draw (-5,1.5) node {$TR$};
   	\draw (5.4,1.5) node {$\sigma_{\mathrm{ssw}}TR$};
		\end{tikzpicture}
       \end{subfigure}
       \begin{subfigure}[b]{0.99\textwidth}
       \begin{tikzpicture}
   	\draw (0, 0) node[inner sep=0] {\includegraphics[width=0.99\linewidth,trim={0.5cm 8.9cm 0.5cm 1.5cm},clip]{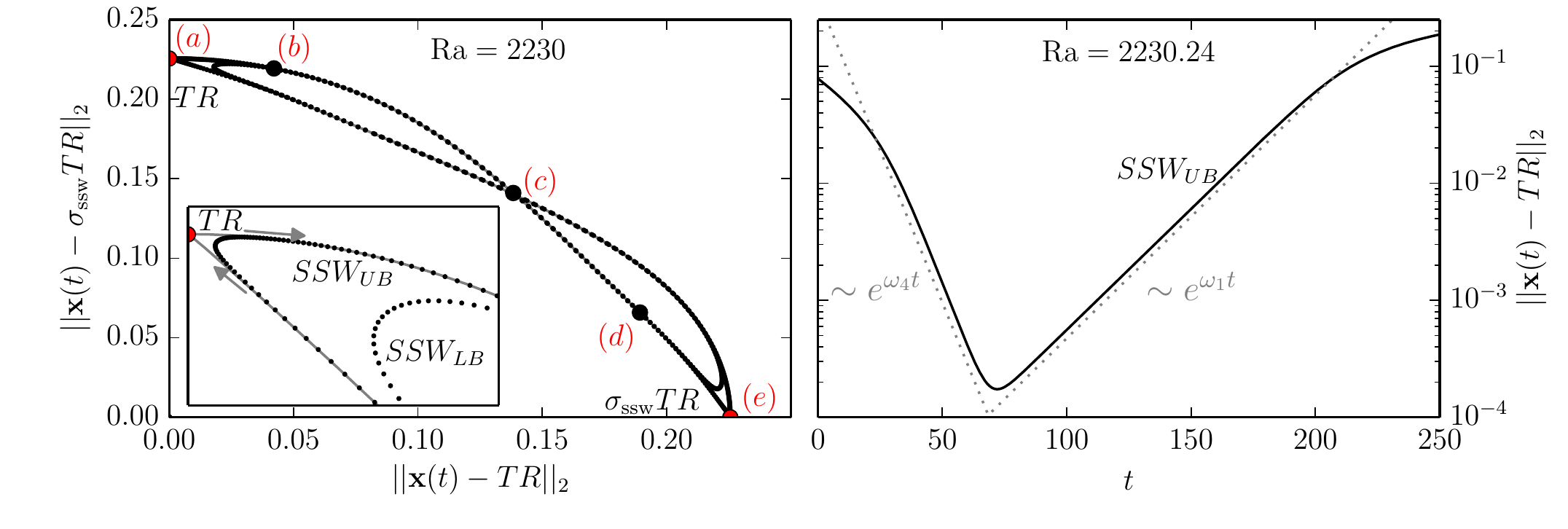}};
   	\draw (-6.2,-1.8) node {\textbf{(f)}};
   	\draw (6.2,-1.8) node {\textbf{(g)}};
		\end{tikzpicture}
       \end{subfigure}
       \begin{subfigure}[b]{0.99\textwidth}
       \begin{tikzpicture}
   	\draw (0, 0) node[inner sep=0] {\includegraphics[width=0.99\linewidth,trim={0cm 0cm 0cm 0cm},clip]{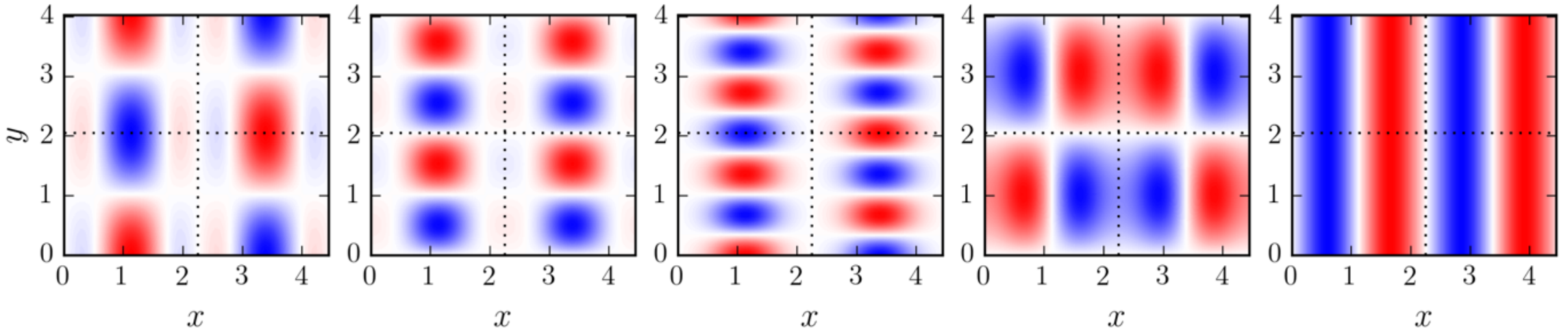}};
   	\draw (-6.4,-1.3) node {\textbf{(h)}};
   	\draw (-4.9,1.5) node {$\bm{e}_1^u$};
   	\draw (-2.3,1.5) node {$\bm{e}_2^u$};
   	\draw (0.3,1.5) node {$\bm{e}_3^s$};
   	\draw (2.8,1.5) node {$\bm{e}_4^s$};
   	\draw (5.4,1.5) node {$\bm{e}_5^s$};
		\end{tikzpicture}
       \end{subfigure}
\caption{\label{fig:hetCycle} The pre-periodic orbit $SSW$ approaches a global bifurcation at $\gamma=10^{\circ}$ and $\mathrm{Ra}_c=2230.25$ ($\epsilon_c=0.286$) where it collides with a robust heteroclinic cyle between $TR$ and the symmetry related equilibrium $\sigma_{ssw}TR$ with $\sigma_{\mathrm{ssw}}=\pi_y \tau(0.25,0.25)$. \textbf{(a-e)} Sequence of midplane temperature contours along the dynamical connection $TR \rightarrow \sigma_{ssw} TR$ at $\mathrm{Ra}=2230$. \textbf{(f)} State space projection of the lower ($LB$) and the upper branch ($UB$) of $SSW$ (cf. Figure \ref{fig:globBif}a). Inset enlarges the orbit trajectories (black dots) near $TR$ and the heteroclinic cycle (grey lines). \textbf{(g)} $L_2$-distance of the $SSW_{UB}$ orbit trajectory from $TR$ at $\mathrm{Ra}=2230.24$ very close to $\mathrm{Ra}_c$. The dynamics of $SSW_{UB}$ is exponential for most of the pre-period and governed by two eigenvalues of $TR$. \textbf{(h)} Midplane temperature contours of two unstable and three stable eigenmodes $\bm{e}^{u/s}_i$ of $TR$ associated to the five leading eigenvalues $[\omega_1,\omega_2,\omega_3,\omega_4,\omega_5]$ $=$ $[0.048,0.045,-0.090,-0.120,-0.138]$ of $TR$.}
\end{figure}
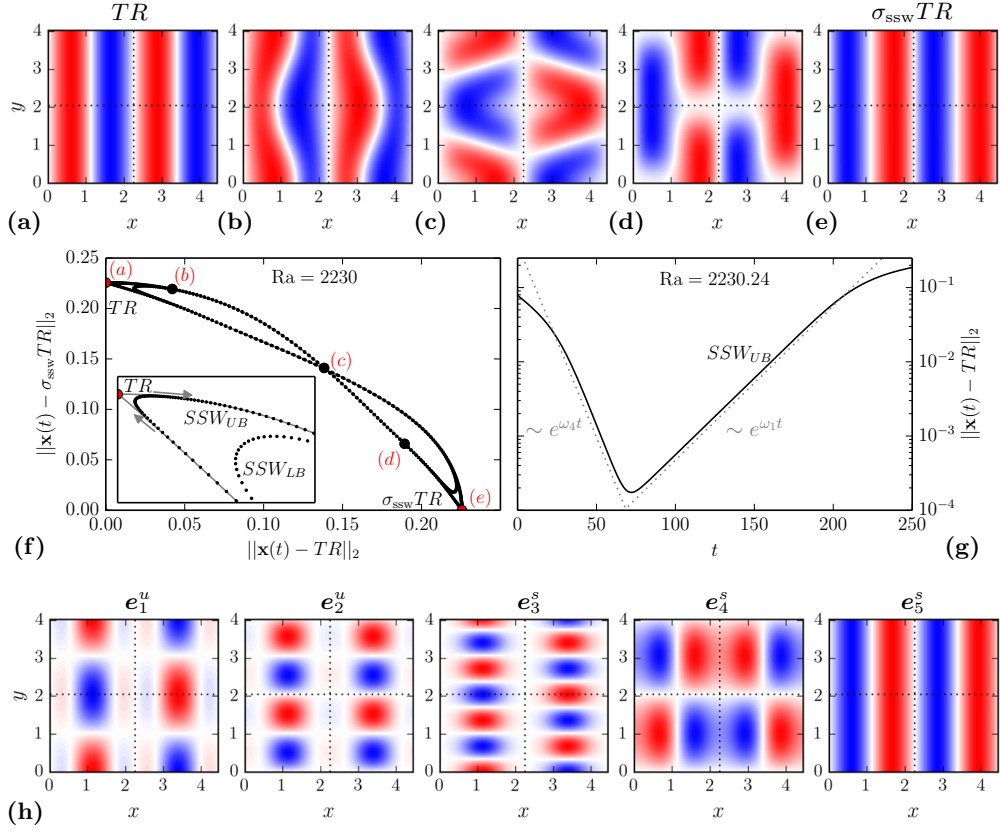

The global bifurcation at $\gamma=10^{\circ}$ occurs at $\mathrm{Ra}_c=2230.25$ ($\epsilon_c=0.286$) where the pre-periodic orbit $SSW$, satisfying (\ref{eq:map}) with $\sigma_{\mathrm{ssw}}=\pi_y \tau(0.25,0.25)$ and a pre-period of $T'=T/4$, collides with a heteroclinic cycle between two symmetry related saddle states $TR$ and $\sigma_{\mathrm{ssw}}TR$ (Figure \ref{fig:hetCycle}). At the global bifurcation point, the spectrum of eigenvalues of $TR$ is computed in a symmetry subspace $\Sigma_0$ given by the $[2\lambda_x,2\lambda_y]$-periodic domain with imposed symmetries of $SSW$, namely $\tau(0.5,0.5)$ and $\pi_{xyz}$. The five leading eigenvalues are real and read $[\omega_1,\omega_2,\omega_3,\omega_4,\omega_5]$ $=$ $[0.048,0.045,-0.090,-0.120,-0.138]$. The midplane temperature contours of the associated eigenmodes $[\bm{e}^u_1,\bm{e}^u_2,\bm{e}^s_3\bm{e}^s_4,\bm{e}^s_5]$ are given in Figure \ref{fig:hetCycle}h. Eigenvalues and eigenmodes of $TR$ do not change significantly when $\mathrm{Ra}$ crosses $\mathrm{Ra}_c$. In contrast to the heteroclinic cycle discussed in RS20, where each of the two symmetry related instances of equilibrium state $OWR$ (further discussed in Section \ref{sec:biffromwr}) has a single unstable eigenmode, the present cycle connects symmetry related instances of $TR$ with two unstable eigenmodes each. Perturbations of $TR$ with the eigenmode $\bm{e}^u_1$ trigger a state transition $TR\rightarrow \sigma_{\mathrm{ssw}}TR$ while perturbations with $\bm{e}^u_2$ lead to $LR$ which is dynamically stable in $\Sigma_0$ at these control parameter values. The symmetry relation between $TR$ and $\sigma_{\mathrm{ssw}} TR$ guarantees that $\sigma_{\mathrm{ssw}} TR$ has the same eigenvalues as $TR$ and symmetry related eigenmodes $\sigma_{\mathrm{ssw}}[\bm{e}^u_1,\bm{e}^u_2,\bm{e}^s_3,\bm{e}^s_4,\bm{e}^s_5]$ allowing for the returning transition $\sigma_{\mathrm{ssw}} TR\rightarrow TR$ to close the heteroclinic cycle.

Direct numerical simulations indicate that states close to the heteroclinic cycle are eventually attracted to $LR$. To show that this heteroclinic cycle is dynamically unstable but structurally stable, we identify two symmetry subspaces of $\Sigma_0$ in which either $TR\rightarrow \sigma_{\mathrm{ssw}}TR$ or $\sigma_{\mathrm{ssw}}TR\rightarrow TR$ exists as heteroclinic connection between an equilibrium with a single unstable eigenmode and a dynamically stable equilibrium. By doing so, the heteroclinic cycle is shown to satisfy all conditions of a structurally stable, or robust, heteroclinic cycle between two symmetry related equilibrium states \citep{Krupa1997}, also discussed in RS20. Subspace $\Sigma_1$ is given by imposing the symmetries in the group $\langle \pi_y,\pi_{xz},\tau(0.5,0.5)\rangle$ and contains the connection $TR\rightarrow \sigma_{\mathrm{ssw}}TR$. Of the five initially considered eigenmodes in $\Sigma_0$, $TR$ in $\Sigma_1$ has still $[\bm{e}^u_1,\bm{e}^s_3,\bm{e}^s_5]$ and $\sigma_{\mathrm{ssw}}TR$ in $\Sigma_1$ has still $\sigma_{\mathrm{ssw}}[\bm{e}^s_4,\bm{e}^s_5]$. Subspace $\Sigma_{\tau}$ is given by imposing the symmetries in the group $\langle \pi_y\tau(0.5,0),\pi_{xz}\tau(0.5,0),\tau(0.5,0.5)\rangle$ and contains the connection $\sigma_{\mathrm{ssw}} TR\rightarrow TR$. Of the five initially considered eigenmodes in $\Sigma_0$, $\sigma_{\mathrm{ssw}}TR$ in $\Sigma_{\tau}$ has still $\sigma_{\mathrm{ssw}}[\bm{e}^u_1,\bm{e}^s_3,\bm{e}^s_5]$ and $TR$ in $\Sigma_{\tau}$ has still $[\bm{e}^s_4,\bm{e}^s_5]$. Using the classification of eigenvalues and associated stability theorem in \citet{Krupa1995}, we identify $\omega_1$ as expanding, $\omega_2$ as transverse, $\omega_4$ as contracting and $\omega_5$ as radial eigenvalue. Eigenvalue $\omega_3$ exists in the same subspace as the expanding eigenvalue $\omega_1$ and therefore does not affect the dynamical stability. Since $\min(-\omega_4,\omega_1-\omega_2)\ngtr \omega_1$, the heteroclinic cycle is not asymptotically stable \citep[][Theorem 2.7]{Krupa1995}.

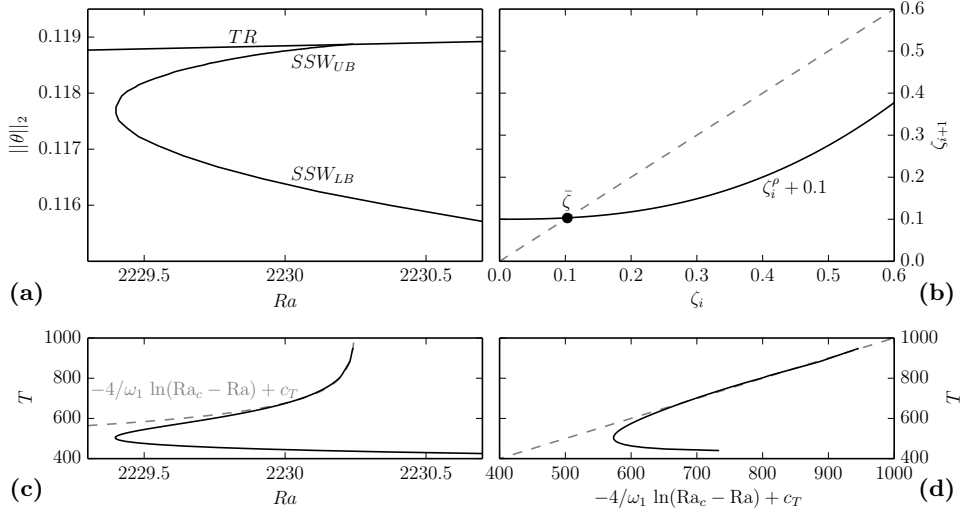
\begin{figure}
       \begin{subfigure}[b]{0.99\textwidth}
       \begin{tikzpicture}
   	\draw (0, 0) node[inner sep=0] {\includegraphics[width=0.99\linewidth,trim={0.5cm 5.0cm 0.5cm 1cm},clip]{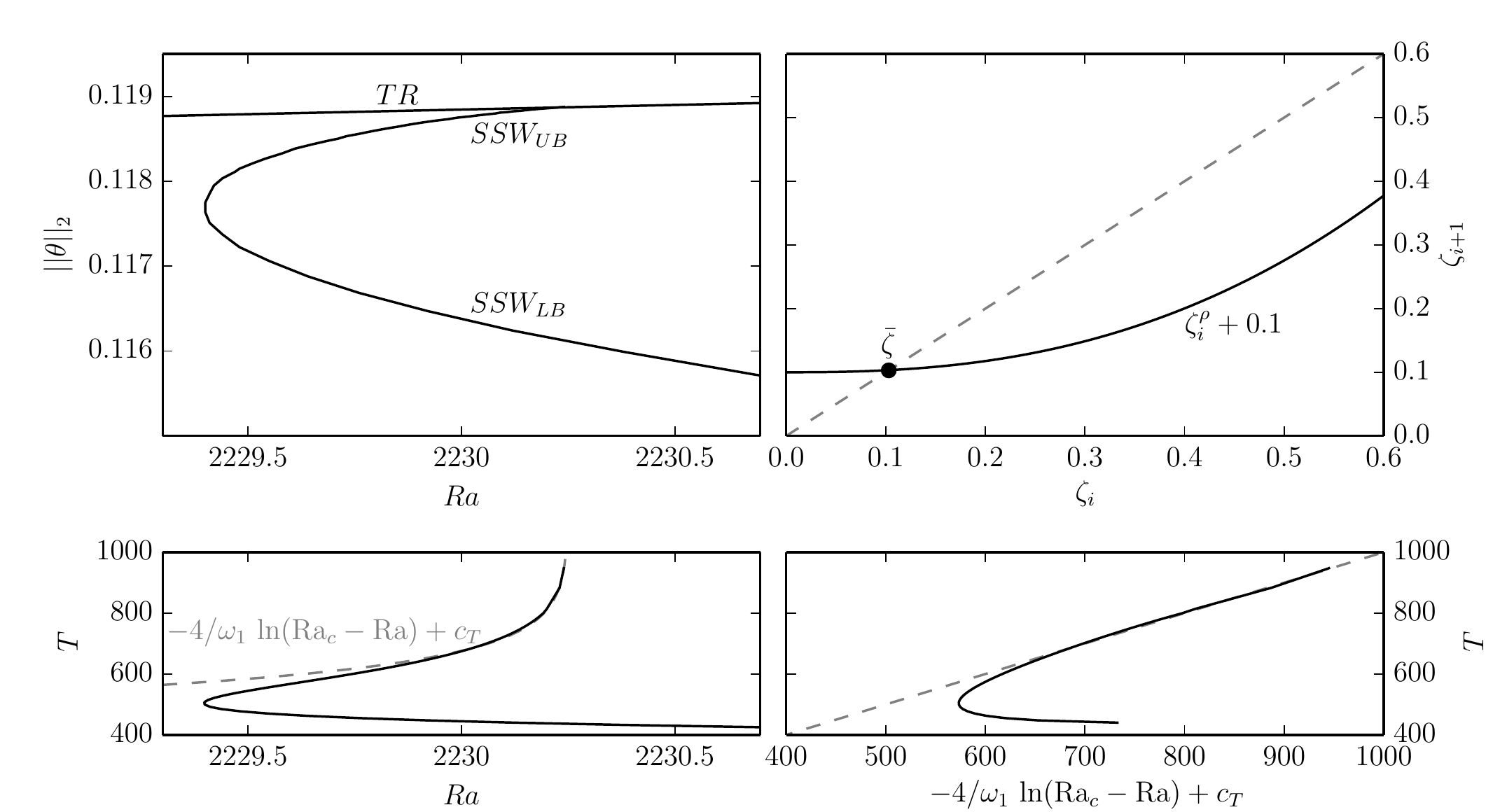}};
   	\draw (-6.0,-0.7) node {\textbf{(a)}};
   	\draw (6.1,-0.7) node {\textbf{(b)}};
   \draw (-6.0,-3.3) node {\textbf{(c)}};
   	\draw (6.1,-3.3) node {\textbf{(d)}};
		\end{tikzpicture}
       \end{subfigure}
\caption{\label{fig:globBif} Before $SSW$ disappears in a global heteroclinic bifurcation at $\gamma=10^{\circ}$ and $\mathrm{Ra}_c=2230.25$ ($\epsilon_c=0.286$), the branch undergoes a saddle node bifurcation. \textbf{(a)} Enlarged bifurcation diagram showing the branch of $SSW$ indicating the maximum $||\theta||_2$ over one cycle (cf. panel $\gamma=10^{\circ}$ in Figure \ref{fig:bifeps3}). \textbf{(b)} Graphical representation of the linearised Poincar\'e map (\ref{eq:hcmap}) for $\mu=0.1$ around the heteroclinic cycle between symmetry related instances of $TR$. The fixed point $\bar{\zeta}$ is stable as the slope of the map at $\bar{\zeta}$ is less than one (dashed line). \textbf{(c,d)} Towards the global bifurcation, the orbit period of $SSW$ approaches an infinite period according to a logarithmic scaling law shown as a function of $\mathrm{Ra}$ in \textbf{(c)} and for rescaled $\mathrm{Ra}$ in \textbf{(d)}. This law follows from the linearised map (\ref{eq:hcmap}), as described in the text. The constant $c_T=560$ is determined by fitting the data close to $\mathrm{Ra}_c$.}
\end{figure}

Before $SSW$ disappears in the global bifurcation at $\mathrm{Ra}_c$, the solution branch undergoes a fold at $\mathrm{Ra}<\mathrm{Ra}_c$ (Figure \ref{fig:globBif}a). The existence of such a fold near a global bifurcation follows from the dynamical stability of the bifurcating periodic orbit relative to the dynamical stability of the heteroclinic cycle. To analyse the stability, we consider the linearised dynamics around the heteroclinic cycle $TR\rightarrow \sigma_{\mathrm{ssw}} TR\rightarrow TR$ at $Ra_c$ and obtain the following Poincar\'e map \citep[see][for a derivation]{Bergeon2002}
\begin{equation}\label{eq:hcmap}
\zeta_{i+1}=c\zeta_i^{\rho}+\mu,\quad \rho=-\frac{\omega_4}{\omega_1}=2.51 \ ,
\end{equation}
with constant $c>0$ and control parameter $\mu \propto \mathrm{Ra}_c-\mathrm{Ra}$. Variable $\zeta_i\ll 1$ describes a local coordinate in a Poincar\'e section located at a distance $\varepsilon\ll 1$ from $TR$ and defining a small perturbation around the state vector of $TR$ as $\bm{x}'=\bm{x}_{TR}+\zeta_i \bm{e}^u_1+\varepsilon\bm{e}^s_4$. The heteroclinic cycle corresponds to $\zeta_i=0$ and is reached at $\mu=0$. The bifurcating periodic orbit is a fixed point $\bar{\zeta}$ of the map such that $\bar{\zeta}=c\bar{\zeta}^{\rho}+\mu$. Since $\rho>1$, a nearby fixed point $\bar{\zeta}$ exists only for $\mu>0$ and $\mathrm{Ra}<\mathrm{Ra}_c$, respectively. The graph in Figure \ref{fig:globBif}b illustrates the map (\ref{eq:hcmap}) and shows that the fixed point $\bar{\zeta}$ is dynamically stable. The stability of $\bar{\zeta}$ assumes that no additional transverse eigendirections are unstable. However, the symmetry subspace that contains $SSW$ also contains $\bm{e}_2^u$ which must be taken into account. Thus, the above analysis predicts that $SSW$ bifurcates with a single unstable eigendirection from the heteroclinic cyle, namely $\bm{e}_2^u$. Since $SSW$ for $\mathrm{Ra}>\mathrm{Ra}_c$ has two unstable eigenmodes, the periodic orbit must undergo a fold prior to the global bifurcation to stabilise the extra unstable eigendirection. Such a fold also exists at $\gamma=20^{\circ}$ (Figure \ref{fig:bifeps3}) but further away from the global bifurcation. Note that unlike the global bifurcation discussed in \citet{Bergeon2002}, the present bifurcation involves first, a heteroclinic cycle between two symmetry related equilibrium states and second, a dynamically unstable periodic orbit such that the fold may have a stabilising effect.

The period of $SSW$ must increases towards an infinite time period as the periodic orbit approaches the heteroclinic cycle. The map (\ref{eq:hcmap}) suggests an asymptotic scaling law for $T$ as a function of $\mathrm{Ra}$ close to the global bifurcation. Since $\zeta\ll1$ and $\rho>1$, the periodic orbit is given by the approximation $\bar{\zeta}\approx \mu$. Over a full period of $SSW$, the orbit trajectory visits both $TR$ and $\sigma_{\mathrm{ssw}} TR$ twice. The time the orbit trajectory spends in the $\varepsilon$-neighbourhood of $TR$ or $\sigma_{\mathrm{ssw}} TR$ dominates the entire orbit period $T$ (Figure \ref{fig:hetCycle}g) such that $T$ satisfies the approximation $\bar{\zeta}\approx \varepsilon \exp(-\omega_1 T/4)$. Hence, the period of $SSW$ is expected to increase as $T\approx -4/\omega_1\ln(\mathrm{Ra}_c-\mathrm{Ra}) + c_T$ with constant $c_T=4\ln(\varepsilon)/\omega_1$ so that asymptotically the period scales as $T\sim -4/\omega_1\ln(\mathrm{Ra}_c-\mathrm{Ra})$. Using a multi-shooting method, $SSW$ is continued close to the global bifurcation. The increasing orbit period confirms the predicted scaling law (Figure \ref{fig:globBif}c,d).

\subsection{Wavy rolls with defects - connecting coexisting state branches}
\label{sec:wavy}
%%%%%%%%%%%%
%key observation to explain by bifurcation analysis
%\textbf{Contrast observation and key result:}\\
%\subsubsection{Subsection summary}
Convection patterns of wavy rolls are observed to quickly incorporate defects in large experimental domains \citep{Daniels2002a,Daniels2008}. These defects may form interfaces between spatially coexisting wavy rolls at different orientations against the base flow. Here, a bifurcation and stability analysis of $WR$ reveals four new equilibrium states, including obliquely oriented states and rolls with defects, that all coexist with $WR$ for the same control parameter values. The multiple states can give rise to the observed spatial coexistence.  

%overview on the WR bifurcations
%\textbf{General discussion of bifurcations:}\\
\subsubsection{Pitchfork bifurcations from longitudinal rolls}
Equilibrium states $WR$ emerge either in pitchfork bifurcations from $LR$ at inclinations $20^{\circ}\le\gamma\le80^{\circ}$ or in saddle-node bifurcations in the absence of $LR$ at $90^{\circ}\le\gamma\le100^{\circ}$. The fact that $WR$ can exist without bifurcating from $LR$ is known from thermal Couette flow \citep{Clever1992}. Almost all computed pitchfork bifurcations from $LR$ to $WR$ are either $\epsilon$-forward or $\gamma$-backward. This observation holds even when the $WR$-branches develop additional folds, like in the bifurcation diagrams at $\gamma=80^{\circ}$ or at $\epsilon=1.5$. The only $\gamma$-forward bifurcation of $WR$ from $LR$ is observed at $\epsilon=0.1$ (Figure \ref{fig:bifgamma3}, $\epsilon=0.1$).

%stability analysis of WR branch at gamma=40 -> DWR -> OR
%\textbf{Detailed discussion of $\gamma=40$:}\\
\subsubsection{Additional bifurcations from wavy rolls}\label{sec:biffromwr}
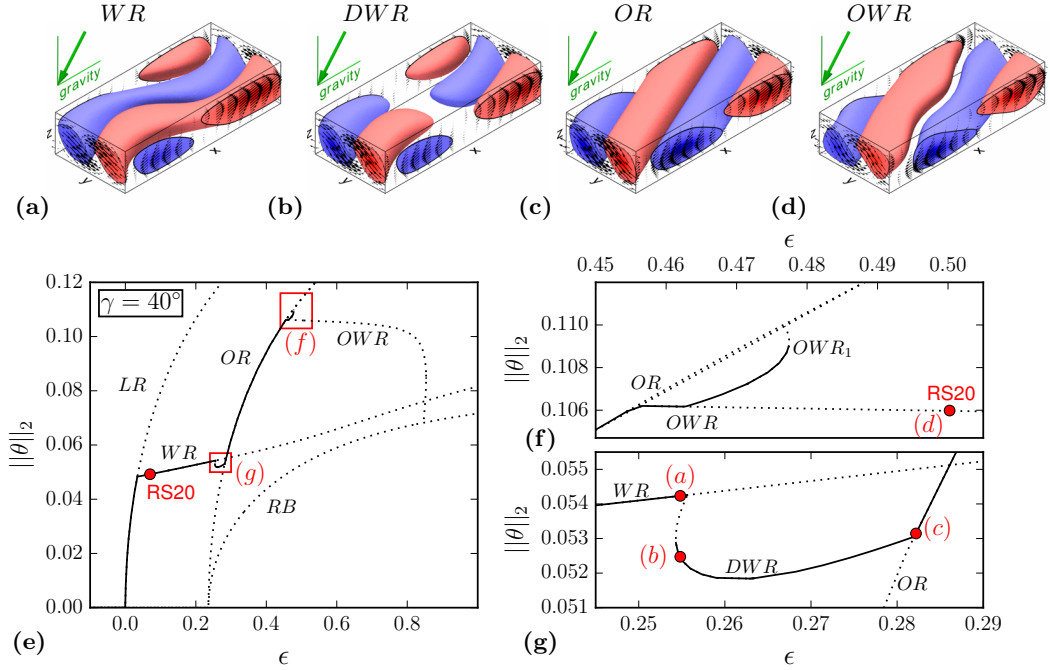
\begin{figure}
\center
       \begin{subfigure}[b]{0.24\textwidth}
       \begin{tikzpicture}
   	\draw (0, 0) node[inner sep=0] {\includegraphics[width=0.99\linewidth,trim={6.8cm 8.0cm 6.8cm 9.0cm},clip]{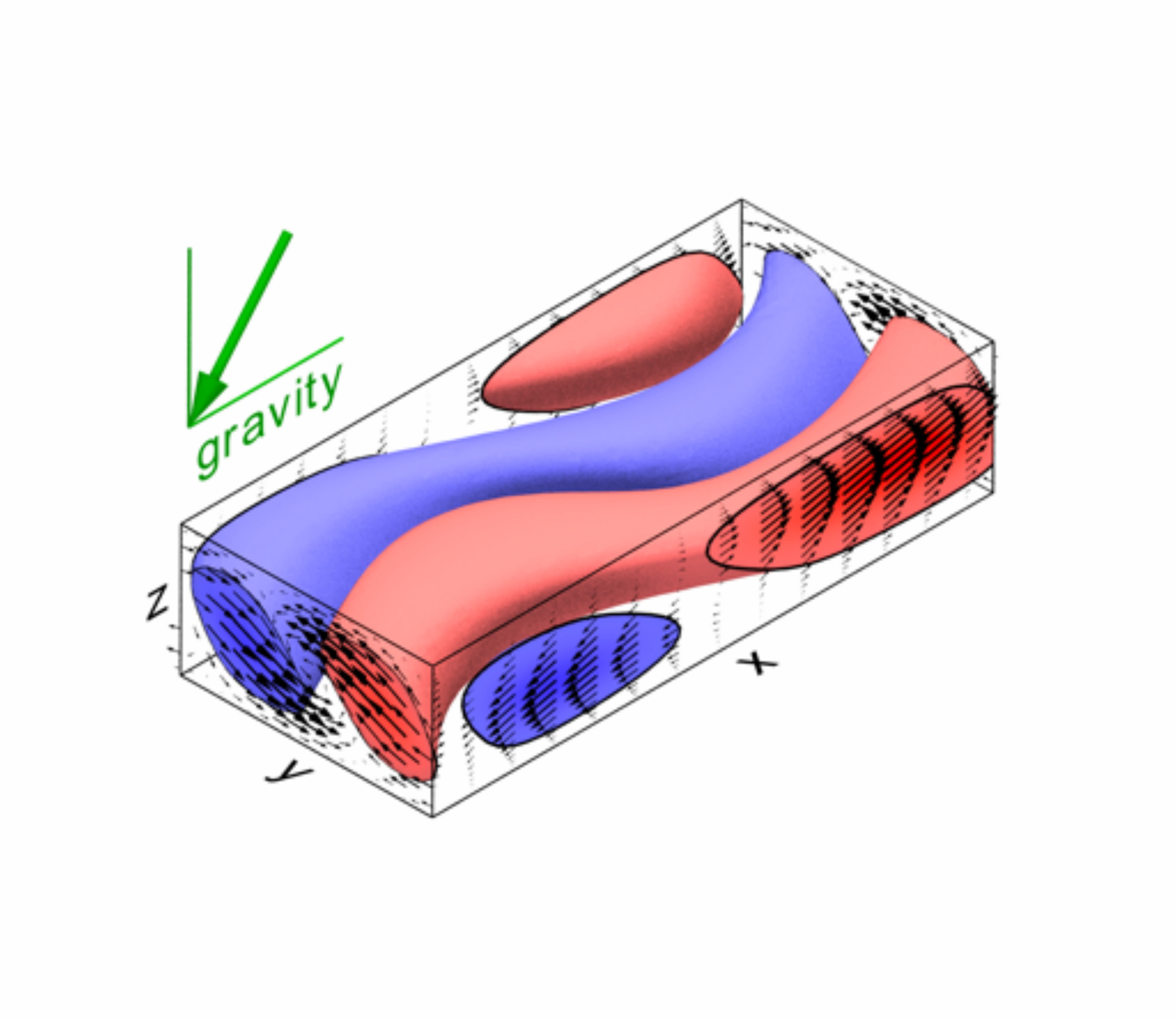}};
   	\draw (-1.8,-1.3) node {\textbf{(a)}};
   	\draw (-0.6,1.3) node {$WR$};
		\end{tikzpicture}
       \end{subfigure}
       \begin{subfigure}[b]{0.24\textwidth}
       \begin{tikzpicture}
   	\draw (0, 0) node[inner sep=0] {\includegraphics[width=0.99\linewidth,trim={6.8cm 8.0cm 6.8cm 9.0cm},clip]{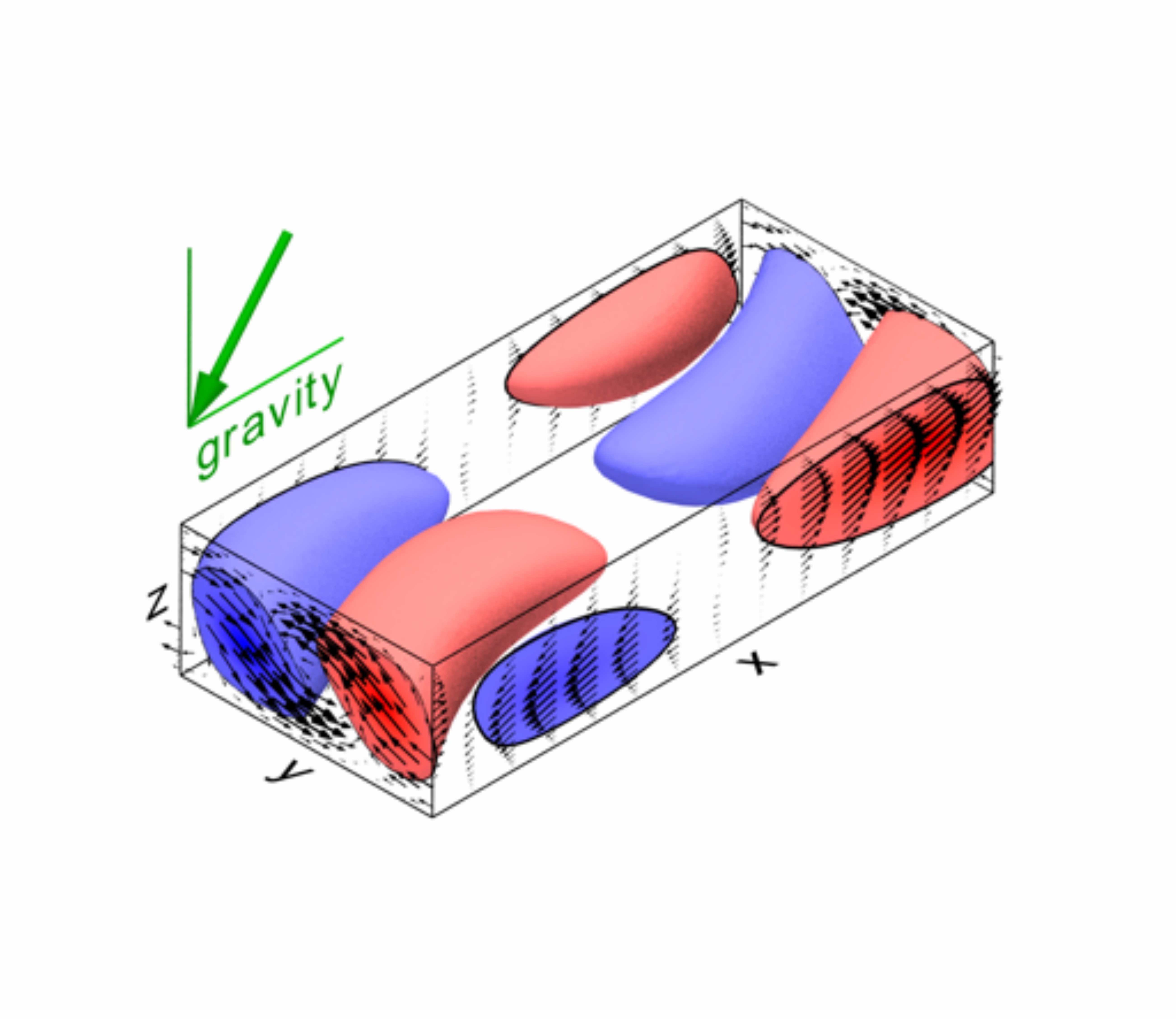}};
   	\draw (-1.9,-1.3) node {\textbf{(b)}};
   	\draw (-0.7,1.3) node {$DWR$};
		\end{tikzpicture}
       \end{subfigure}
       \begin{subfigure}[b]{0.24\textwidth}
       \begin{tikzpicture}
   	\draw (0, 0) node[inner sep=0] {\includegraphics[width=0.99\linewidth,trim={6.8cm 8.0cm 6.8cm 9.0cm},clip]{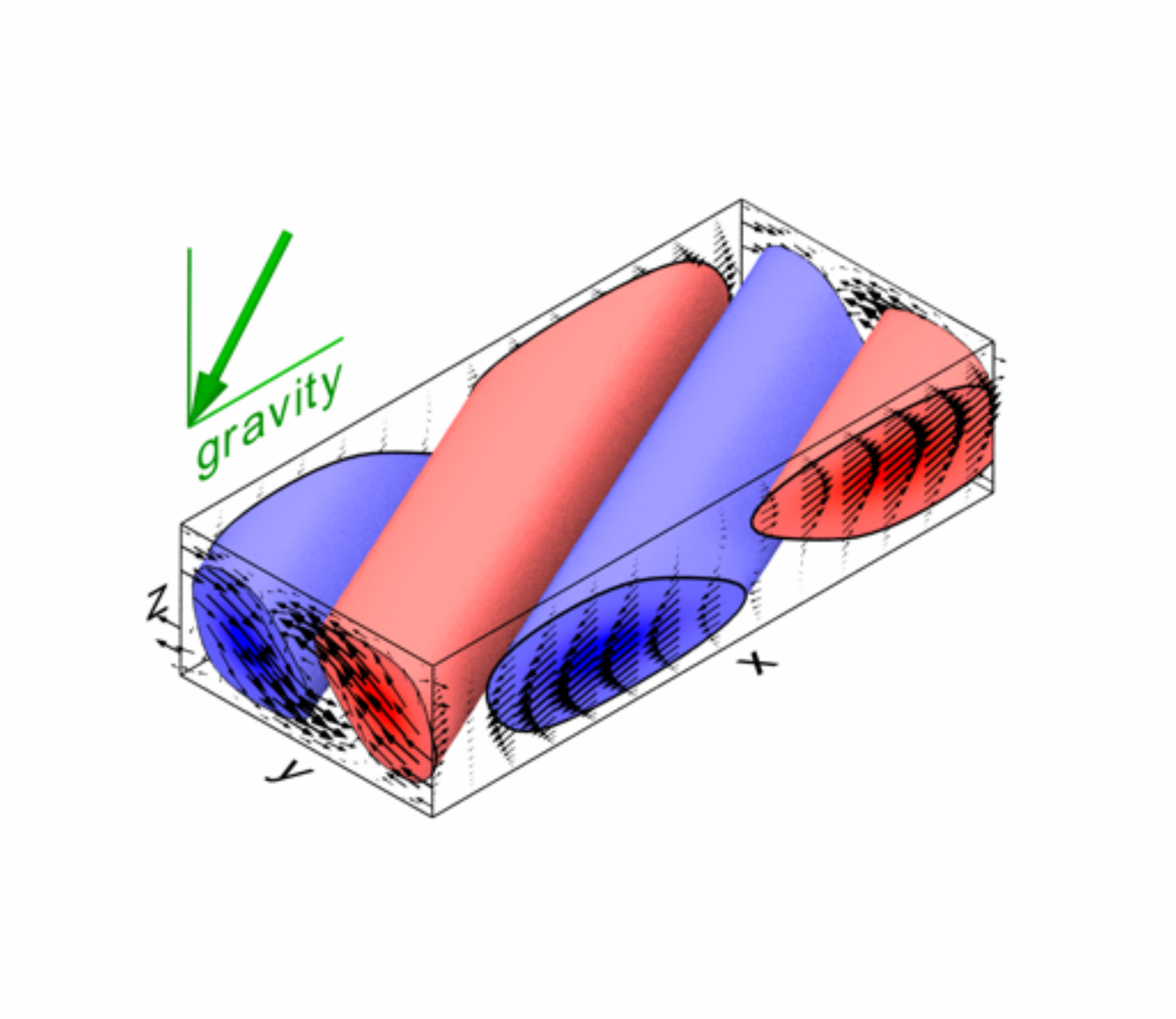}};
   	\draw (-1.9,-1.3) node {\textbf{(c)}};
   	\draw (-0.6,1.3) node {$OR$};
		\end{tikzpicture}
       \end{subfigure}
       \begin{subfigure}[b]{0.24\textwidth}
       \begin{tikzpicture}
   	\draw (0, 0) node[inner sep=0] {\includegraphics[width=0.99\linewidth,trim={6.8cm 8.0cm 6.8cm 9.0cm},clip]{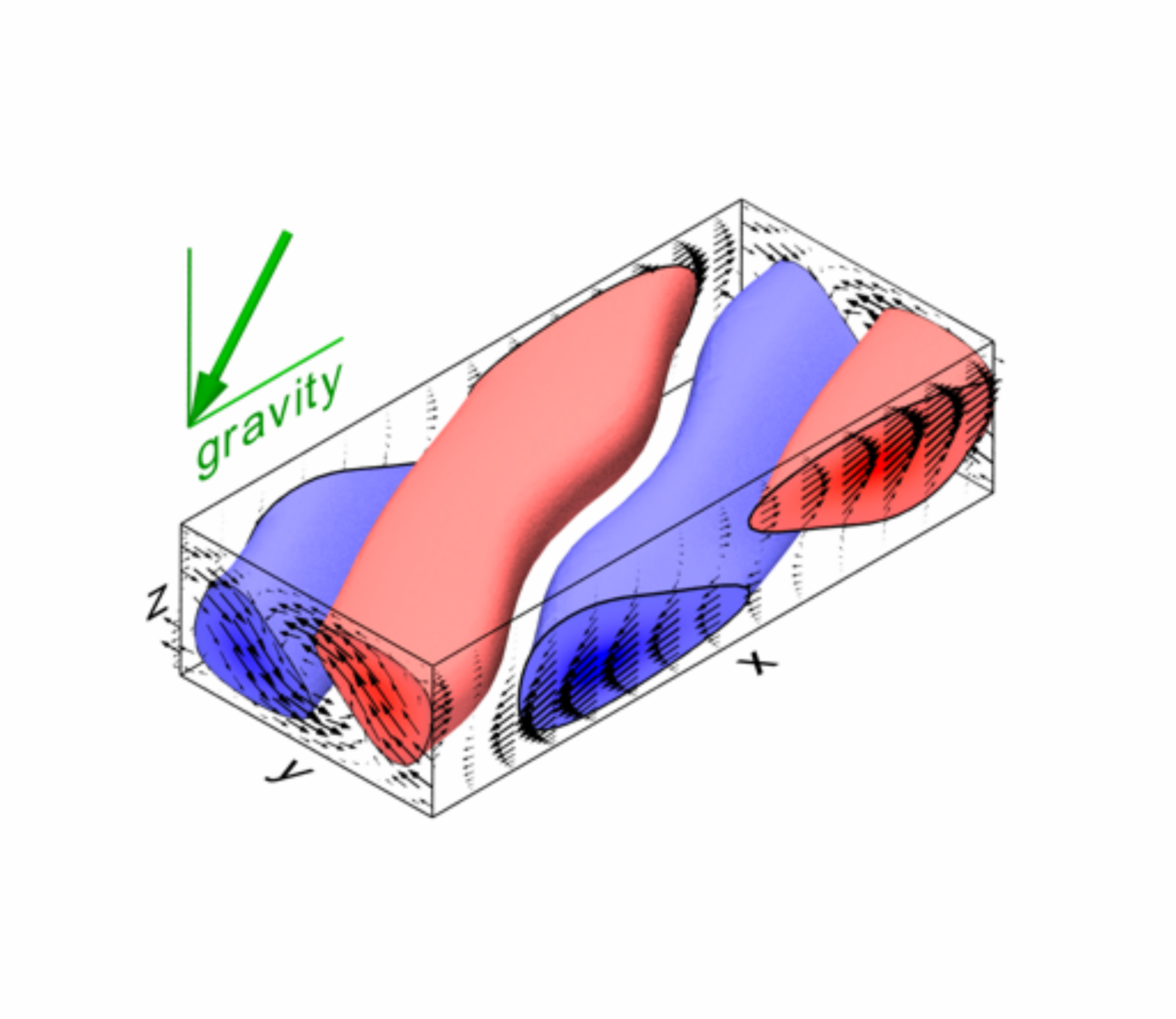}};
   	\draw (-1.9,-1.3) node {\textbf{(d)}};
   	\draw (-0.7,1.3) node {$OWR$};
		\end{tikzpicture}
       \end{subfigure}
       \begin{subfigure}[b]{0.99\textwidth}
       \begin{tikzpicture}
   	\draw (0, 0) node[inner sep=0] {\includegraphics[width=0.99\linewidth,trim={1.5cm 1.0cm 4.2cm 0.0cm},clip]{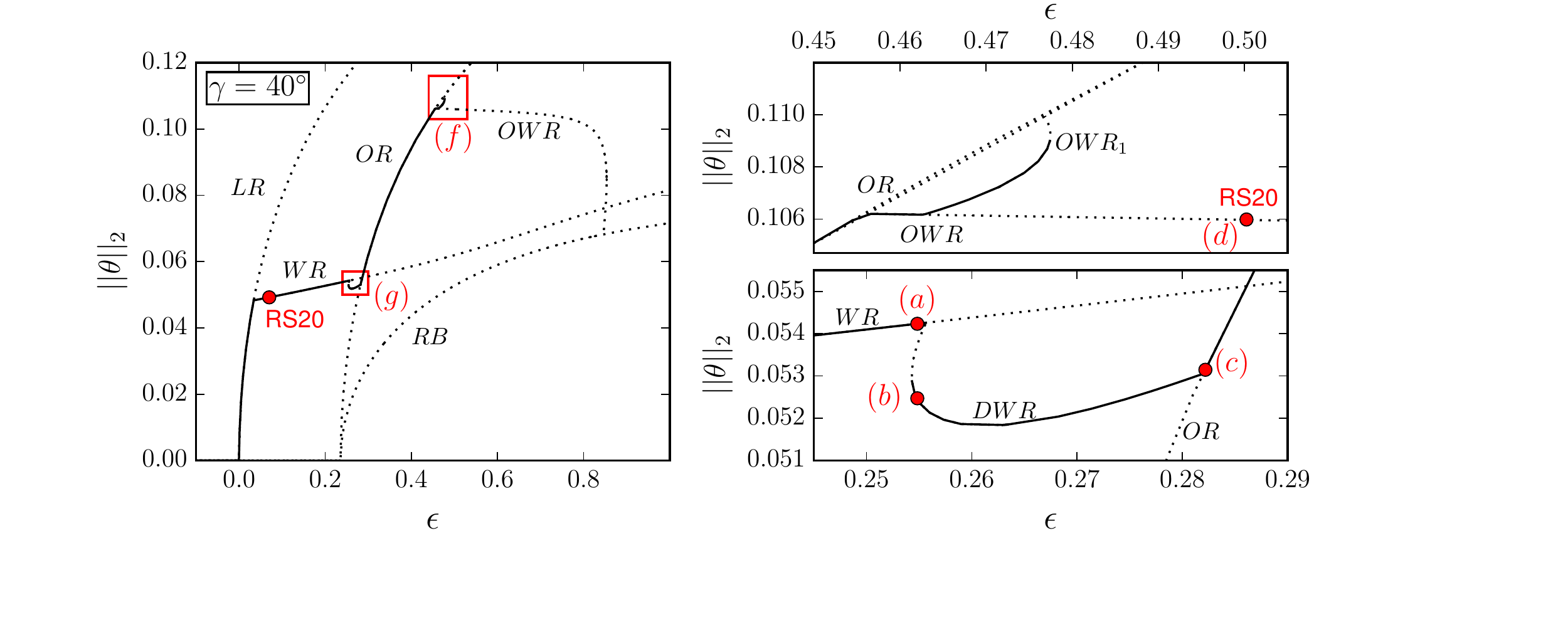}};
   	\draw (-6.3,-2.4) node {\textbf{(e)}};
   	\draw (0.5,0.3) node {\textbf{(f)}};
   	\draw (0.5,-2.4) node {\textbf{(g)}};
		\end{tikzpicture}
       \end{subfigure}
\caption{\label{fig:wrbif} Bifurcation sequence from wavy rolls to oblique wavy rolls \textbf{(a-d)}. Bifurcation diagram \textbf{(e)} shows how seven equilibrium states are connected over $\epsilon$ and indicates stable (solid) and unstable (dotted) parts of the branches. Dynamical stability is computed by Arnoldi iteration in a $(2\lambda_x,\lambda_y)$-periodic domain. Panel \textbf{(g)} enlarges the parameter region where $DWR$ connects $WR$ with $OR$. Panel \textbf{(f)} enlarges the parameter region where $OWR_1$ with pattern wavenumber $m=1$ connects $OWR$ with pattern wavenumber $m=2$ with $OR$. `RS20' labels the control parameter value at which the temporal evolution has been studied in RS20.}
\end{figure}

In most cases, $WR$-branches continue to large $\epsilon$. This does not imply the absence of connections to other invariant states. When $WR$ lose stability, additional bifurcations occur. 
We demonstrate the increasing number of invariant states and their patterns by following the higher order instabilities of $WR$ at $\gamma=40^{\circ}$. Arnoldi iteration for $WR$ indicates the stability of the $WR$-branch up to $\epsilon=0.256$ in a $[2\lambda_x,\lambda_y]$-periodic domain (Figure \ref{fig:wrbif}g). At this point, a subcritical pitchfork bifurcation breaks the $\pi_{xz}$- and $\pi_y$-symmetry and gives rise to an equilibrium state showing disconnected wavy rolls and named $DWR$. $DWR$ is invariant under $\pi_{xyz}$. Following the $DWR$-branch from the pitchfork bifurcation, it undergoes a saddle-node bifurcation at $\epsilon=0.254$, becomes bistable with $WR$ and connects to an equilibrium state of oblique rolls ($OR$) that continues as a stable branch to larger values of $\epsilon$. $DWR$ represents stationary roll defects that along its branch breaks the topology of rolls in a double periodic domain (Figure \ref{fig:wrbif}b), and connects convection rolls at different orientations. 

%WAVY ROLLS

%discuss OR and OWR
$OR$ is a secondary state bifurcating from the laminar flow $B$. The oblique orientation at angle $\tan^{-1}(\lambda_y/2\lambda_x)$ against the laminar flow direction coincides with the diagonal of the $[2\lambda_x,\lambda_y]$-periodic domain (Figure \ref{fig:wrbif}c). $OR$ is invariant under transformations of $S_{\mathrm{or}}=\langle \pi_{xyz},\tau(a_x,\pm a_x)\rangle$ with $a_x\in [0,1)$ corresponding to an $O(2)$-symmetric state. The sign in front of the continuous shift factor $a_x$ differs between left $(+)$ and right $(-)$ oblique rolls, with $OR^l=\pi_y OR^r$.  When $OR$ becomes unstable, an $\epsilon$-forward pitchfork bifurcation at $\epsilon=0.456$ gives rise to stable oblique wavy rolls ($OWR$), see Figure \ref{fig:wrbif}d. As $OR$, $OWR$ can have left or right orientation. The symmetry group of $OWR$ is $S_{\mathrm{owr}}=\langle \pi_{xyz},\tau(0.5,\pm 0.5)\rangle$. Equilibrium $OWR$ with a wavy pattern of wavenumber $m=2$ along the domain diagonal loses stability at $\epsilon=0.463$ to an equilibrium $OWR_1$ with a pattern wavenumber of $m=1$ and broken $\tau(0.5,\pm 0.5)$-symmetry. The branch of $OWR_1$ undergoes a saddle-node bifurcation at $\epsilon=0.477$ and terminates on $OR$ at $\epsilon=0.476$ (Figure \ref{fig:wrbif}f). The small $\epsilon$-range with $\Delta \epsilon=0.476-0.456=0.02$ between the two symmetry-breaking bifurcations of $OWR_1$ with $m=1$ and $OWR$ with $m=2$ from $OR$ suggests a nearby codimension-2 point with spatial 1:2 resonance. When stable $OWR_1$ disappear in the saddle-node bifurcation, the temporal dynamics becomes attracted to a robust heteroclinic cycle between unstable instances of $OWR$ that is discussed in RS20. The branch of $OWR$ continues as unstable branch until it terminates at $\epsilon=0.8$ on ribbons ($RB$), an unstable equilibrium state bifurcating together with $OR$ from $B$ in an equivariant pitchfork bifurcation (Figure \ref{fig:wrbif}e). The detailed properties of $RB$ are discussed in the following section, but we already note that $RB$ shows disconnected rolls or plumes, similar to $DWR$. Thus, we find two instances of bifurcation sequences that may be described as ``straight rolls bifurcate to wavy rolls, wavy rolls bifurcate to disconnected rolls''. In one instance the sequence happens for longitudinal orientation and in the other instance for oblique orientation. The fact that all of the above states coexist with the $WR$ branch at equal control parameters explains that all patterns represented by the invariant states can spatially coexist with wavy rolls in large domains.

\subsection{Knots and ribbons - two different types of bimodal states}
\label{sec:bim}
%%%%%%%%%%%%
%\textbf{Contrast observation and key result:}\\
%\subsubsection{Subsection summary}
Observations of knot patterns exist in horizontal convection \citep{Busse1974,Busse1979} and inclined layer convection \citep{Daniels2000}. They have been described as `bimodal convection' in both cases. Here, the properties of equilibrium states for knots are compared to ribbons, a bimodal state identified in the previous section. A decomposition along their bifurcation branches implies that knots and ribbons are bimodal states that fundamentally differ in their bifurcation structure.

%
%\textbf{General discussion of $KN$ bifurcations:}\\
\subsubsection{Bifurcations to states for knots and ribbons}
$KN$ at $\gamma=80^{\circ}$ bifurcates $\epsilon$-forward from $TR$. At smaller $||\theta||_2$ than $TR$, $KN$ continues without folds and terminates in $\epsilon$-backward bifurcations from $LR$ (Figure \ref{fig:bifeps3}, $\gamma=80^{\circ}$). This bifurcation sequence requires $\gamma>\gamma_{c2}$ and was previously analysed using two-mode interactions \citep{Fujimura1993}. At $\gamma=90^{\circ}$, $LR$ does not exist at finite $\epsilon$ and the $KN$-branch terminates in a bifurcation from an equilibrium state we term subharmonic lambda plumes ($SL$) and briefly discuss in Appendix \ref{sec:app:additional}.

%
%\textbf{General discussion of $RB$ bifurcations:}\\
$RB$ is an equilibrium state found via continuing the state branches of $OWR$ and $WR$ that terminate in $\epsilon$-bifurcations from $RB$ at $\gamma=40^{\circ}$ and $\gamma=50^{\circ}$, respectively (Figure \ref{fig:bifeps3}). $RB$ is invariant under transformations of $S_{\mathrm{rb}}=\langle \pi_y,\pi_{xz},\tau(0.5,0.5)\rangle$. Neither experiments nor simulations of ILC observe the pattern of $RB$ as dynamically stable pattern at the considered parameters. However, we refer to experimental observations of ``ribbons'' in Taylor-Couette flow \citep{Tagg1989}. Ribbons in Taylor-Couette flow are analogous to ribbons in ILC. In Taylor-Couette flow, they bifurcate together with oblique spirals \citep{Chossat1994} and are connected via oblique wavy cross-spirals \citep{Pinter2006}, two states that are comparable to $OR$ and $OWR$ in ILC. As in the Taylor-Couette flow, $OR$ and $RB$ in ILC bifurcate robustly together in equivariant bifurcations \citep{Knobloch1986} and are connected via $OWR$. However, $OR$ and $RB$ are stationary states and their bifurcation is an equivariant pitchfork bifurcation, unlike equivariant Hopf bifurcations such as those found in Taylor-Couette flow, and those discussed in Section \ref{sec:so}.

%
%\textbf{Introduce the bimodal concept:}\\
\subsubsection{Decomposition in terms of straight convection rolls}
As a consequence of the stationary equivariant bifurcation, the linear relation $\bm{x}_{RB}=\alpha\, \bm{x}_{OR}^l + \beta\, \bm{x}_{OR}^r$ with $\alpha=\beta$ holds at the bifurcation points, where  $\bm{x}_{RB}$ indicates the state space vector of $RB$, and $\bm{x}_{OR}^l$ and $\bm{x}_{OR}^r$ are the state space vectors of $OR^l$ and $OR^r$, respectively. This linear decomposition is valid for all parameters, where $RB$ and $OR$ bifurcate from laminar flow $B$. Since $RB$ emerges as linear superposition of two differently oriented straight convection rolls, we call $RB$ a `bimodal state'. The term `bimodal' has been used previously to describe knot patterns of straight convection rolls at orthogonal orientations in experiments of Rayleigh-B\'enard convection \citep{Busse1974} and in experiments of ILC for $\gamma>\gamma_{c2}$ \citep{Daniels2000}. In line with previously used terminology, we describe $KN$ and $RB$ both as bimodal states. However, there are fundamental differences between $KN$ and $RB$ bimodal states that are illustrated by the subsequently discussed decomposition analysis.

We consider a bimodal equilibrium state vector $\mathbf{b}(\epsilon)$ depending on continuation parameter $\epsilon$ as decomposition
\begin{align}
\mathbf{b}(\epsilon)=\alpha(\epsilon)\, \mathbf{m}_1(\epsilon) + \beta(\epsilon)\, \mathbf{m}_2(\epsilon) + \mathbf{d}(\epsilon) \label{eq:bimodaldef}
\end{align}
where $\alpha,\beta\in\mathbb{R}$ and $\mathbf{m}_1,\mathbf{m}_2$ are state vectors of two differently oriented straight convection rolls. $\mathbf{d}$ is the difference vector that is necessary to create the composite state $\mathbf{b}$. We simplify the notation by suppressing the dependence of the decomposition on $\epsilon$. We seek the optimal coefficients $\alpha$ and $\beta$ such that $||\mathbf{d}||_2$ is minimal. The optimality condition $0=\partial_\alpha ||\mathbf{d}||_2 = \partial_\beta ||\mathbf{d}||_2$ implies $\langle\mathbf{m}_1,\mathbf{d} \rangle=0$ and $\langle\mathbf{m}_2,\mathbf{d} \rangle=0$, where the inner product $\langle, \rangle$ is induced by the full norm (\ref{eq:statenorm}). As a consequence, $\alpha$ and $\beta$ may be found via inner products of (\ref{eq:bimodaldef}) with $\mathbf{m_1}$ and $\mathbf{m_2}$, respectively,
\begin{align}
\alpha = \frac{\langle \mathbf{m}_1,\mathbf{b} \rangle}{\langle \mathbf{m}_1,\mathbf{m}_1\rangle} - \beta\frac{\langle \mathbf{m}_1,\mathbf{m}_2 \rangle}{\langle \mathbf{m}_1,\mathbf{m}_1 \rangle} , \quad  \beta = \frac{\langle \mathbf{m}_2,\mathbf{b} \rangle}{\langle \mathbf{m}_2,\mathbf{m}_2\rangle} - \alpha\frac{\langle \mathbf{m}_2,\mathbf{m}_1 \rangle}{\langle \mathbf{m}_2,\mathbf{m}_2 \rangle}. \label{eq:bimcoeff} 
\end{align}
These coupled equations determine the optimal coefficients $\alpha$ and $\beta$. The corresponding minimal $\mathbf{d}$ measures nonlinear and non-bimodal effects.

\begin{figure}
       \begin{subfigure}[b]{0.50\textwidth}
               \includegraphics[width=0.99\linewidth,trim={1.7cm 1.5cm 1.0cm 0.5cm},clip]{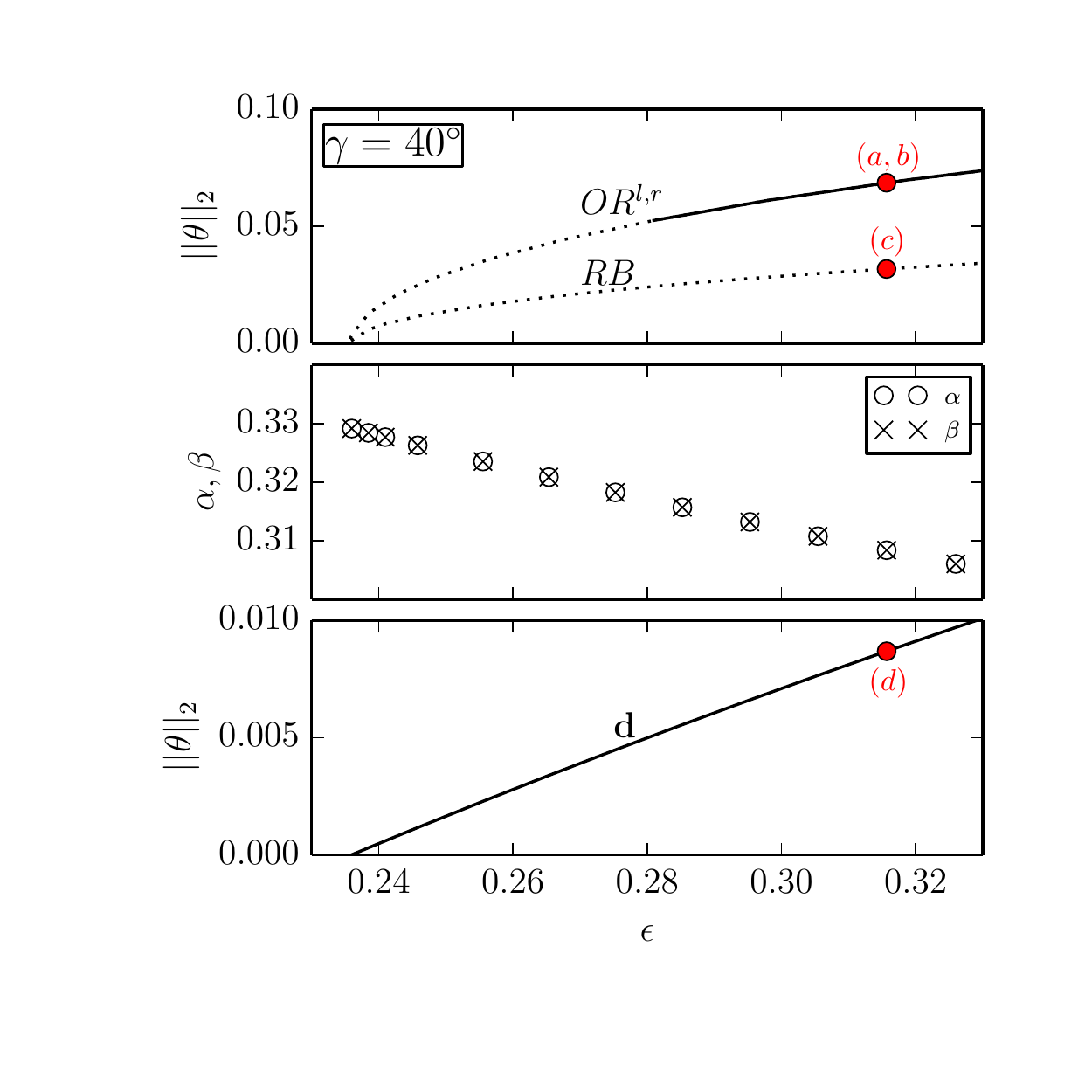}
       \end{subfigure}
       \begin{subfigure}[b]{0.49\textwidth}
               \includegraphics[width=0.99\linewidth,trim={0.0cm -3.0cm 0.0cm 0.0cm},clip]{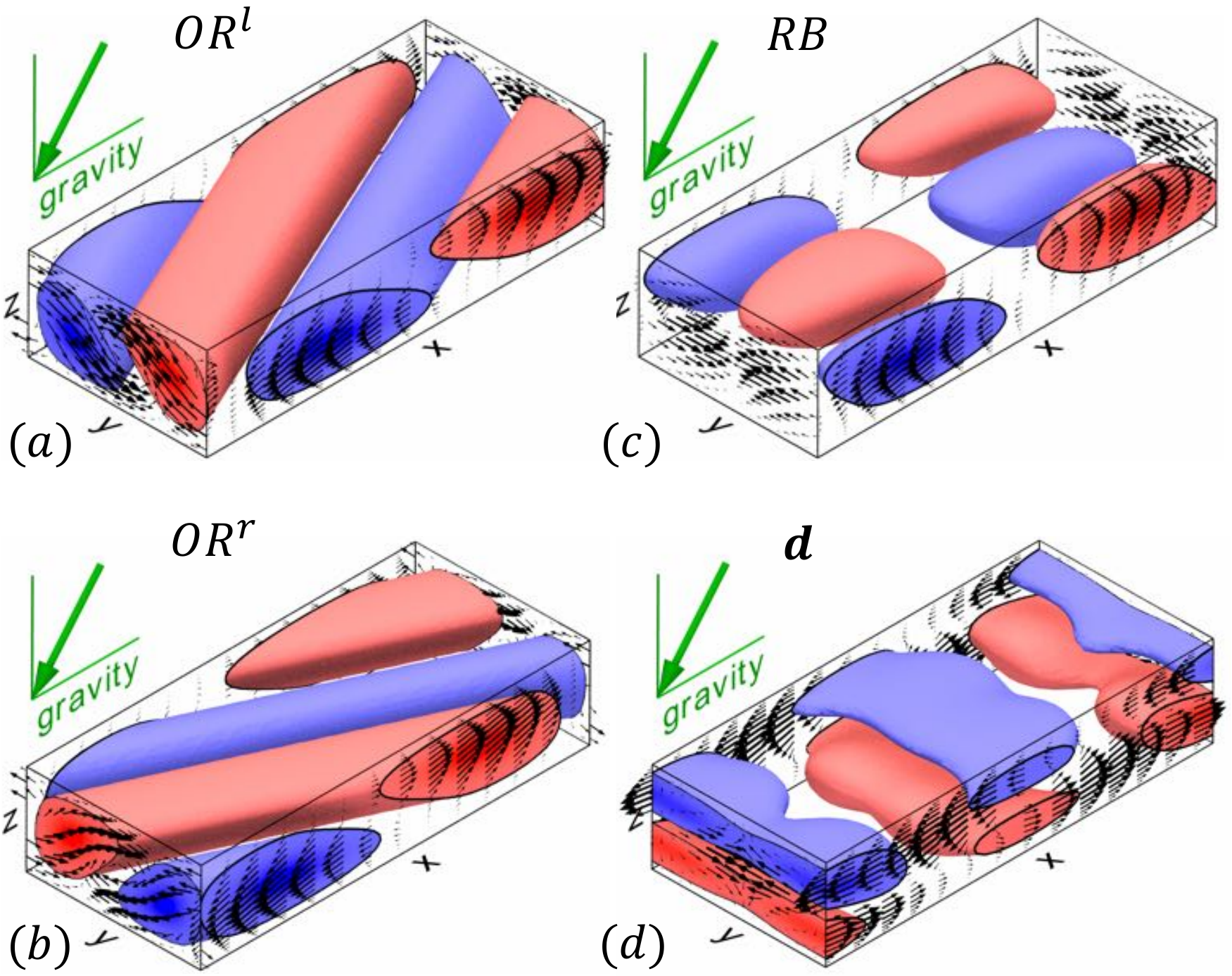}
       \end{subfigure}
\caption{\label{fig:bim_rb} Decomposition of bimodal ribbons ($RB$) into left and right oriented oblique rolls ($OR^{l,r}$). The bifurcation diagram with details of the decomposition (left) is shown together with visualisations of the temperature and velocity fields on the right (red labels in the diagram indicate panels on the right). $RB$ bifurcates together with $OR^{l,r}$ in equivariant pitchfork bifurcations from the laminar base flow. At $\gamma=40^{\circ}$, $RB$ emerges with one additional unstable eigendirection, and $OR^{l,r}$ become dynamically stable in a $[2\lambda_x,\lambda_y]$-periodic domain at $\epsilon=0.282$ (solid line in top left panel). The optimal decomposition $\bm{x}_{RB}(\epsilon)=\alpha(\epsilon)\, \bm{x}_{OR}^{l}(\epsilon) + \beta(\epsilon)\, \bm{x}_{OR}^{r}(\epsilon) + \mathbf{d}(\epsilon)$ implies linearly decreasing equal coefficients $\alpha=\beta$ along the branches (middle panel). The difference vector $\mathbf{d}$ grows linearly in $||\theta ||_2$ from zero at the bifurcation point (bottom panel). Thus, $RB$ can be viewed as a bimodal state combining two equally contributing oblique rolls.}
\end{figure}
%
%\textbf{Apply the bimodal concept to $RB$ at $\gamma=40$:}\\
The optimal bimodal decomposition (\ref{eq:bimodaldef}) with (\ref{eq:bimcoeff}) is calculated for $\mathbf{b}=\bm{x}_{RB}$ and $\mathbf{m}_{1,2}=\bm{x}_{OR}^{l,r}$ along the $\epsilon$-bifurcation branches at $\gamma=40^{\circ}$. The coefficients are found to be equal at all $\epsilon$, and to decrease linearly from $\alpha=\beta=0.3291$ at the bifurcation point (Figure \ref{fig:bim_rb}). The difference vector $\mathbf{d}$ increases linearly in $||\theta||_2$ and mostly accounts for corrections to the flow at the streamwise interfaces between hot and cold plumes (Figure \ref{fig:bim_rb}d).

\begin{figure}
       \begin{subfigure}[b]{0.50\textwidth}
               \includegraphics[width=0.99\linewidth,trim={1.7cm 1.5cm 1.0cm 0.5cm},clip]{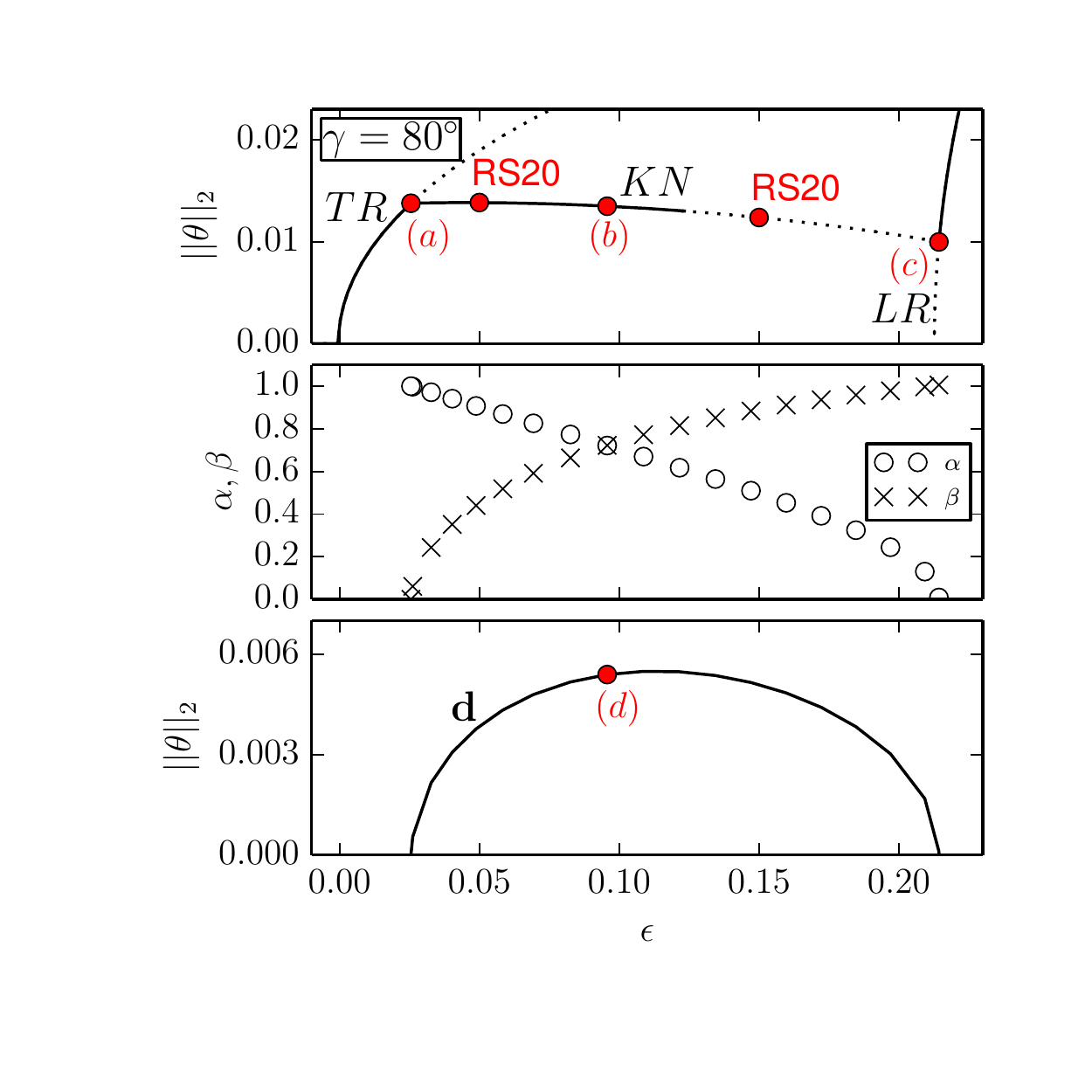}
       \end{subfigure}
       \begin{subfigure}[b]{0.49\textwidth}
               \includegraphics[width=0.99\linewidth,trim={0.0cm -2.0cm 0.0cm 0.0cm},clip]{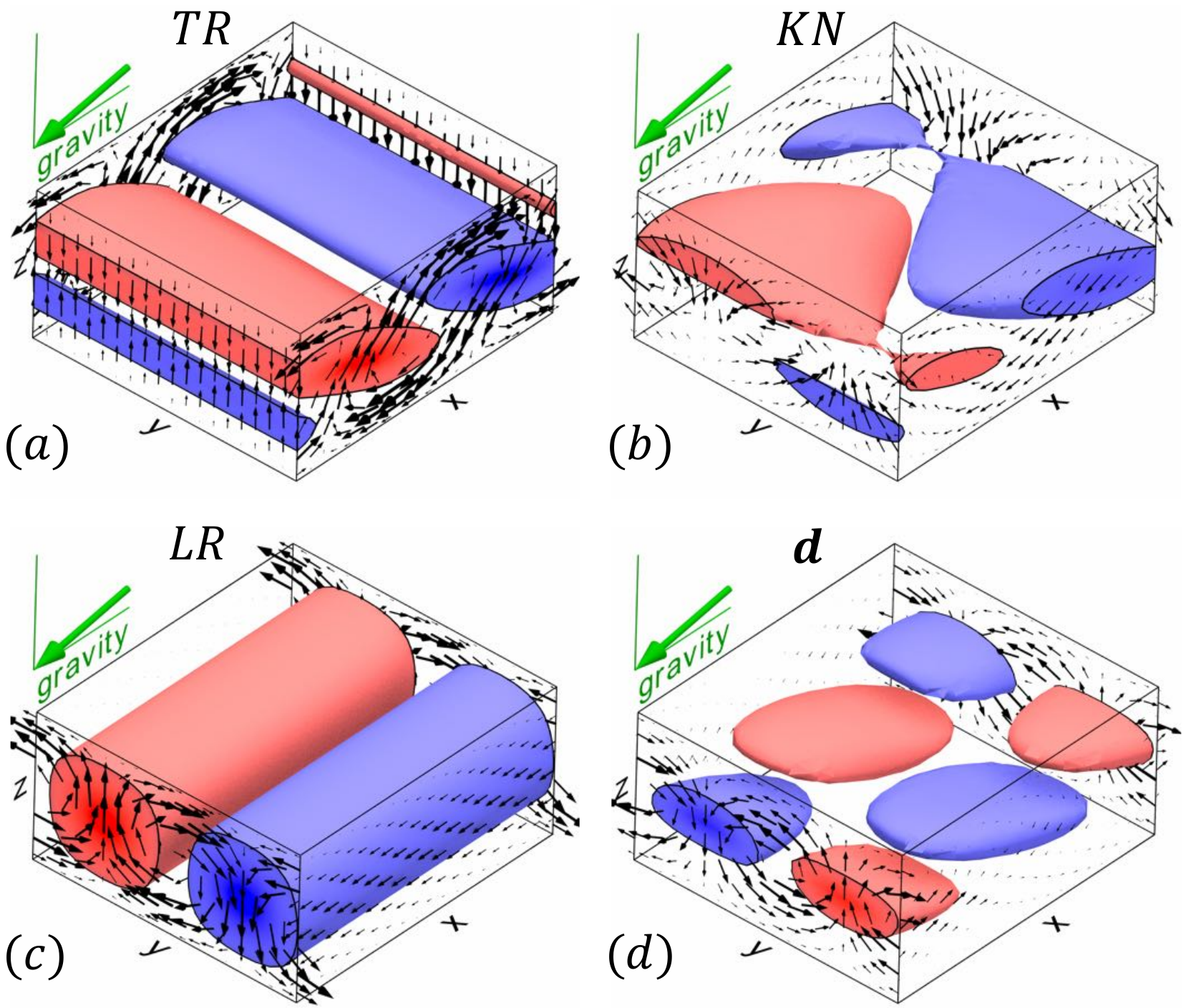}
       \end{subfigure}
\caption{\label{fig:bim_kn} Decomposition of knots ($KN$) into longitudinal and transverse oriented rolls, $LR$ and $TR$, respectively. The bifurcation diagram with details of the decomposition (left) is shown together with visualisations of the temperature and velocity fields on the right (red labels in the diagram indicate panels on the right). $KN$ connect $TR$ and $LR$ at $\gamma=80^{\circ}$ (top). Dynamical stability in a $[\lambda_x,\lambda_y]$-periodic domain is indicated by the solid lines. `RS20' labels the control parameter value where the temporal evolution has been studied in RS20. Between the bifurcation points, the optimal decomposition $KN(\epsilon)=\alpha(\epsilon)\, TR(\epsilon=0.024) + \beta(\epsilon)\, LR(\epsilon=0.22) + \mathbf{d}(\epsilon)$ results in decreasing $\alpha$, increasing $\beta$ (middle), and parabolically varying $\mathbf{d}$ (bottom). Thus, $KN$ can be viewed as a bimodal state combining transverse and longitudinal rolls with monotonically changing relative contributions.}
\end{figure}
%
%
%\textbf{Apply the bimodal concept to $KN$ at $\gamma=80$:}\\
The optimal bimodal decomposition (\ref{eq:bimodaldef}) with (\ref{eq:bimcoeff}) is calculated again for $\mathbf{b}=\bm{x}_{KN}$ along the $KN$-branch at $\gamma=80^{\circ}$. Since $LR$ does not coexist with most of the $KN$-branch (Figure \ref{fig:bim_kn}), the state vectors $\bm{m}_1=\bm{x}_{TR}$ and $\bm{m}_2=\bm{x}_{LR}$ in the decomposition are not considered as $\epsilon$-dependent. We choose the decomposition $\bm{x}_{KN}(\epsilon)=\alpha(\epsilon)\bm{x}_{TR}(\epsilon=0.024)+\beta(\epsilon)\bm{x}_{LR}(\epsilon=0.22)+\bm{d}(\epsilon)$. Here, $\epsilon$ parametrises linear interpolation between two bifurcation points. The resulting optimal coefficients $\alpha$ and $\beta$ in general differ. While the contribution of the longitudinal rolls monotonically increases, the contribution of the transverse rolls decreases (Figure \ref{fig:bim_kn}). A decomposition with $\alpha=\beta$ is found at $\epsilon=0.095$, approximately half-way between the bifurcation points and close to the maximum of $||\mathbf{d}||_2$. Note that $\mathbf{d}$ at the maximum amplitude resembles a ribbon pattern (Figure \ref{fig:bim_kn}d). Towards the bifurcation points, $\mathbf{d}$ decreases parabolically to zero. Since $\mathbf{d}$ combines nonlinear and non-bimodal effects, as well as effects due to interpolation between $TR$ and $LR$ at fixed values of $\epsilon$, the dominant source for the large values of $\partial ||\bm{d}||_2/\partial \epsilon$ at the bifurcations is unclear. Analysis in the weakly nonlinear regime near the knot instability indicates a three-mode interaction\citep{Fujimura1993,Subramanian2016}, suggesting that beside longitudinal and transverse states a third resonant oblique state is involved. Here, the importance of an oblique mode is evidenced by the significant contribution of $\bm{d}$ (Figure \ref{fig:bim_kn}).

$KN$ and $RB$ differ in their bifurcation structure. $KN$ is a connecting state between $TR$ and $LR$, while $RB$ bifurcates together with $OR^{l,r}$ in an equivariant pitchfork bifurcation. As a consequence, their bimodal decomposition into two composing straight convection rolls differs significantly. $RB$ is composed of an equal weight superposition of two symmetry related oblique rolls $OR^l=\pi_y OR^r$. $KN$ in ILC at $\gamma\neq 0^{\circ}$ is a mixed mode state, composed of transverse and longitudinal rolls that are not symmetry related and whose weight continuously changes along the branch. At $\gamma=0^{\circ}$ however, $TR$ and $LR$ become symmetry related via rotation. The knot patterns observed in Rayleigh-Benard convection \citep{Busse1974} are thus expected to bifurcate in equivariant pitchfork bifurcations, like $RB$ in ILC .

\subsection{Transverse oscillations - continuation towards a chaotic state space}
\label{sec:to}
%%%%%%%%%%%%
%\textbf{Contrast observation and key result:}\\
%\subsubsection{Subsection summary}
The pattern of obliquely modulated transverse rolls, called `switching diamond panes', shows complex dynamics with chaotically switching pattern orientations \citep{Daniels2000}. A periodic orbit $TO$ underlying transverse oscillations has been identified in RS20 at moderate $\epsilon$. The pattern of transverse oscillations seems to capture some aspects of the observed complex dynamics. $\epsilon$-continuations of $TO$ show that the orbit period of $TO$ is subject to large and non-monotonic changes, and the number of unstable eigendirections of $TO$ increases quickly with $\epsilon$. This suggests the existence of complex state space structures that support the chaotic dynamics of switching diamond panes. 

%
%\textbf{General discussion of bifurcations:}\\
%start with gamma
\subsubsection{Bifurcations to transverse oscillations}
In all but one analysed parameter continuations, the pre-periodic orbit $TO$ bifurcates from $TR$ in a supercritical Hopf bifurcation. The bifurcations are either $\epsilon$-forward at $\epsilon\approx 0.07$, found at inclination angles $\gamma=[80^{\circ},90^{\circ},100^{\circ},110^{\circ}]$, or $\gamma$-backward, found at $\epsilon=0.1$ and $\gamma=132.2^{\circ}$. The latter case represents the upper inclination limit of existence of $TO$ at $\epsilon=0.1$. At the lower limit, just below $\gamma_{c2}$, $TO$ bifurcates as quaternary state from the transverse subharmonic varicose state $TSV$ (see inset panel in Figure \ref{fig:bifgamma3}). $TSV$ is an equilibrium state discussed briefly in Appendix \ref{sec:app:additional}. A common feature of all $TO$-branches is that the $||\theta||_2$-maximum over the orbit period remains close to the $||\theta||_2$-value of $TR$. This agrees with the observation that $TO$ modulations are sinusoidal oscillations around strictly transverse rolls with the maximum deflection associated to the minimum in $||\theta||_2$ over the orbit period (RS20).

\begin{figure}
     \begin{subfigure}[b]{0.47\textwidth}
       \begin{tikzpicture}
   	\draw (0, 0) node[inner sep=0] {\includegraphics[width=0.99\linewidth,trim={-0.7cm 0.0cm 0.2cm 0.0cm},clip]{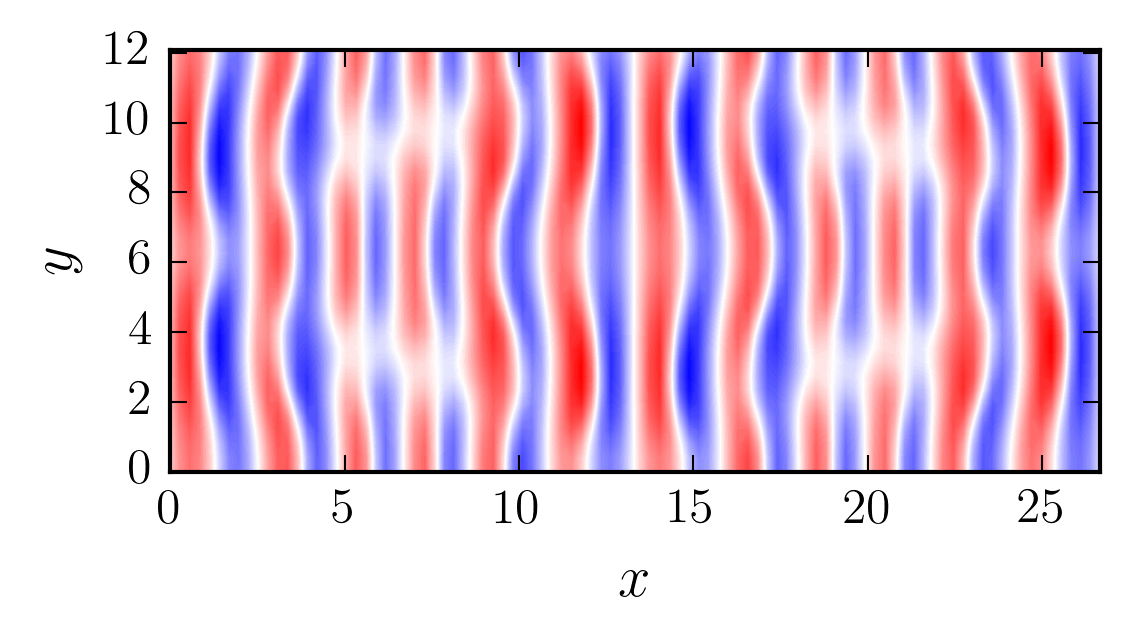}};
   	\draw (-2.4,-1.1) node {\textbf{(a)}};
		\end{tikzpicture}
     \end{subfigure}
     \begin{subfigure}[b]{0.52\textwidth}
       \begin{tikzpicture}
   	\draw (0, 0) node[inner sep=0] {\includegraphics[width=0.99\linewidth,trim={0.2cm 0.0cm -1.7cm 0.0cm},clip]{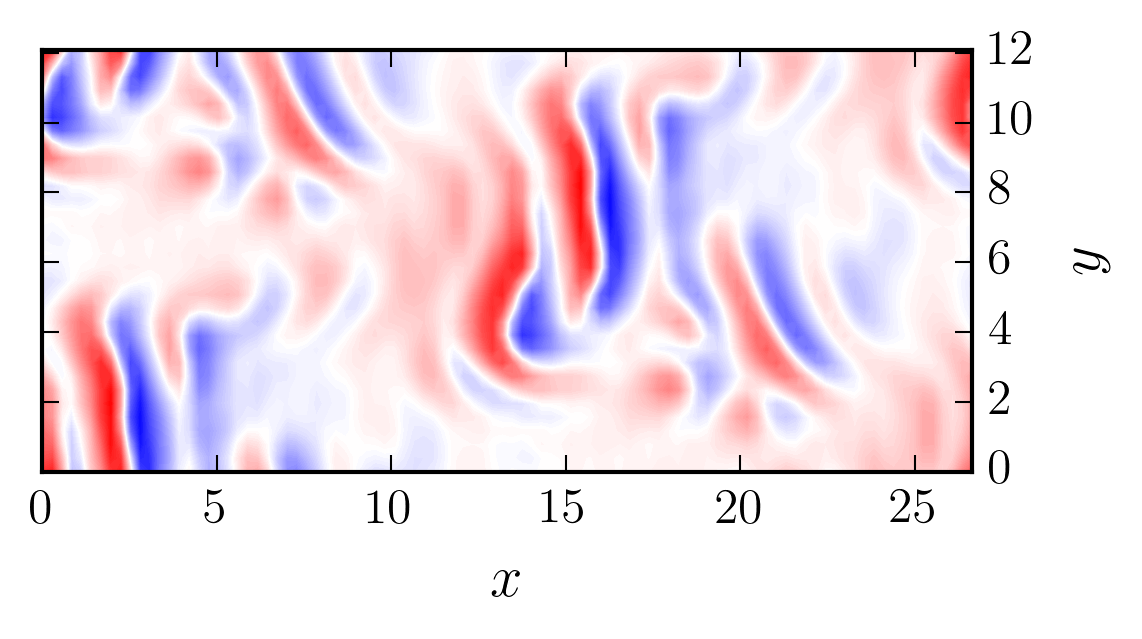}};
   	\draw (2.5,-1.1) node {\textbf{(b)}};
		\end{tikzpicture}
     \end{subfigure}
     \begin{subfigure}[b]{0.99\textwidth}
       \begin{tikzpicture}
   	\draw (0, 0) node[inner sep=0] {\includegraphics[width=0.99\linewidth,trim={0.7cm 0.9cm 1.4cm 1cm},clip]{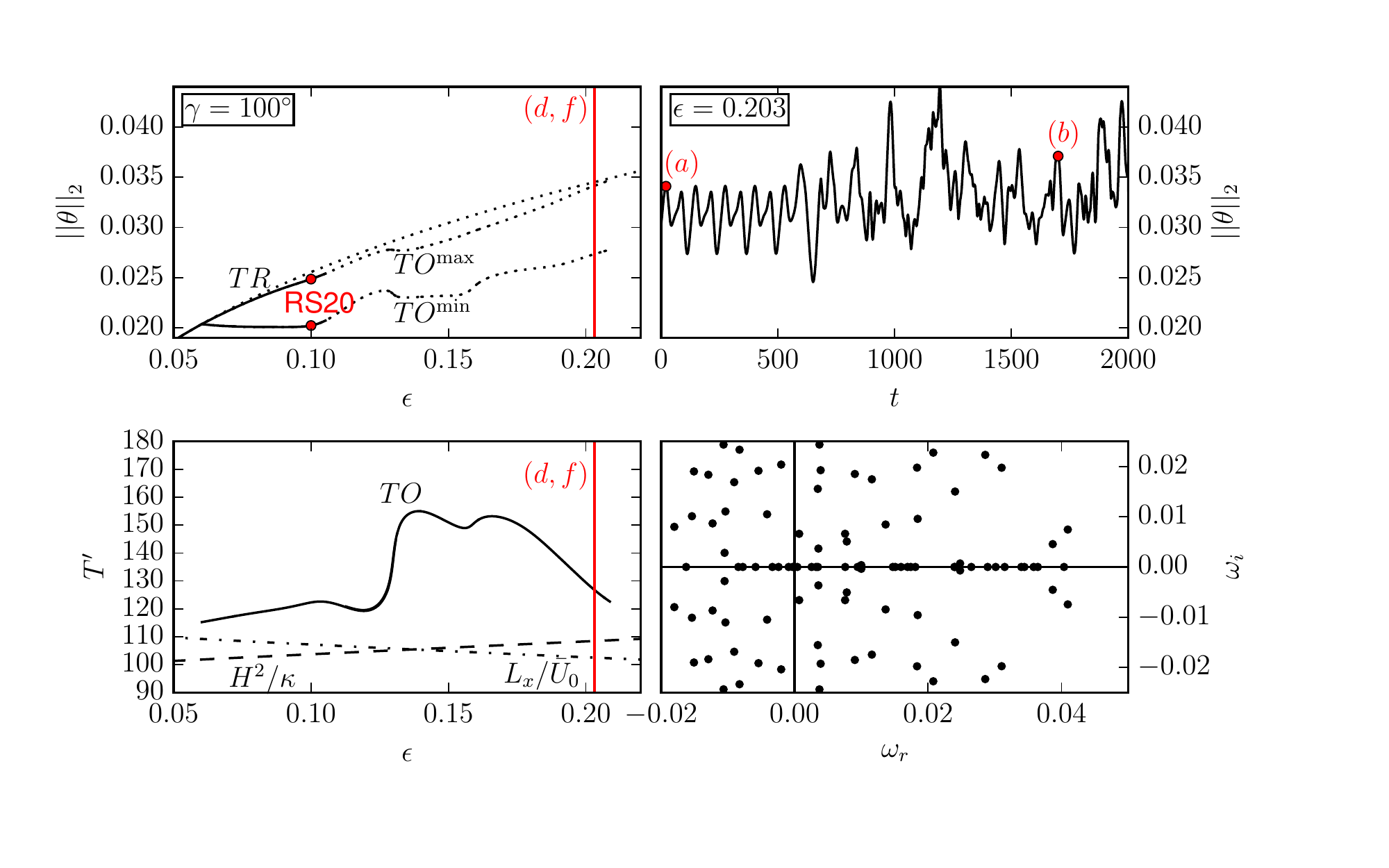}};
   	\draw (-5.9,0.7) node {\textbf{(c)}};
   	\draw (5.9,0.7) node {\textbf{(d)}};
   	\draw (-5.9,-3.2) node {\textbf{(e)}};
   	\draw (5.9,-3.2) node {\textbf{(f)}};
		\end{tikzpicture}
     \end{subfigure}
\caption{\label{fig:to} Continuation of $TO$ at $\gamma=100^{\circ}$ towards a chaotic state space. \textbf{(a)} Midplane temperature of $TO$ at $\mathrm{Ra}=11\,000$ ($\epsilon=0.203$) before instabilities create a turbulent flow \textbf{(b)}. The corresponding simulated time series of $TO$ \textbf{(d)} shows the transition from a periodic to a turbulent signal. The transition is a consequence of the many linear instabilities of $TO$ \textbf{(f)} that have emerged along the bifurcation branch \textbf{(c)}. Solid/dotted bifurcation branches indicate stable and unstable states in the symmetry subspace of $S=\langle \pi_y,\pi_{xz},\tau(0.5,0.5) \rangle $. `RS20' labels the control parameter value at which the temporal evolution has been studied in RS20. The changing relative period $T'$ of $TO$ along the bifurcation branch is shown in \textbf{(e)} and compared to the time scale of thermal diffusion $H^2/\kappa$ and laminar mean advection $L_x/\bar{U}_0$ (dashed and dashed-dotted lines). }
\end{figure}
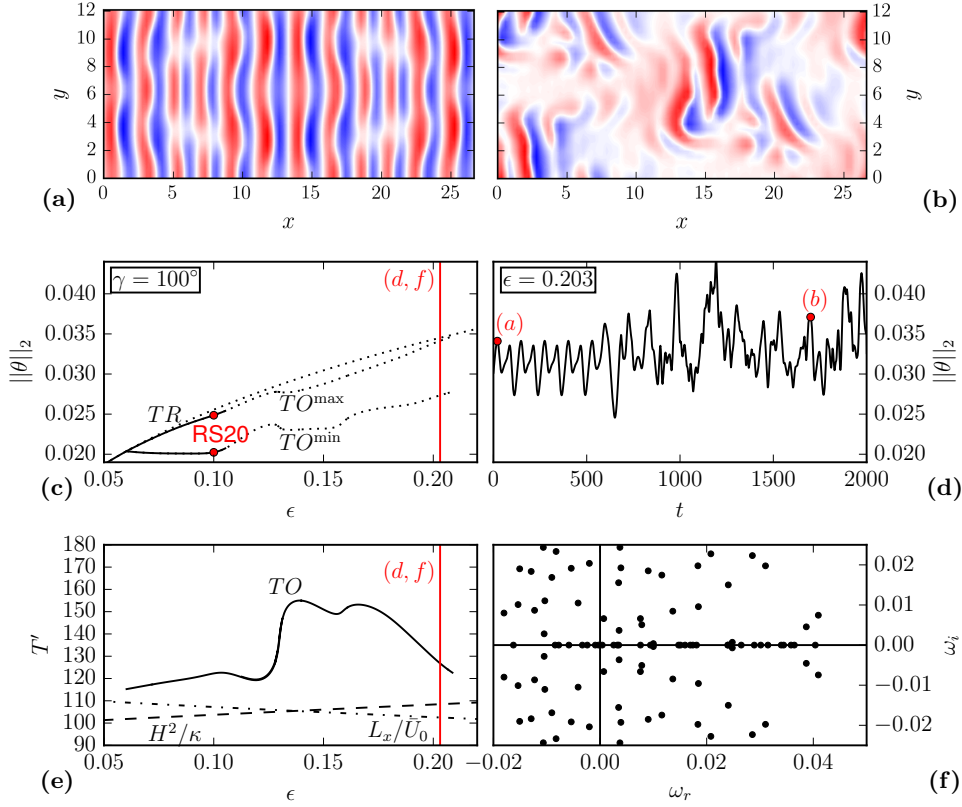

%\textbf{Detailed discussion of $\gamma=100^{\circ}$:}\\
\subsubsection{Numerical continuation of transverse oscillations}
$\gamma$-continuation of $TO$ at $\epsilon=0.1$ is numerically straight forward and yields periodic orbits showing weak bending modulations around a purely transverse orientation (RS20, Section 4.2.2). $\epsilon$-continuations are found to be numerically challenging for increasing $\epsilon$. We could not continue $TO$ much beyond $\epsilon=0.2$ (Figure \ref{fig:to}c). The reason for the computational difficulty is two-fold. First, the time period of the orbit drastically changes with $\epsilon$, which causes challenges for our shooting method. The pre-periodic orbit $TO$ satisfying (\ref{eq:map}) with $\sigma=\tau(0.5,0)$ oscillates slowly with a relative period $T'\sim\mathcal{O}(10^2)$ close to the heat diffusion time $H^2/\kappa$ and the laminar mean advection time $L_x/\bar{U}_0$. Along the continuation, the large orbit period is subject to significant and non-monotonic changes over small $\epsilon$-intervals (Figure \ref{fig:to}e). These changes in the orbit period are numerically difficult to trace. Secondly, the iterative solver of the Newton algorithm converges better if the target state is dynamically stable or weakly unstable \citep{Sanchez2004}. Computing the spectrum of eigenvalues of $TO$ in the symmetry subspace of $[12\lambda_x,6\lambda_y]$-periodicity indicates that the state branch at $\gamma=100^{\circ}$ and $\epsilon=0.203$ has collected $63$ unstable eigenvalues with a broad range of frequencies $\omega_i$ (Figure \ref{fig:to}f). At these parameters, the single-shooting Newton algorithm converged $TO$ to a residual of $||\mathcal{G}(\mathbf{x})||_2<3\times 10^{-11}$ (see Equation \ref{eq:map}). When integrating the converged orbit forward in time, unstable directions trigger a transition to a turbulent state after $t=500$ (Figure \ref{fig:to}d). This turbulent state has been described as longitudinal bursts within switching diamond panes \citep{Daniels2000}. We conclude that continuation of $TO$ for $\epsilon>0.2$ is challenging due to the numerical condition of a temporally slow, spatially large and very unstable periodic orbit that competes with many fast and small-scale modes in a chaotic turbulent state space.

\section{Discussion}
\label{sec:discussion}
%%%%%%%%%%%%
%Connect to part 1
Towards understanding how temporal and spatio-temporally complex dynamics arises in ILC, we have computed three-dimensional invariant states underlying several observed spatially periodic convection patterns in ILC at $\mathrm{Pr}=1.07$. Numerical continuation of these invariant states in two control parameters, the normalised Rayleigh number $\epsilon$ and the inclination angle $\gamma$, yields 15 bifurcation diagrams covering systematically selected parameter sections in the intervals $\epsilon \in [0,2]$ and $\gamma \in [0,120)$. For some selected bifurcating state branches, we have characterised their stability properties and pattern features along the branches. These state branches were selected for a more detailed discussion in the present article for two reasons. First, each selected branch bifurcates at a different secondary instability. Second, they cover the control parameter values at which the temporal dynamics along dynamical connections between stable and unstable invariant states have previously been described (RS20).

The relevance of the computed invariant states for observed spatio-temporally complex dynamics in ILC depends in general on the type of bifurcation creating the states, the range in control parameters over which state branches exist, and the stability properties of the invariant states along their branches. The dynamical relevance of invariant states in the context of the entire bifurcation structure is discussed below by answering the three specific questions posed in the introduction ($Q1$-$Q3$). To describe the role of individual invariant states for temporal pattern dynamics, we can distinguish three different cases: \\
In case 1, a stable invariant state represents a dynamical attractor at specific control parameter values. This case corresponds for example to supercritical $\epsilon$-forward bifurcations where the stable bifurcating invariant state is an attractor for the dynamics above the critical values of the control parameters for the bifurcation. For invariant states that have been identified because they represent dynamical attractors at specific control parameter values (RS20), the present bifurcation analysis indeed confirms supercritical $\epsilon$-forward bifurcations (Sections \ref{sec:so}-\ref{sec:to}). \\
In case 2, invariant states exist at specific control parameter values but are dynamically unstable. State branches are only stable over a finite range in control parameters. This range is limited by instabilities along state branches. Invariant states which the present study indicates as dynamically unstable at specific values of the control parameters, may still be relevant for the observed temporal dynamics at these control parameter values. One reason is that the range of stability along state branches depends on the considered pattern wavelength. Thus, invariant states might be dynamically stable at other pattern wavelengths not considered here. Another reason is that weakly unstable invariant states may be building blocks for the dynamics supported by a more complex state space attractor. Here, the evolving state vector may transiently visit weakly unstable invariant states by approaching and escaping along their stable and unstable manifolds, respectively \citep[e.g.][]{Suri2017}. The simplest example for such complex state space attractors is the robust heteroclinic cycle between two weakly unstable instances of symmetry related $OWR$ described in RS20. \\
In case 3, invariant states do not exist at specific control parameter values but their pattern is reminiscent in some state space regions that may be transiently visited by the dynamics. Folds or symmetry-breaking bifurcations may limit the existence of invariant states in parameter space. However, the pattern of the invariant state may still emerge transiently at control parameter values beyond the existence limits. We have observed this case for the transient skewed varicose pattern along a dynamical connection from unstable to stable straight convection rolls in Rayleigh-B\'enard convection (Section \ref{sec:sv}), as well as for transient subharmonic oscillations at $[\epsilon,\gamma]=[1.5,17^{\circ}]$ (see Section 4.2.1 in RS20) where the $SSW$-branch does not exist anymore due to a fold (Figure \ref{fig:lsw}). The state space structure supporting such transient dynamics seems related to a state space structure supporting intermittency \citep{Pomeau1980}.

Consequently, the patterns of invariant states are often observed because invariant states are stable and attracting, but neither stability nor existence of invariant states is required for observing their pattern.

\subsection{Bifurcation types (Q1)}
Bifurcations create or destroy invariant states and change the stability along state branches. Thus, bifurcation structures describe how state space structures change across control parameters. In response to question $Q1$, stated in the introduction, we list all the different types of bifurcations found in the present study and refer to particular examples. Identified bifurcation types include:\\
Pitchfork bifurcation, e.g. from $TR$ or $LR$ to $KN$ along $\epsilon$ at $\gamma=80^{\circ}$ (Figure \ref{fig:bim_kn}). Equivariant pitchfork bifurcation, e.g. from $B$ to $RB$ and $OR$ along $\epsilon$ at $\gamma=40^{\circ}$ (Figure \ref{fig:bim_rb}). Hopf bifurcation, e.g. from $TR$ to $TO$ along $\epsilon$ at $\gamma=100^{\circ}$ (Figure \ref{fig:to}). Equivariant Hopf bifurcation, e.g. from $LR$ to $SSW$ and $STW$ along $\gamma$ at $\epsilon=1.5$ (Figure \ref{fig:lsw}). Saddle-node bifurcation, e.g. $WR$ along $\epsilon$ at $\gamma=90^{\circ}$ (Figure \ref{fig:bifeps3}, panel $\gamma=90^{\circ}$). Mutual annihilation of two periodic orbits, e.g. the two folds bounding the $SSW$ isola along $\gamma$ at $\epsilon=0.5$ (Figure \ref{fig:bifgamma3}). The global bifurcation of a periodic orbit colliding with a structurally robust heteroclinic cycle, e.g. the $SSW$ collision with $TR\rightarrow \tau_x TR \rightarrow TR$ along $\mathrm{Ra}$ at $\gamma=10^{\circ}$ (Figure \ref{fig:globBif}).\\
The symmetry-breaking pitchfork and Hopf bifurcations are found as $\epsilon$- or $\gamma$-forward or backward bifurcations. The orientation of bifurcations can change when control parameter values are changed, e.g. $WR$ bifurcates $\gamma$-forward from $LR$ at $\epsilon=0.1$, but $\gamma$-backward at $\epsilon=0.5$ (Figure \ref{fig:bifgamma3}). Moreover, pitchfork and Hopf bifurcations can be supercritical or subcritical independent of their orientation. The $\epsilon$-backward pitchfork bifurcation from $R_{\lambda 2}$ to $SV$ at $\gamma=0^{\circ}$ is subcritical (Figure \ref{fig:sv}) but the $\epsilon$-backward pitchfork bifurcation from $OR$ to $DWR$ at $\gamma=40^{\circ}$ is supercritical (Figure \ref{fig:wrbif}f). \\
The sequential order in which bifurcations occur may depend on the considered path through parameter space. $RB$ at $\epsilon=0.5$ for example can bifurcate in primary or secondary bifurcations along $\gamma$. When decreasing $\gamma$ towards $\gamma=46^{\circ}$, $RB$ bifurcate from $B$ in a primary bifurcation. When increasing $\gamma$ towards $\gamma=24^{\circ}$, $RB$ bifurcate from $TR$ in a secondary bifurcation (Figure \ref{fig:bifgamma3}, panel $\epsilon=0.5$). Thus, describing for example $WR$ as tertiary state implies a particular parameter path. Since $WR$ can bifurcate from $RB$ that may be described as tertiary state (Figure \ref{fig:bifeps3}, panel $\gamma=50^{\circ}$), $WR$ may also be described as quartenary state. 

The relation between bifurcation structures and spatio-temporally complex dynamics is in general complicated. The various local and global bifurcations can modify the coexisting invariant states and their dynamical connections in various ways. Coexistence of invariant states may result from supercritical or subcritical bifurcations as well as from folds. These bifurcation types exist in ILC at all angles of inclinations. For example, the subcritical coexistence of stable straight convection rolls with unstable $SV$ (Figure \ref{fig:sv}) or with unstable $SSW$ (Figure \ref{fig:bifeps3}, panels $\gamma=10^{\circ},20^{\circ}$), supports the experimental observation of spatially localized variants of these spatially periodic states \citep{Bodenschatz2000,Daniels2000}. The supercritical coexistence of $WR$ with $DWR$, $OR$ or $OWR$ (Figure \ref{fig:wrbif}) supports the observed pattern defects within the spatially coexisting wavy rolls of different orientations \citep{Daniels2002a}. The details of these relations are non-trivial as they require to consider spatial dynamics \citep[e.g.][]{Knobloch2015}. 

For a specific bifurcation structure we see a generic relation to complex temporal dynamics. All computed sequences of primary and secondary supercritical $\epsilon$-forward pitchfork or Hopf bifurcations give rise to one of the four sequences of dynamical connections. These are $B\rightarrow LR\rightarrow SSW,WR$ and $B\rightarrow TR\rightarrow KN,TO$ as observed in RS20 and illustrated in Figure \ref{fig:overview}. Consequently, a `sequence of bifurcations' \citep{Busse1996}, that consists of supercritical $\epsilon$-forward bifurcations, gives rise to a corresponding `sequence of dynamical connections'.

\subsection{Connection to instabilities (Q2)}

\begin{table}
 \begin{center}
   \caption{Critical thresholds and wavelengths $[l_x,l_y]$ of secondary instabilities determined by Floquet analysis \citep{Subramanian2016} and the critical bifurcation points determined by the present bifurcation analysis for prescribed box sizes $[L_x,L_y]$. The comparison requires to state all critical frequencies $\omega^{\dagger}_c$ in diffusion time scales. Frequencies in free fall time units, used throughout this paper series, are obtained via $\omega=\omega^{\dagger}/\sqrt{PrRa}$.}
   \label{tab:table1}
   \begin{tabular}{| r | c | c | c | c | c | c | c | c |} 
     \toprule
     \multicolumn{1}{|c|}{} & \multicolumn{4}{|c|}{\textbf{Floquet analysis}} & \multicolumn{4}{|c|}{\textbf{bifurcation analysis}}   \\ \midrule
     $\gamma$ &instability  &  $[l_x,l_y]$ & $\epsilon_c$ & $\omega^{\dagger}_c$ & invariant state & $[L_x,L_y]$ & $\epsilon_c$ & $\omega^{\dagger}_c$  \\ \midrule
       $0^{\circ}$   & skewed varicose & [10.6, 8.07] & 1.100 & 0 & $SV$ & [8.88, 8.06] & 1.020 & 0 \\
     $10^{\circ}$  & long. subh. oscil. & [4.89, 4.03] & 1.360 & 6.211 & $SSW/STW$ & [4.44, 4.03] & 1.454 & 6.269 \\
     $20^{\circ}$  & long. subh. oscil. & [4.89, 4.03] & 0.900 & 11.66 & $SSW/STW$ & [4.44, 4.03] &  0.929 & 11.64 \\
     $20^{\circ}$  & wavy & [62.8, 2.02] & 0.018 & 0 &$WR$ & [4.44, 2.02] & 0.054 & 0 \\
     $30^{\circ}$  & wavy & [62.8, 2.02] & 0.014 & 0 &$WR$ & [4.44, 2.02] & 0.033 & 0 \\
     $40^{\circ}$  & wavy & [62.8, 2.02] & 0.013 & 0 &$WR$ & [4.44, 2.02] & 0.034 & 0 \\
     $50^{\circ}$  & wavy & [62.8, 2.02] & 0.013 & 0 &$WR$ & [4.44, 2.02] & 0.043 & 0 \\
     $60^{\circ}$  & wavy & [62.8, 2.02] & 0.013 & 0 &$WR$ & [4.44, 2.02] & 0.072 & 0 \\
     $70^{\circ}$  & wavy & [62.8, 2.02] & 0.013 & 0 &$WR$ & [4.44, 2.02] & 0.159 & 0 \\
     $80^{\circ}$  & knot &  [2.23, 2.03] & 0.026 & 0 &$KN$ &  [2.22, 2.02] & 0.024 & 0 \\
     $90^{\circ}$  & trans. oscil. & [26.9, 13.4] & 0.063 & 1.733 &$TO$ & [26.7, 12.1] & 0.061 & 2.527 \\
   $100^{\circ}$  & trans. oscil. & [27.2, 15.7] & 0.060 & 1.484 &$TO$ & [26.7, 12.1] & 0.060 & 2.776 \\
   $110^{\circ}$  & trans. oscil. & [27.4, 17.0] & 0.057 & 1.312 &$TO$ & [26.7, 12.1] & 0.059 & 3.043 \\ 
   \end{tabular}
 \end{center}
\end{table}

The patterns of the tertiary invariant states $SV$, $SSW/STW$, $WR$, $KN$ and $TO$ are similar to the pattern motifs associated to the five secondary instabilities in ILC at $\mathrm{Pr}=1.07$ \citep{Subramanian2016}. The similarity suggests that the invariant states bifurcate at corresponding secondary instabilities. To confirm this, we compare the bifurcation points of the nonlinear state branches with the critical threshold parameters of the secondary instabilities determined by \citet{Subramanian2016} using Floquet analysis (compare with question $Q2$ stated in the introduction). Floquet analysis solves for the pattern wavelengths that first become unstable at the critical threshold $\epsilon_c$ when $\epsilon$ is increased towards $\epsilon_c$ for fixed $\gamma$. For numerical continuation of invariant states, the pattern wavelength is prescribed. The critical threshold $\epsilon_c$ is determined by continuing the state branch down in $\epsilon$ towards the bifurcation at $\epsilon_c$ for fixed $\gamma$. Consequently, Floquet analysis yields the minimal $\epsilon_c$ of the instability, while branches of invariant states at prescribed wavelengths bifurcate at higher $\epsilon_c$. We expect comparable critical thresholds between the two methods if the associated pattern wavelengths are comparable.

Table \ref{tab:table1} compares the results of Floquet analysis and bifurcation analysis in terms of pattern wavelengths $L_x$ and $L_z$, critical thresholds $\epsilon_c$, and critical frequency $\omega_c$ for Hopf bifurcations. We find clear agreement between the results for skewed varicose, longitudinal subharmonic oscillatory, and knot instabilities. Note that Floquet analysis finds the skewed varicose instability for $\gamma=0^{\circ}$ at a slightly higher $\epsilon_c$ than the bifurcation analysis. This suggests that the Floquet analysis did not capture the most unstable wavelengths of the skewed varicose instability. \\
For the wavy instabilities, the $\epsilon_c$ obtained from the bifurcation analysis is significantly larger. This discrepancy results from the difference in wavelength $L_x$. Floquet analysis indicates $L_x$ one order of magnitude larger than the $L_x$ prescribed in the bifurcation analysis. We confirmed that $WR$ bifurcates at identical $\epsilon_c$ when identical pattern wavelengths are prescribed. Thus, the equilibrium state $WR$ bifurcates at the previously characterised wavy instability. \\
For the transverse oscillatory instability, the two methods agree in $\epsilon_c$ but differ in the critical frequency $\omega_c$. The reasons for this discrepancy are not clear. Continuing the periodic orbit $TO$ to identical pattern wavelengths does not change the critical frequency much. Thus, we hypothesise that the instability characterised by Floquet analysis corresponds to a different bifurcating periodic orbit as $TO$. This hypothesis is supported by two observations. First, a Galerkin projection near the transverse oscillatory instability suggests a subcritical $\epsilon$-backward bifurcation \citep{Subramanian2015}. $TO$ however, is always found to bifurcate supercritically and $\epsilon$-forward. Second, the pattern of $TO$ can be described as spatially subharmonic standing wave oscillations. Like the subharmonic standing wave state $SSW$, also $TO$ oscillates on the time scale of the laminar mean advection across the pattern $L_x/\bar{U}_0$ (Section \ref{sec:ilc}). Thus, these subharmonic standing waves satisfy the approximate resonance condition
\begin{equation}
m\,L_x\,\omega_c \approx n\,\bar{U}_02\pi \ , \label{eq:approxTO}
\end{equation}
with $(m,n)\in \mathbb{N}$. For bifurcations to $SSW$ at small $\gamma$, this approximation holds for $(m,n)=(2,1)$ with relative errors of about $\pm 15\%$. The nonlinear time scales along the $SSW$-branch are shown in Figure \ref{fig:lsw}d. For bifurcations to $TO$, this approximation holds for $(m,n)=(2,1)$ with relative errors of less than $\pm 10\%$. The nonlinear time scales along the $TO$-branch are shown in Figure \ref{fig:to}e. The transverse oscillatory instability from Floquet analysis however satisfies (\ref{eq:approxTO}) best for $(m,n)=(4,1)$. Due to the different resonance numbers, we suspect other physics than those of subharmonic standing waves to govern the instability described by Floquet analysis. Future research should investigate the possibility for other periodic orbits than $TO$ to bifurcate, possibly subcritically, at or near the transverse oscillatory instability. Except in the case of $TO$, the bifurcating invariant states match the characteristics of the secondary instabilities described previously in \citet{Subramanian2016}.

\subsection{Range of existence (Q3)}

\begin{figure}
\includegraphics[width=0.99\linewidth,trim={1.0cm 6.5cm 2cm 1.5cm},clip]{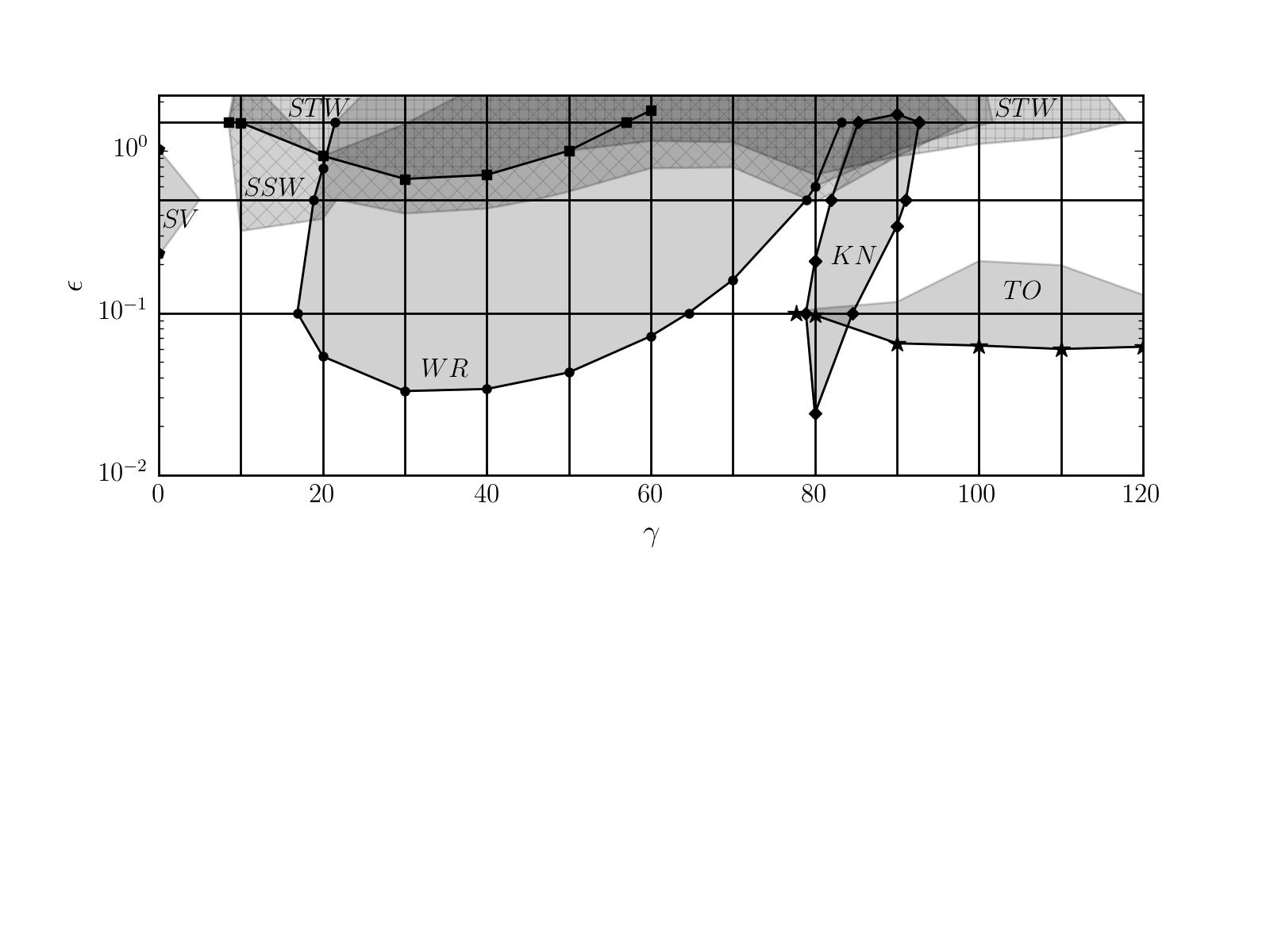}
\caption{\label{fig:existence} Existence regions of invariant states in the considered parameter space at $\mathrm{Pr}=1.07$. Local symmetry breaking bifurcations are marked and connected to guide the eye: $SV$ ($\pentagon$), $WR$ ($\circ$) and $KN$ ($\Diamond$) bifurcate in pitchfork bifurcations. $SSW/STW$ ($\square$) and $TO$ ($\star$) bifurcate in Hopf bifurcations. Where existence is not limited by symmetry breaking bifurcations, the limits are due to folds, global bifurcations or, in the case of $TO$, due to numerical challenges. Hatched regions of existence allow to better distinguish $SSW$ ($\times$) and $STW$ ($+$).}
\end{figure}

The third specific question is about the limits of existence of invariant solutions as control parameters are varied (\emph{Q3} state in the introduction). This problem has been approached by continuing invariant states as far as possible along \emph{a priori} defined sections across the $[\gamma,\epsilon]$-parameter space at $\mathrm{Pr}=1.07$. Since the continuation methods allow tracing invariant states beyond critical threshold parameters of additional instabilities, it was possible to follow bifurcation branches over large intervals of control parameters. The travelling wave $STW$, for example, is found to exist over a large range of inclinations $10^{\circ}\le \gamma \le 110^{\circ}$, covering different flow regimes with small and large laminar shear forces. We identify three invariant states $SSW/STW$ and $WR$ whose solution branches persist across the angle of the codimension-2 point $\gamma_{c2}$, and for $\gamma>90^{\circ}$ where their parent state $LR$ has disappeared. With the exception of $SV$, all tertiary invariant states are found to exist for the case of vertical convection with $\gamma=90^{\circ}$. All invariant states existing at $[\gamma,\epsilon]=[90^{\circ},1.5]$ are briefly discussed and compared with turbulent vertical convection in Appendix \ref{sec:app:vertical}. We visually summarise the regions of existence and coexistence of the computed invariant states in Figure \ref{fig:existence}.

Continuation and stability analysis along state branches revealed bifurcations to or from other invariant states. These states are neither clearly observed in experiments or simulations, nor do they correspond to instabilities found by Floquet analysis. The stability analysis along the branch of $WR$ (Section \ref{sec:wavy}) introduced four additional equilibrium states, namely $DWR$, $OR$, $OWR$, and $RB$. Other invariant states were obtained because continuations terminated at bifurcations from these states. Specifically, $TO$ may bifurcate from the $TSV$ equilibrium (Figure \ref{fig:bifgamma3}, $\epsilon=0.1$), $KN$ may bifurcate from $SL$ (Figure \ref{fig:bifeps3}, $\gamma=90^{\circ}$), and a global bifurcation of $SSW$ may involve $LSV$ as parent state (Figure \ref{fig:lsw}c). These three additional states are described in Appendix \ref{sec:app:additional}. We do not distinguish invariant states connected via folded bifurcation branches as upper and lower branch states. Folds exist at all angles of inclinations. See e.g., the bifurcation branches of $SV$ at $\gamma=0^{\circ}$, and of $STW$ at $\gamma=110^{\circ}$. However, we observe that state branches tend to become more folded towards inclinations around vertical (compare panels in Figure \ref{fig:bifeps3}).

\section{Conclusions}
The present bifurcation analysis has identified an extensive network of parametrically connected invariant state branches in inclined layer convection. Overall, 16 different nonlinear three-dimensional invariant states have been discussed in the present article. Many of them are related to spatio-temporally complex dynamics observed in experiments and simulations. Seven different types of bifurcations were found, including common types like Hopf bifurcations or saddle-node bifurcations, and including less common types like equivariant bifurcations or global collisions between periodic orbits and robust heteroclinic cycles. Computing this many different invariant states and branches just for this work and RS20 has been straightforward relative to recent integrated efforts of the research community to compute similar numbers of invariant states and branches in other canonical shear flows like plane Couette or pipe flow. Inclined layer convection covers flows from horizontal Rayleigh-B\'enard convection to vertical layer convection which are relevant for engineering applications and which have been widely studied using experiments and simulations. This article demonstrates that these flows are numerically accessible to nonlinear dynamical systems concepts.

\section*{Declaration of Interests}
The authors report no conflict of interest.

\section*{Acknowledgements}
\begin{acknowledgments}
The authors are indebted to Laurette Tuckerman and Edgar Knobloch for in-depth discussions on the formalisation of symmetries, equivariant bifurcations and the structural stability of heteroclinic cycles, as well as for valuable comments on the manuscript. PS acknowledges discussions with Alastair Rucklidge. This work was supported by the Swiss National Science Foundation (SNF) under grant no. 200021-160088, and in parts by a L'Or{\'e}al UK and Ireland Fellowship for Women in Science (PS).
\end{acknowledgments}

%\bibliographystyle{jfm}
%\bibliography{ilc_litera}

\appendix

\section{Additional invariant states participating in bifurcations}\label{sec:app:additional}
This section briefly discusses three additional invariant states that do not represent previously observed convection patterns at $\mathrm{Pr}=1.07$ but that participate in the bifurcation network by bifurcating to the above discussed invariant states.

\begin{figure}
        \begin{subfigure}[b]{0.32\textwidth}    
        \begin{tikzpicture}
    	\draw (0, 0) node[inner sep=0] {\includegraphics[width=\linewidth,trim={0.2cm 0.1cm 0.2cm 0.2cm},clip]{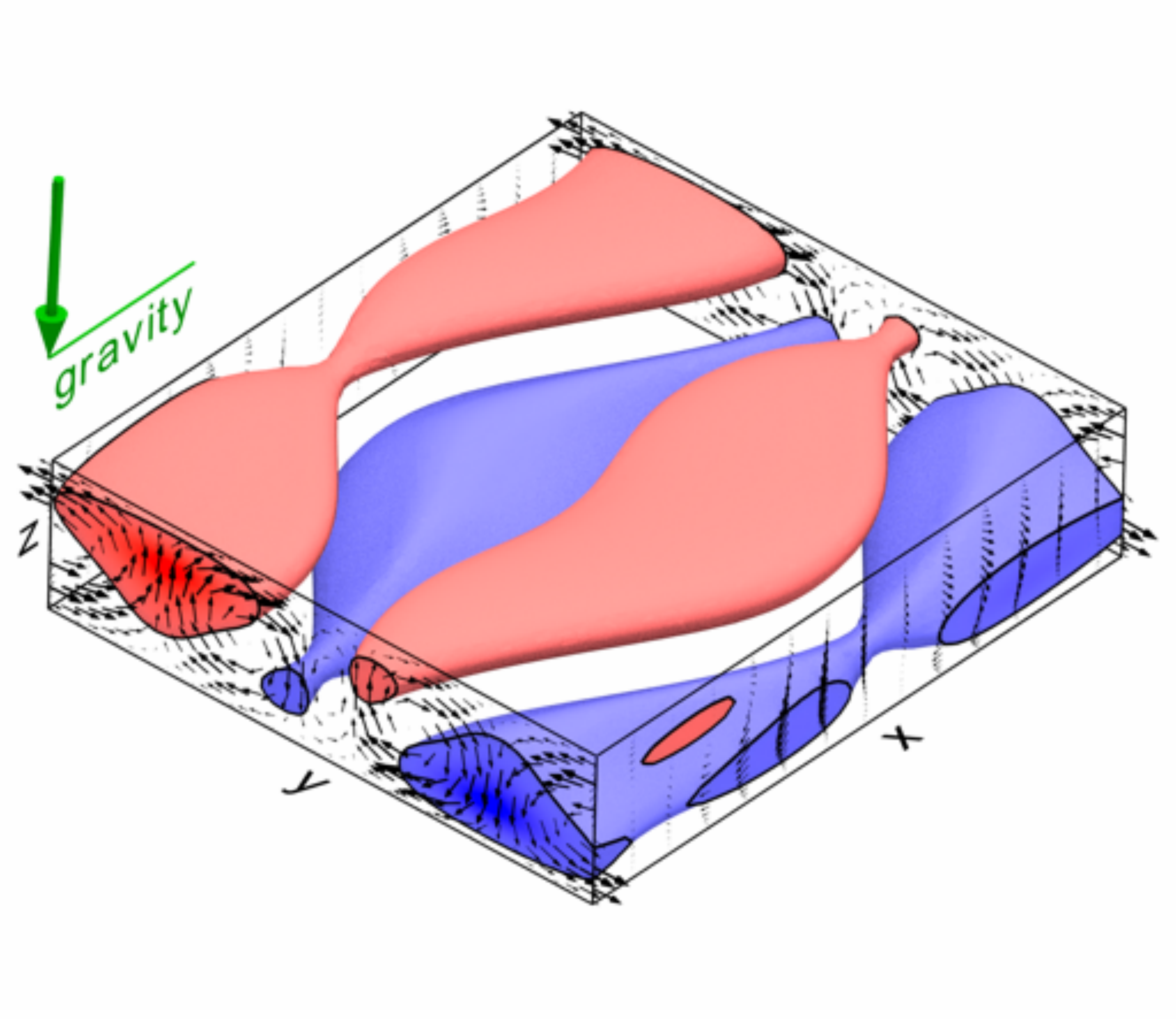}};
    	%\draw (-1.9,-1.8) node {(a)};
		\end{tikzpicture}
        \end{subfigure}
        \begin{subfigure}[b]{0.32\textwidth}
        \begin{tikzpicture}
    	\draw (0, 0) node[inner sep=0] {\includegraphics[width=\linewidth,trim={0.2cm 0.1cm 0.2cm 0.2cm},clip]{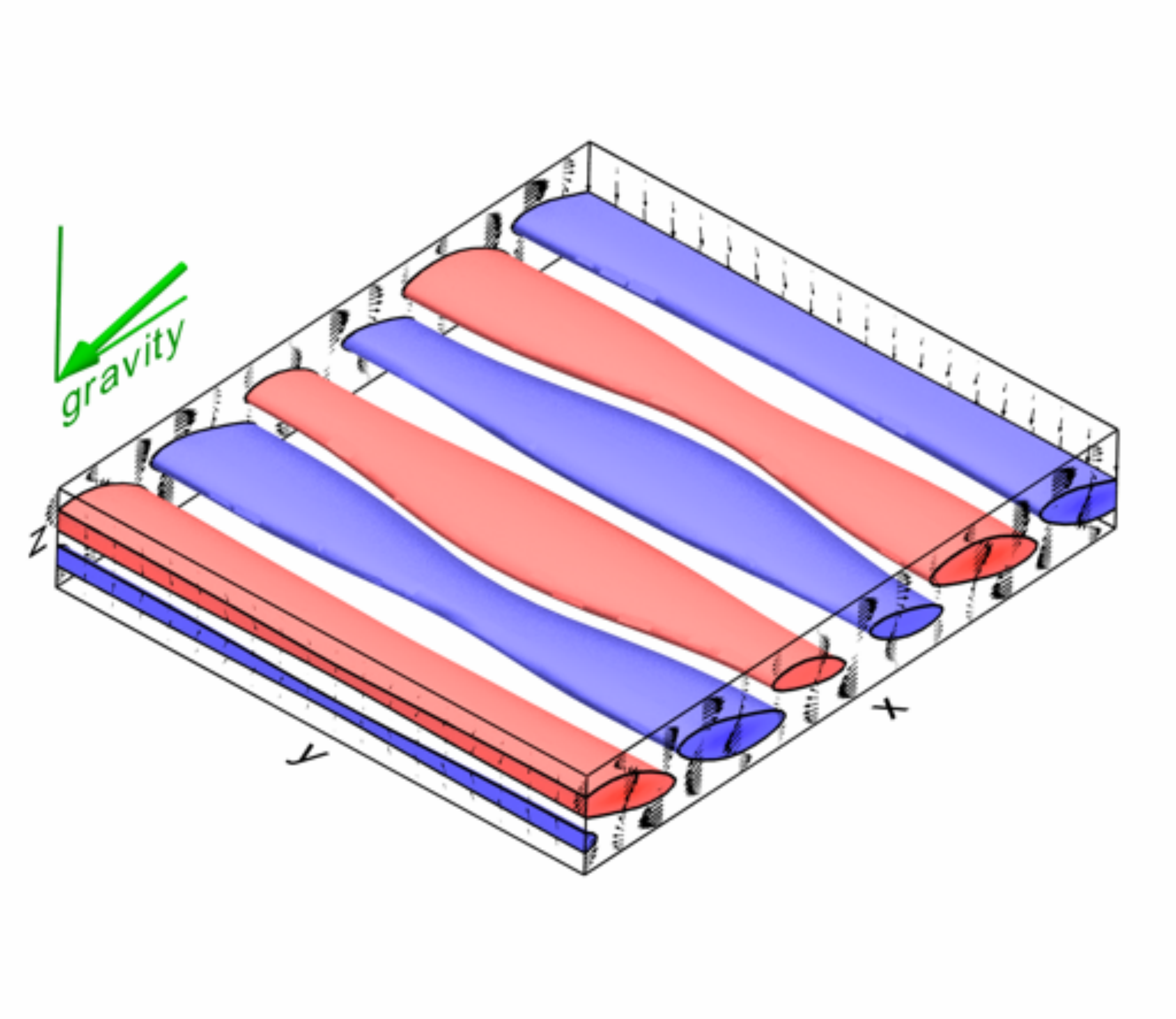}};
    	%\draw (-1.9,-1.8) node {(b)};
		\end{tikzpicture}
        \end{subfigure}
        \begin{subfigure}[b]{0.32\textwidth}
        \begin{tikzpicture}
    	\draw (0, 0) node[inner sep=0] {\includegraphics[width=\linewidth,trim={0.2cm 0.1cm 0.2cm 0.2cm},clip]{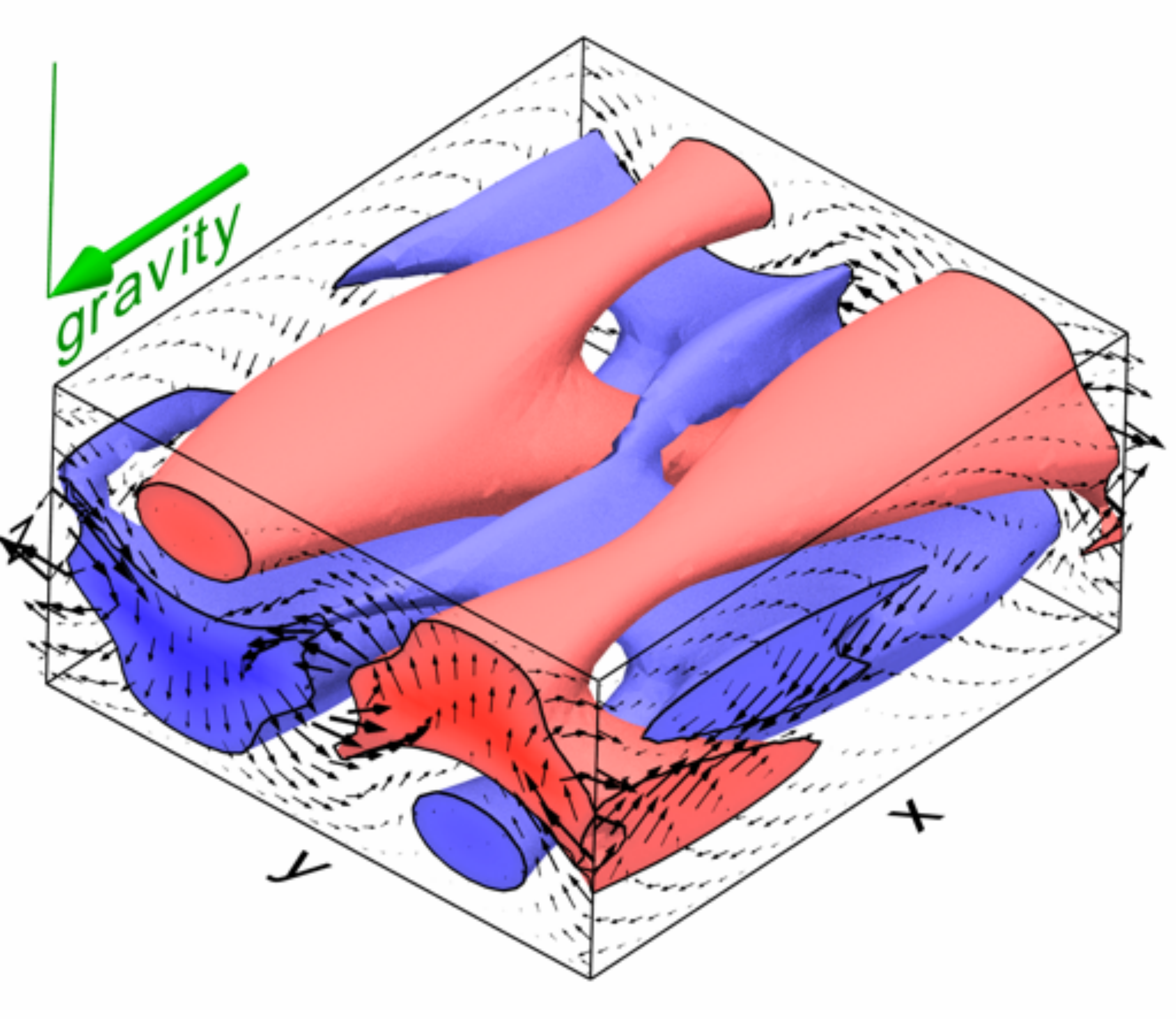}};
    	%\draw (-1.9,-1.8) node {(c)};
		\end{tikzpicture}
        \end{subfigure}
        \begin{subfigure}[b]{0.32\textwidth}    
        \begin{tikzpicture}
    	\draw (0, 0) node[inner sep=0] {\includegraphics[width=\linewidth,trim={0.2cm 0.1cm 0.2cm 0.2cm},clip]{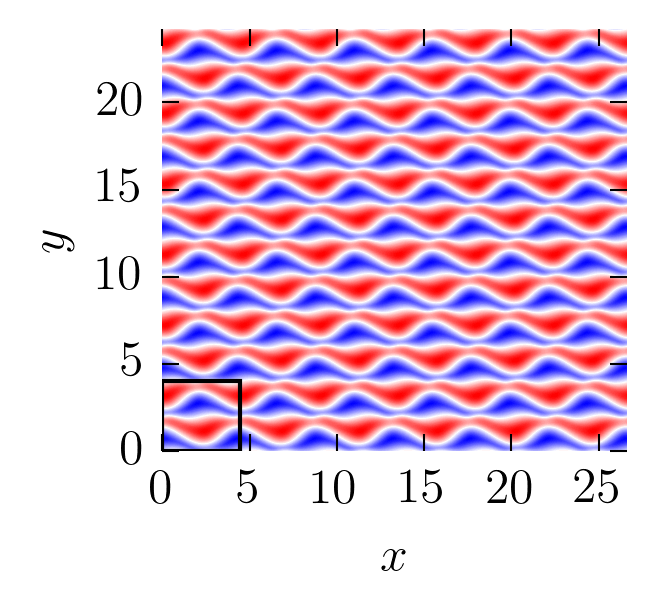}};
    	\draw (-1.9,-1.8) node {\textbf{(a)}};
		\end{tikzpicture}
        \end{subfigure}
        \begin{subfigure}[b]{0.32\textwidth}
        \begin{tikzpicture}
    	\draw (0, 0) node[inner sep=0] {\includegraphics[width=\linewidth,trim={0.2cm 0.1cm 0.2cm 0.2cm},clip]{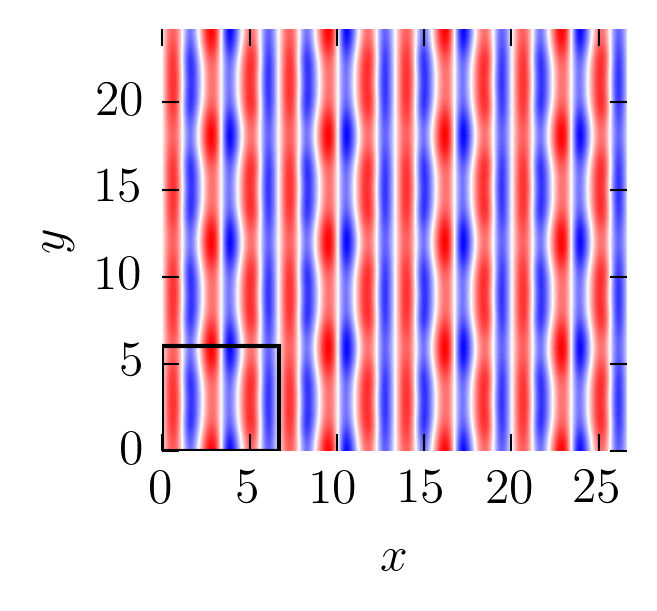}};
    	\draw (-1.9,-1.8) node {\textbf{(b)}};
		\end{tikzpicture}
        \end{subfigure}
        \begin{subfigure}[b]{0.32\textwidth}
        \begin{tikzpicture}
    	\draw (0, 0) node[inner sep=0] {\includegraphics[width=\linewidth,trim={0.2cm 0.1cm 0.2cm 0.2cm},clip]{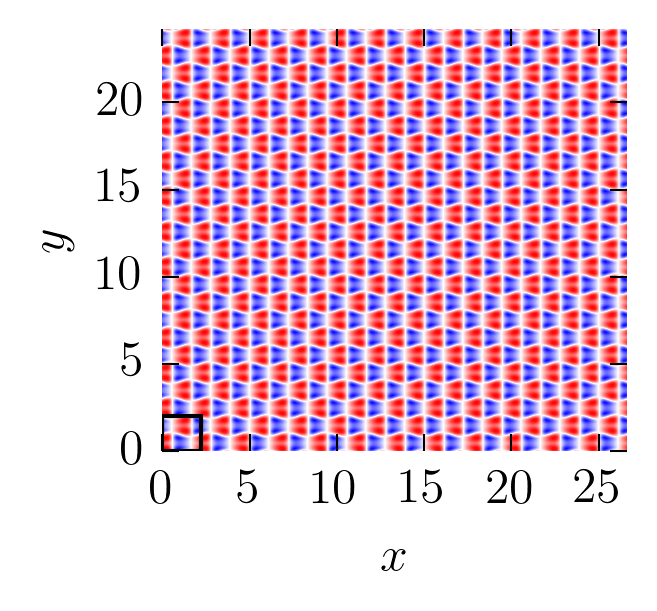}};
    	\draw (-1.9,-1.8) node {\textbf{(c)}};
		\end{tikzpicture}
        \end{subfigure}
       	\caption{\label{fig:app:additional} 3D flow structure and midplane temperature contours of three steady equilibrium states participating in bifurcations. \textbf{(a)} longitudinal subharmonic varicose $LSV$, \textbf{(b)} transverse subharmonic varicose $TSV$, \textbf{(c)} subharmonic lambda plumes $SL$. }
\end{figure}
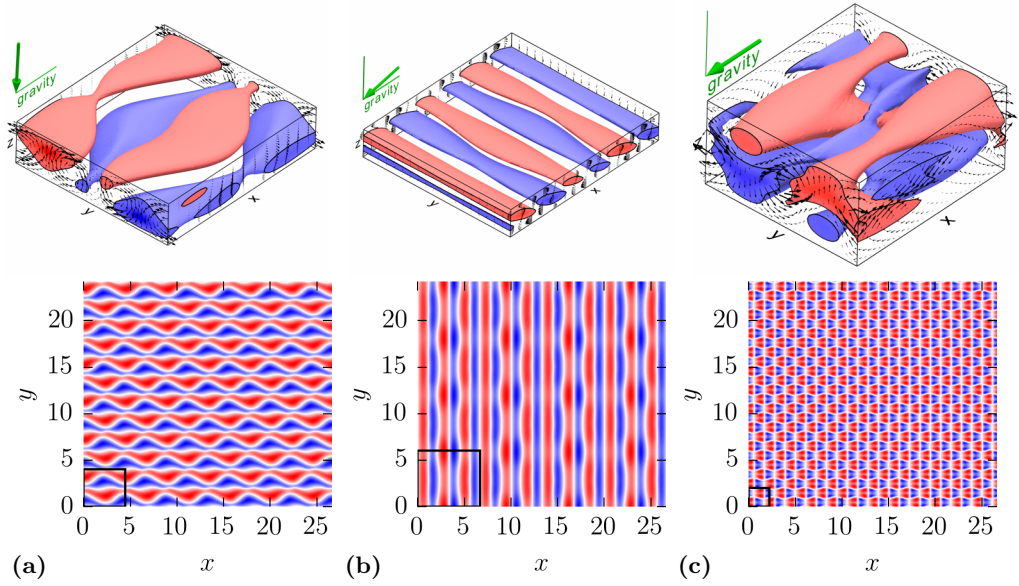

The $\gamma$-continuation of $SSW$ at $\epsilon=1.5$ approaches an equilibrium state at $\gamma=1.9^{\circ}$ that we name $LSV$, short for longitudinal subharmonic varicose state. The pattern of the equilibrium resembles an instance in time along the orbit $SSW$ at $\gamma=15^{\circ}$ (Figure \ref{fig:app:additional}a). $LSV$ is invariant under transformations of $S_{\mathrm{ssw}}=\langle \pi_{xyz},\tau(0.5,0.5)\rangle$ and can be continued from $\gamma=0^{\circ}$ to $\gamma=30^{\circ}$ along which the relative position of the hot and cold plumes changes continuously.

Just below $\gamma_{c2}$, a $\gamma$-forward Hopf bifurcation generates $TO$ from an equilibrium state, previously named transverse subharmonic varicose ($TSV$), at $\epsilon=0.1$ (inset panel in Figure \ref{fig:bifgamma3}). This invariant state represents stationary varicose modulations of the $TR$ pattern with $[3\lambda_x,3\lambda_y]$-periodicity and invariance under transformations of $S_{\mathrm{tv}}=\langle\pi_{xz},\pi_y\rangle$ (Figure \ref{fig:app:additional}b). $TSV$ is similar to the spatially subharmonic state discussed in \citet{Clever1995} but is not strictly subharmonic here since $TSV$ is not invariant under $\tau(0.5,0.5)$.

Due to the absence of $LR$ at finite $\mathrm{Ra}$ at $\gamma=90^{\circ}$, $KN$ do not terminate in a bifurcation from $LR$ as described in Section \ref{sec:bim} but from $SL$, an equilibrium that we name subharmonic lambda plumes. The $[\lambda_x,\lambda_y]$-periodic $SL$ emerges in a saddle-node bifurcation at $\epsilon=1.670$. From the upper branch of $SL$, $KN$ bifurcate $\epsilon$-backward at $\epsilon=1.672$ (Figure \ref{fig:bifeps3}). The subharmonic flow structure of $SL$ is invariant under transformations of $S_{\mathrm{sl}}=\langle\pi_{xz}\tau(0,0.5),\pi_y\tau(0.5,0),\tau(0.5,0.5)\rangle$ and resembles lambda-shaped plumes at scales of half the gap height (Figure \ref{fig:app:additional}c).

\section{Invariant states in vertical layer convection}
\label{sec:app:vertical}
\begin{figure}
        \begin{subfigure}[b]{0.24\textwidth}
        		\begin{tikzpicture}
    				\draw (0, 0) node[inner sep=0] {\includegraphics[width=\linewidth,trim={0.2cm 0.1cm 0.2cm 0.2cm},clip]{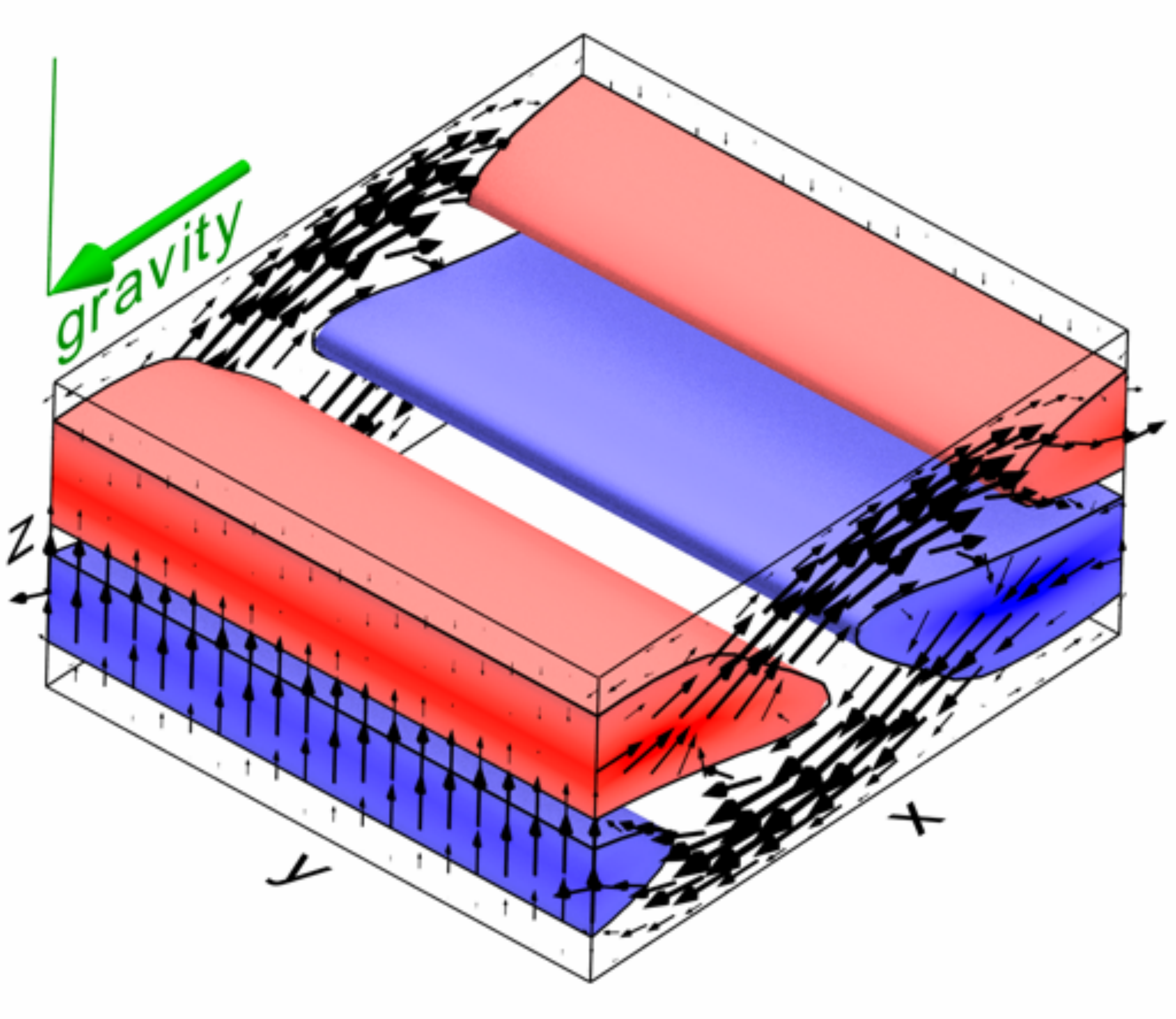}};
    	\draw (-1.4,-1.2) node {\textbf{(a)}};
			\end{tikzpicture}
        \end{subfigure}
        \begin{subfigure}[b]{0.24\textwidth}
        		\begin{tikzpicture}
    				\draw (0, 0) node[inner sep=0] {\includegraphics[width=\linewidth,trim={5.2cm 5.1cm 6.2cm 4.2cm},clip]{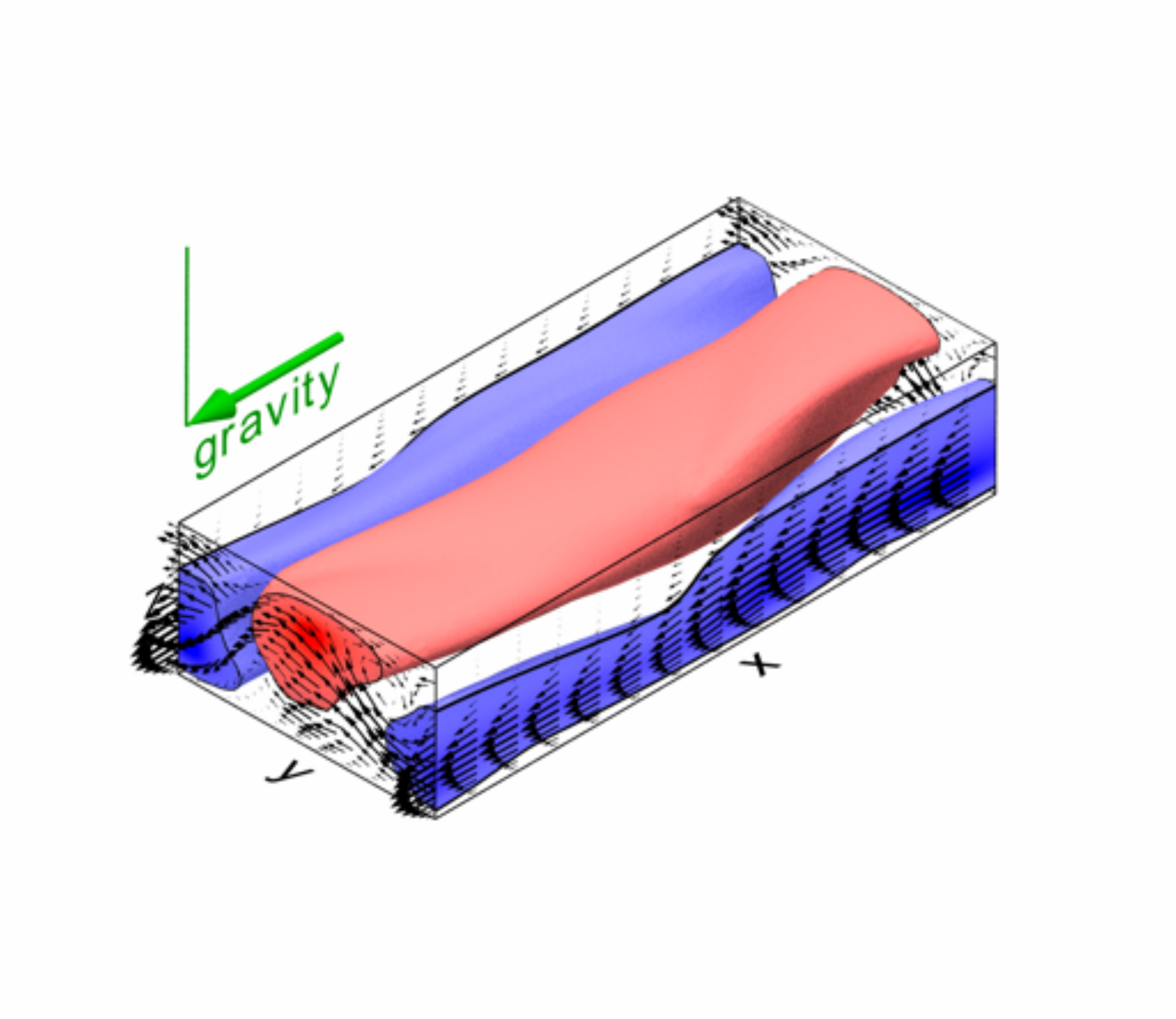}};
    	\draw (-1.4,-1.2) node {\textbf{(c)}};
			\end{tikzpicture}
        \end{subfigure}
        \begin{subfigure}[b]{0.24\textwidth}
        		\begin{tikzpicture}
    				\draw (0, 0) node[inner sep=0] {\includegraphics[width=\linewidth,trim={0.2cm 0.1cm 0.2cm 0.2cm},clip]{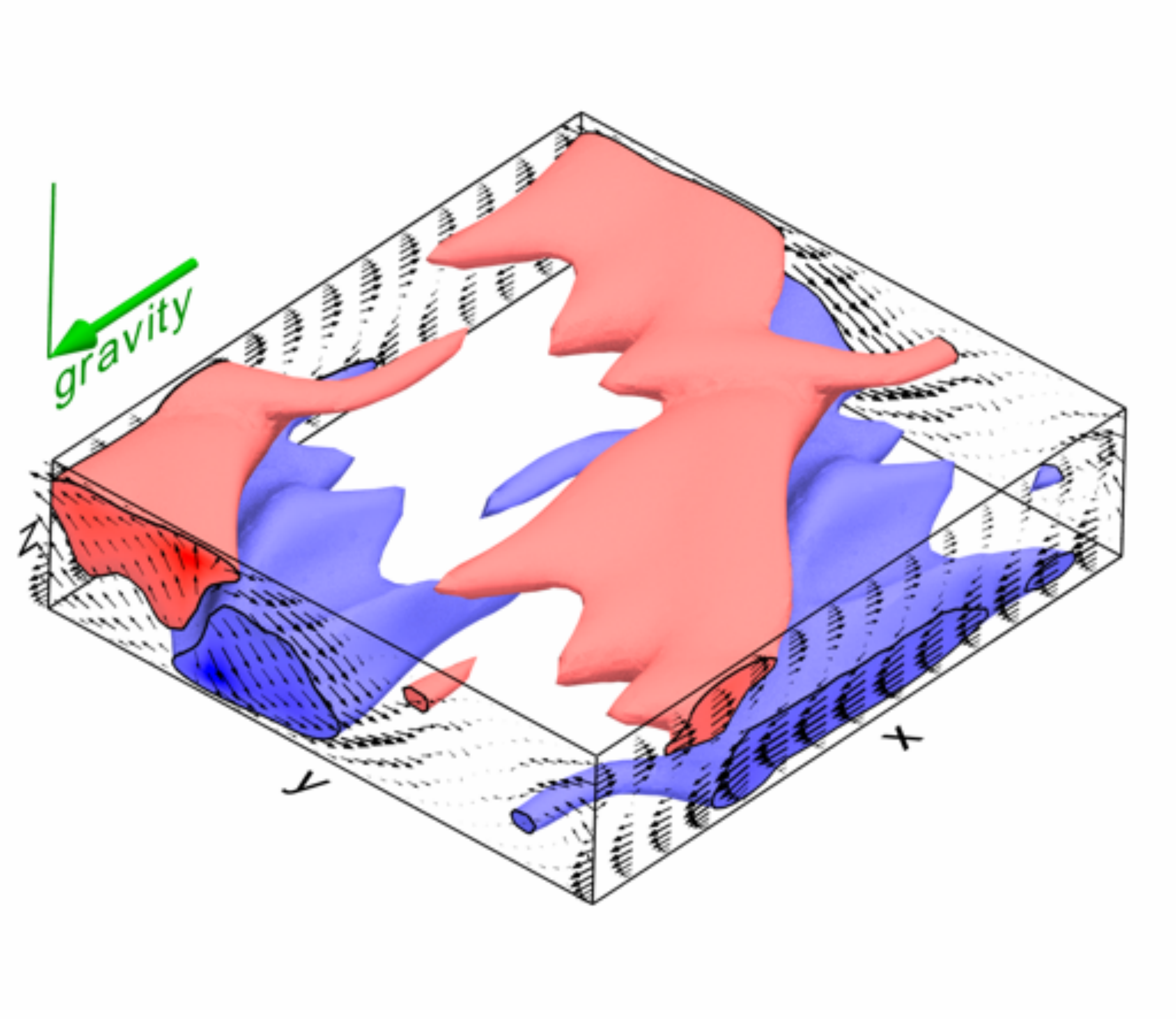}};
    	\draw (-1.4,-1.2) node {\textbf{(e)}};
			\end{tikzpicture}
        \end{subfigure}
        \begin{subfigure}[b]{0.24\textwidth}
        		\begin{tikzpicture}
    				\draw (0, 0) node[inner sep=0] {\includegraphics[width=\linewidth,trim={0.2cm 0.1cm 0.2cm 0.2cm},clip]{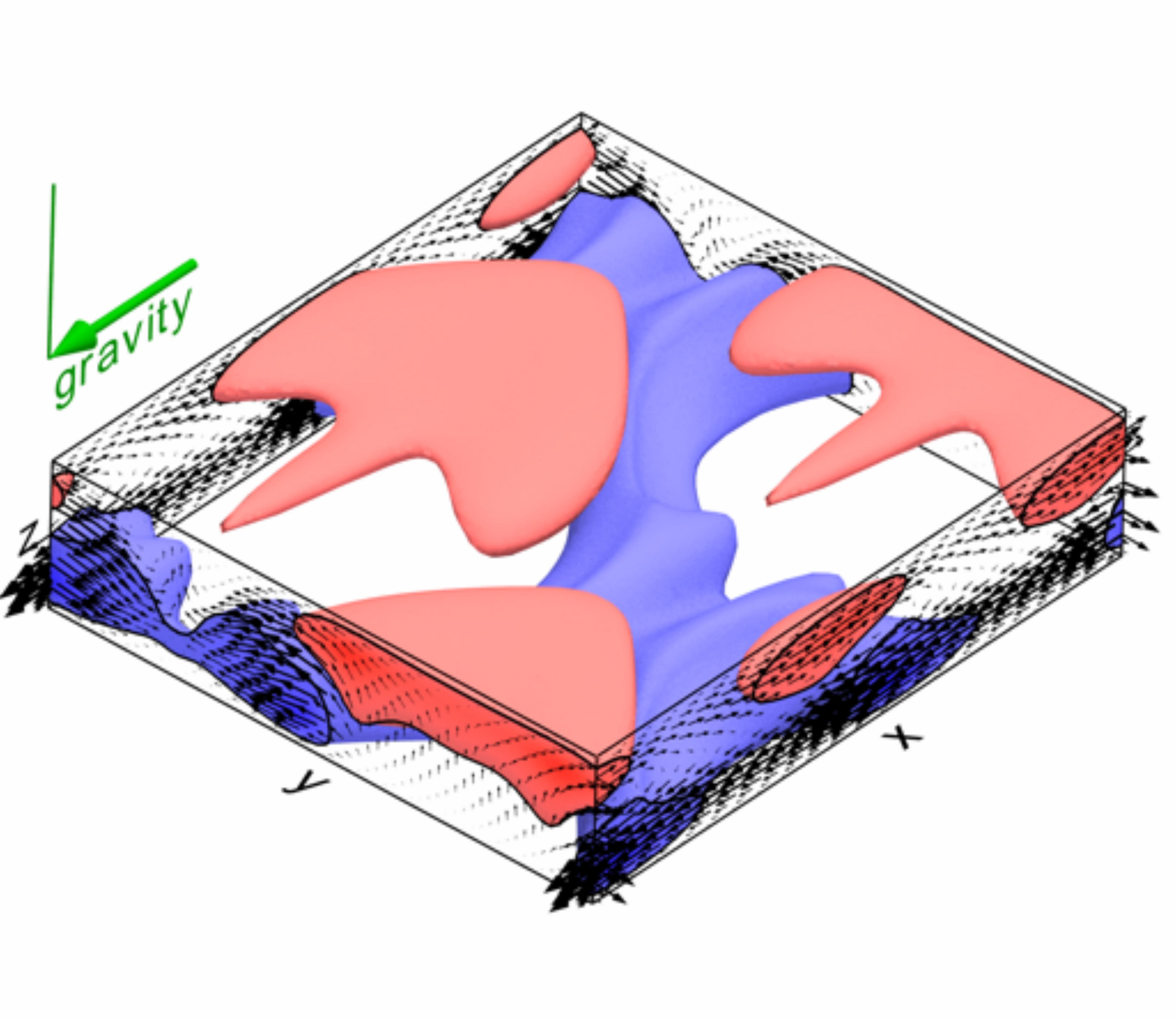}};
    	\draw (-1.4,-1.2) node {\textbf{(g)}};
			\end{tikzpicture}
        \end{subfigure}
        \begin{subfigure}[b]{0.24\textwidth}
        		\begin{tikzpicture}
    				\draw (0, 0) node[inner sep=0] {\includegraphics[width=\linewidth,trim={0.2cm 0.1cm 0.2cm 0.2cm},clip]{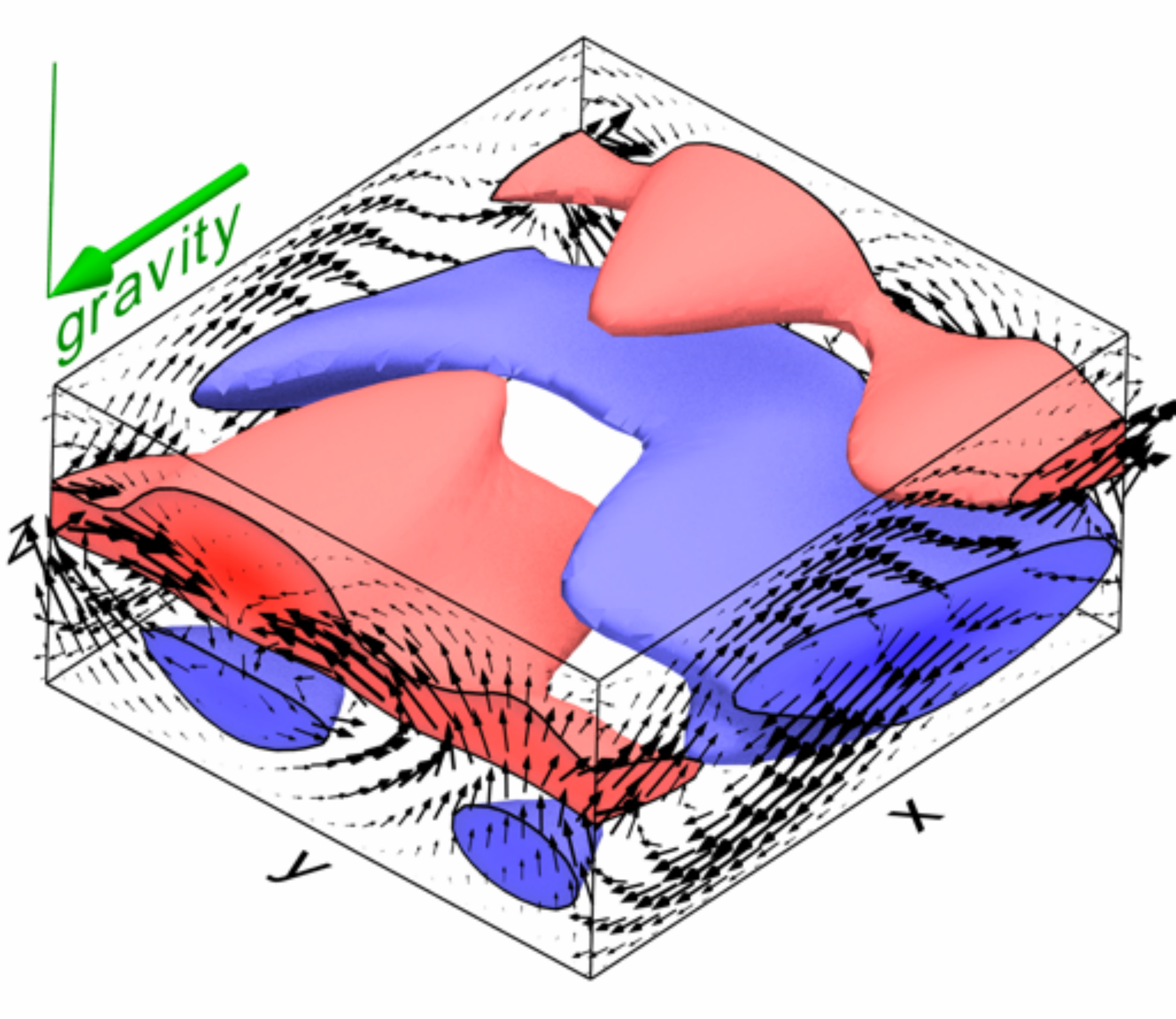}};
    	\draw (-1.4,-1.2) node {\textbf{(b)}};
			\end{tikzpicture}
        \end{subfigure}
        \begin{subfigure}[b]{0.24\textwidth}
        		\begin{tikzpicture}
    				\draw (0, 0) node[inner sep=0] {\includegraphics[width=\linewidth,trim={5.2cm 5.1cm 6.2cm 4.2cm},clip]{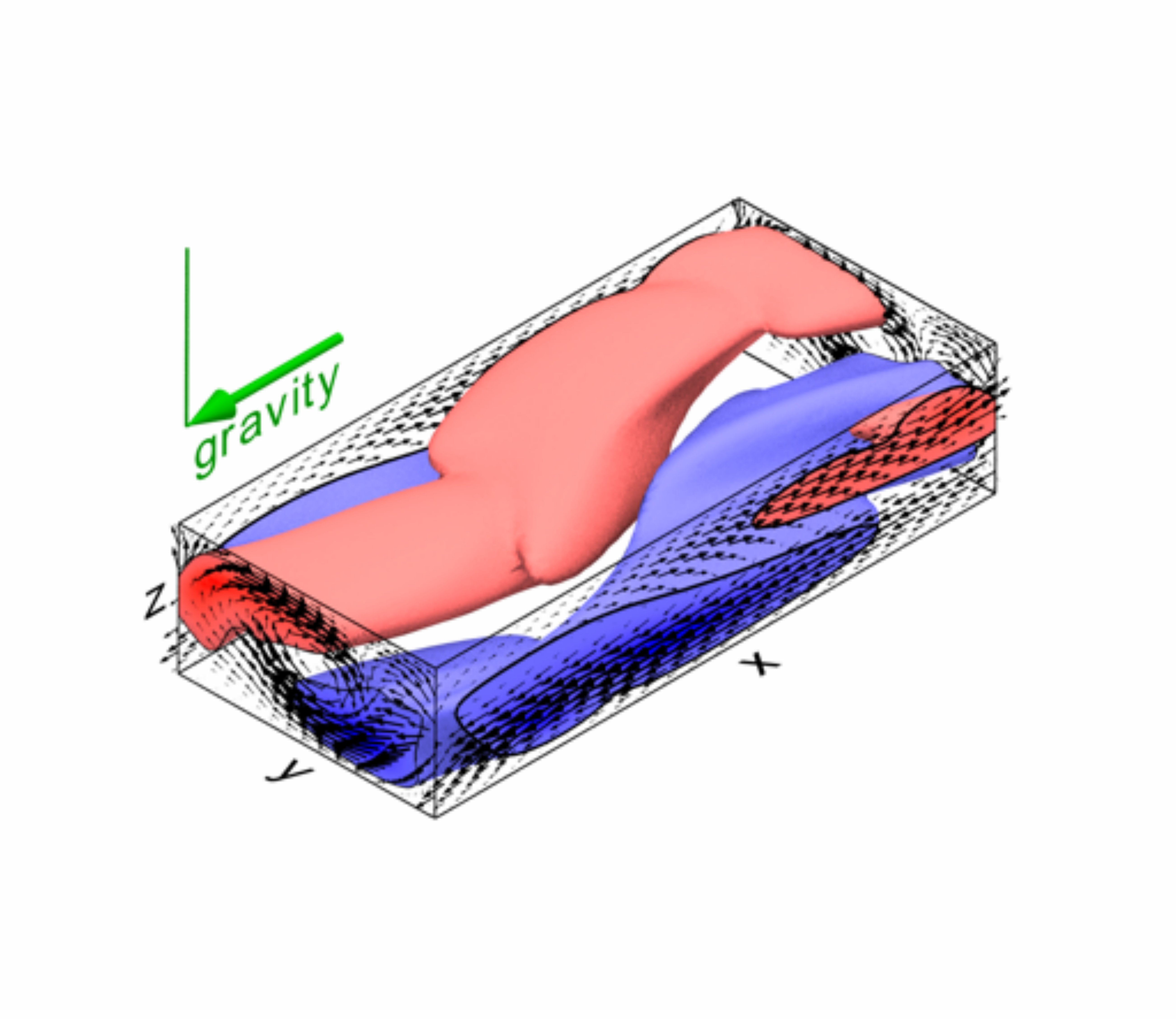}};
    	\draw (-1.4,-1.2) node {\textbf{(d)}};
			\end{tikzpicture}
        \end{subfigure}
        \begin{subfigure}[b]{0.24\textwidth}
        		\begin{tikzpicture}
    				\draw (0, 0) node[inner sep=0] {\includegraphics[width=\linewidth,trim={0.2cm 0.1cm 0.2cm 0.2cm},clip]{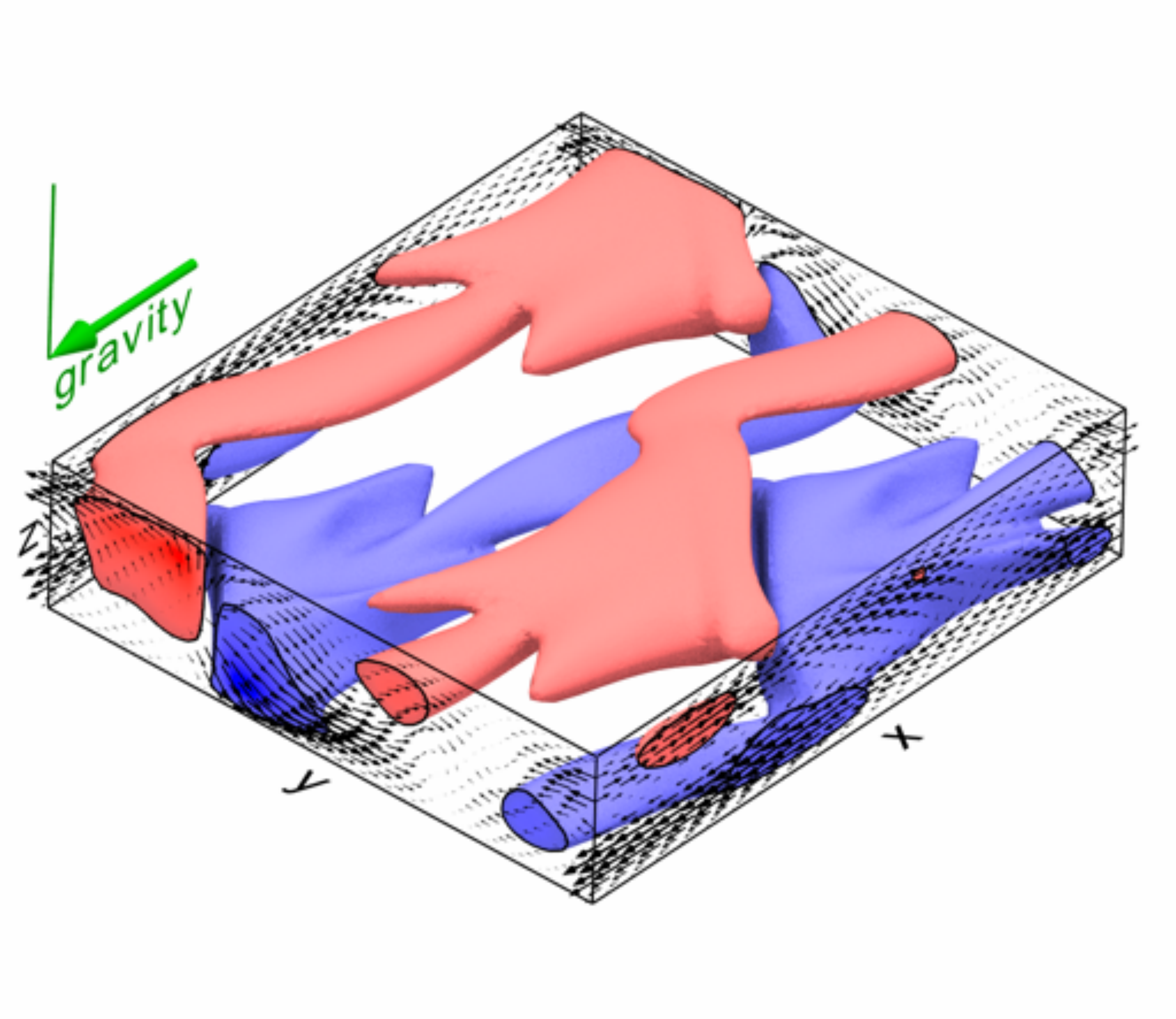}};
    	\draw (-1.4,-1.2) node {\textbf{(f)}};
			\end{tikzpicture}
        \end{subfigure}
        \begin{subfigure}[b]{0.24\textwidth}
        		\begin{tikzpicture}
    				\draw (0, 0) node[inner sep=0] {\includegraphics[width=\linewidth,trim={0.2cm 0.1cm 0.2cm 0.2cm},clip]{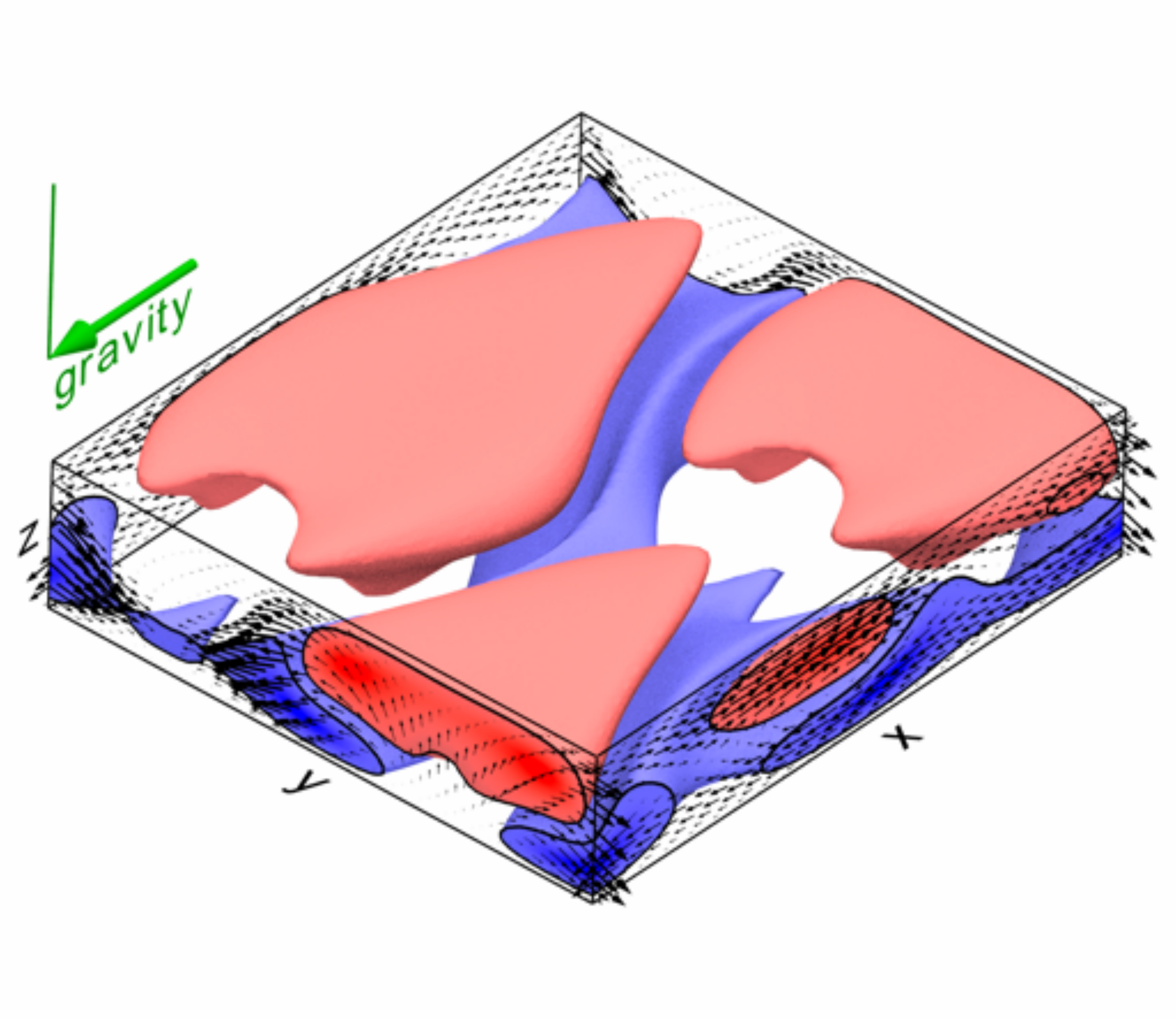}};
    	\draw (-1.4,-1.2) node {\textbf{(h)}};
			\end{tikzpicture}
        \end{subfigure}
        \begin{subfigure}[b]{0.99\textwidth}
        		\begin{tikzpicture}
    				\draw (0, 0) node[inner sep=0] {\includegraphics[width=0.23\linewidth,trim={0.2cm 0.1cm 0.2cm 0.2cm},clip]{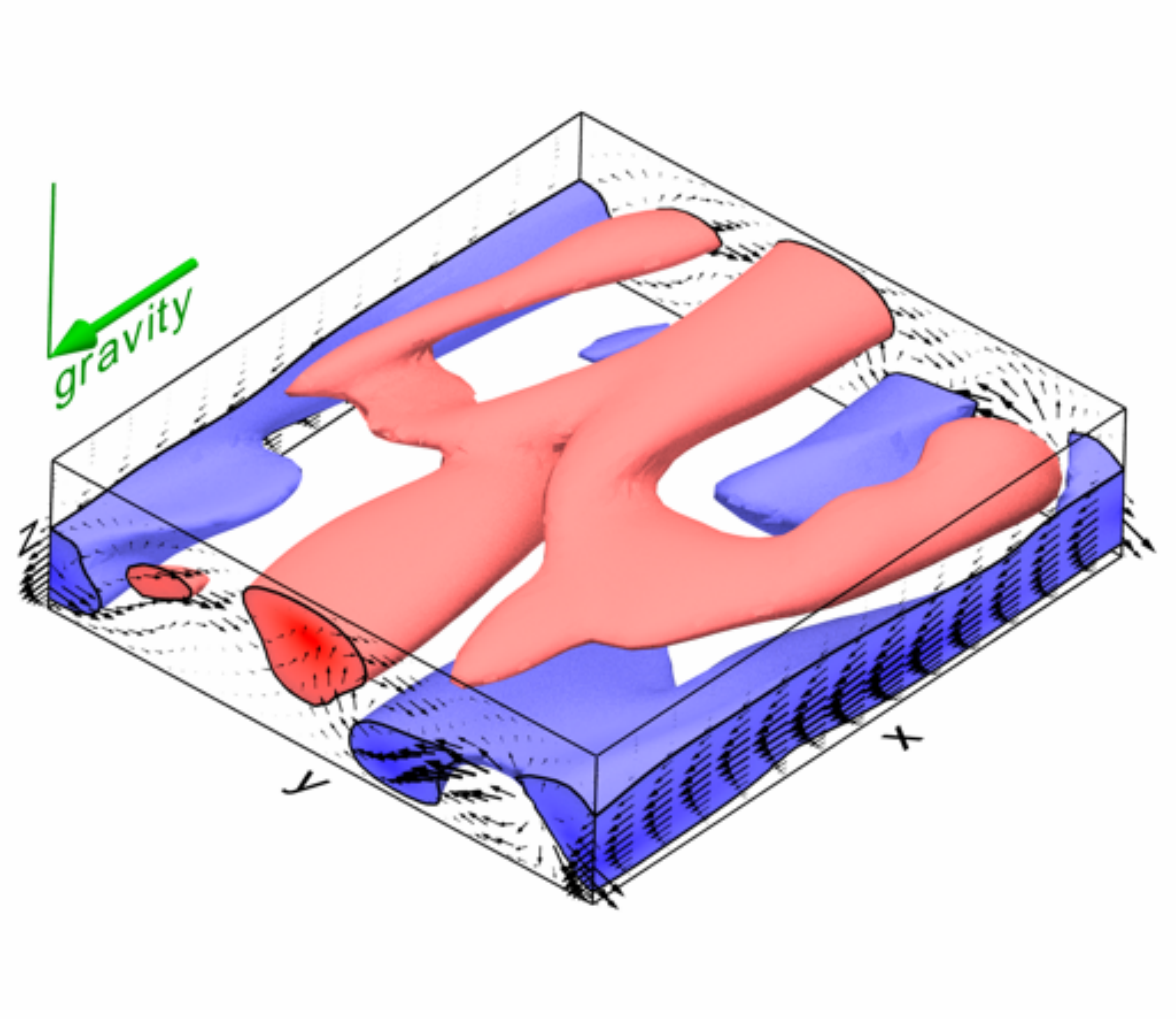}};
    	\draw (-1.4,-1.2) node {\textbf{(i)}};
			\end{tikzpicture}
        		\begin{tikzpicture}
    				\draw (0, 0) node[inner sep=0] {\includegraphics[width=0.23\linewidth,trim={0.2cm 0.1cm 0.2cm 0.2cm},clip]{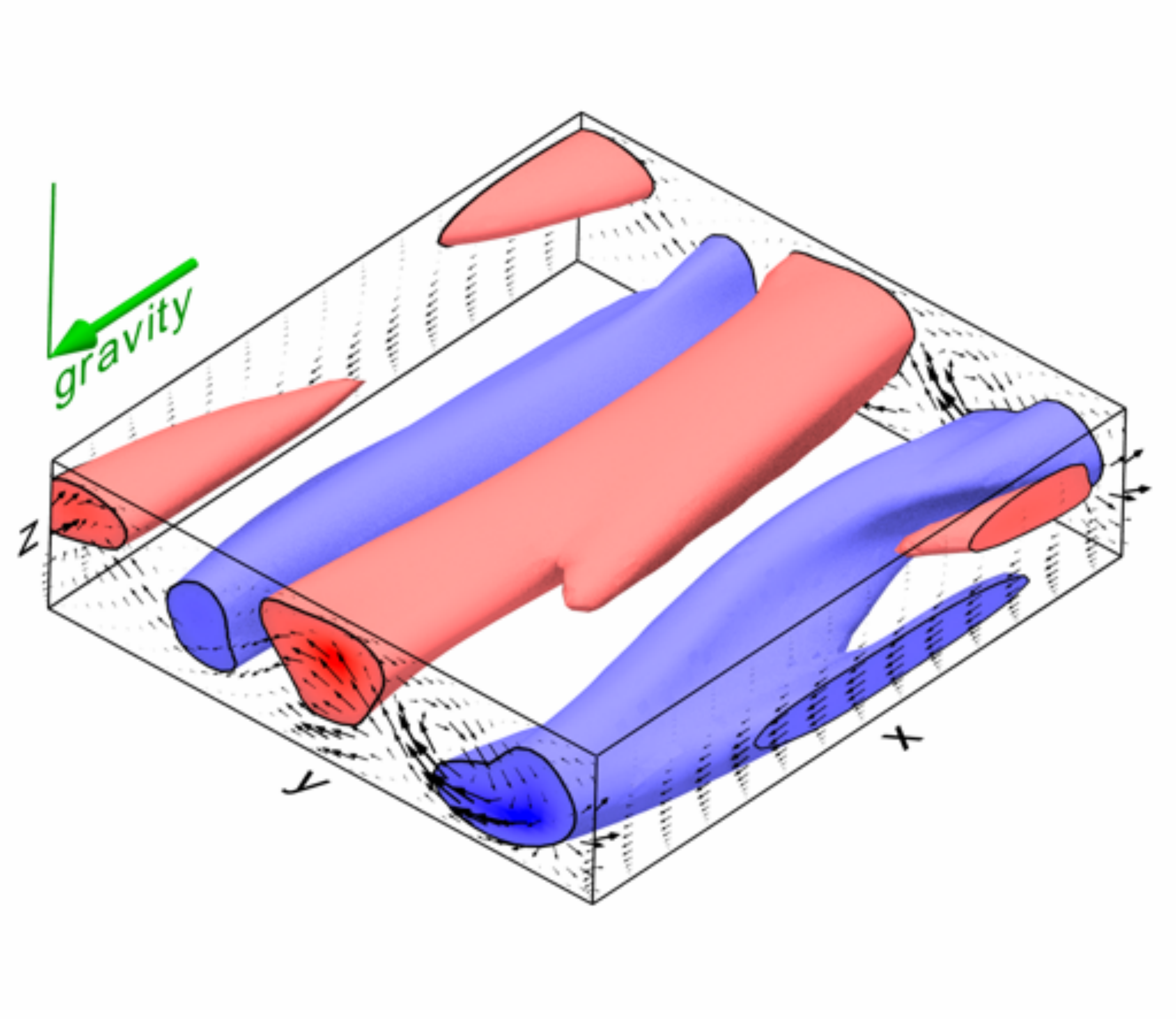}};
    	\draw (-1.4,-1.2) node {\textbf{(j)}};
			\end{tikzpicture}
        		\begin{tikzpicture}
    				\draw (0, 0) node[inner sep=0] {\includegraphics[width=0.23\linewidth,trim={0.2cm 0.1cm 0.2cm 0.2cm},clip]{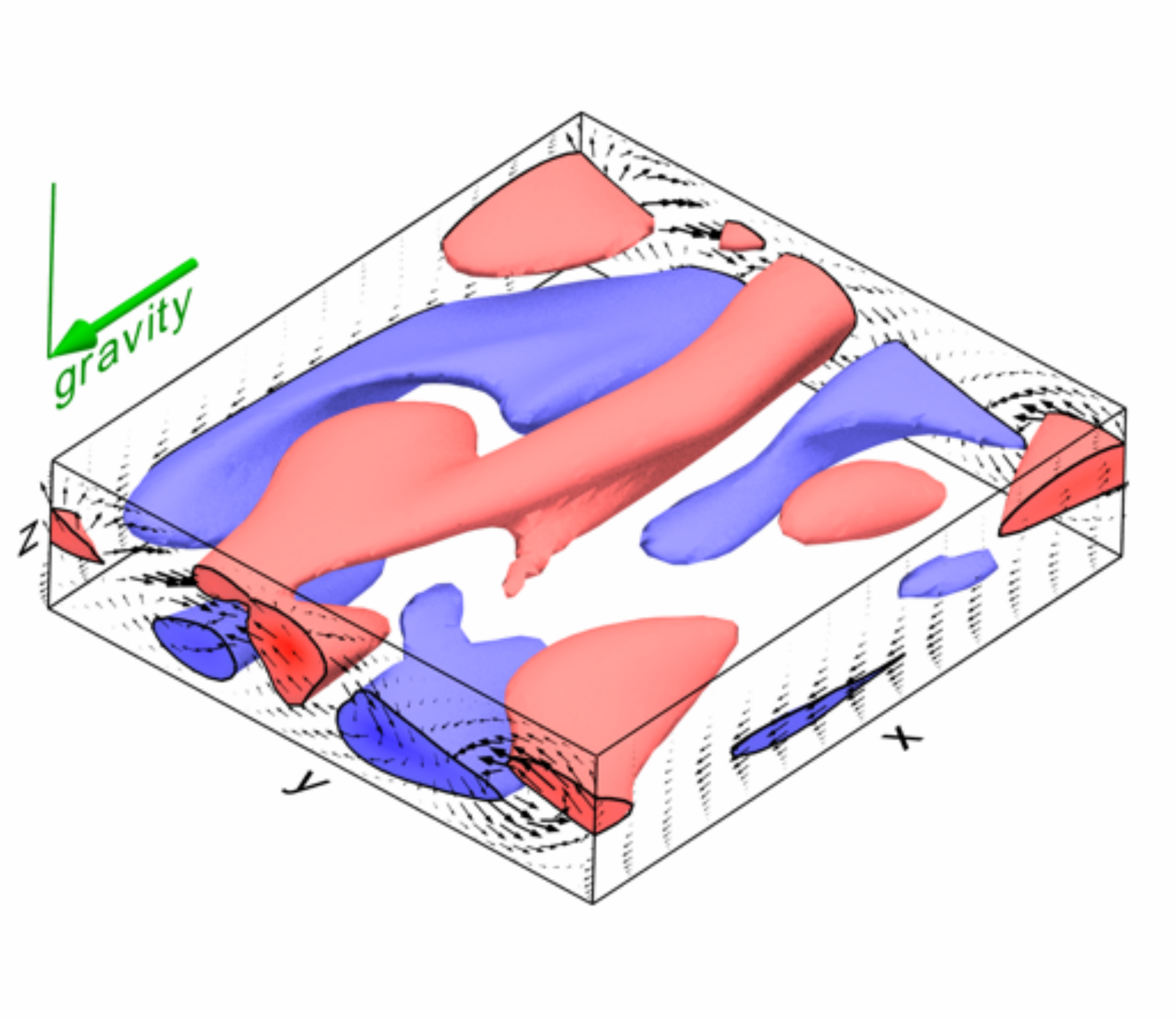}};
    	\draw (-1.4,-1.2) node {\textbf{(k)}};
			\end{tikzpicture}
        		\begin{tikzpicture}
    				\draw (0, 0) node[inner sep=0] {\includegraphics[width=0.23\linewidth,trim={0.2cm 0.1cm 0.2cm 0.2cm},clip]{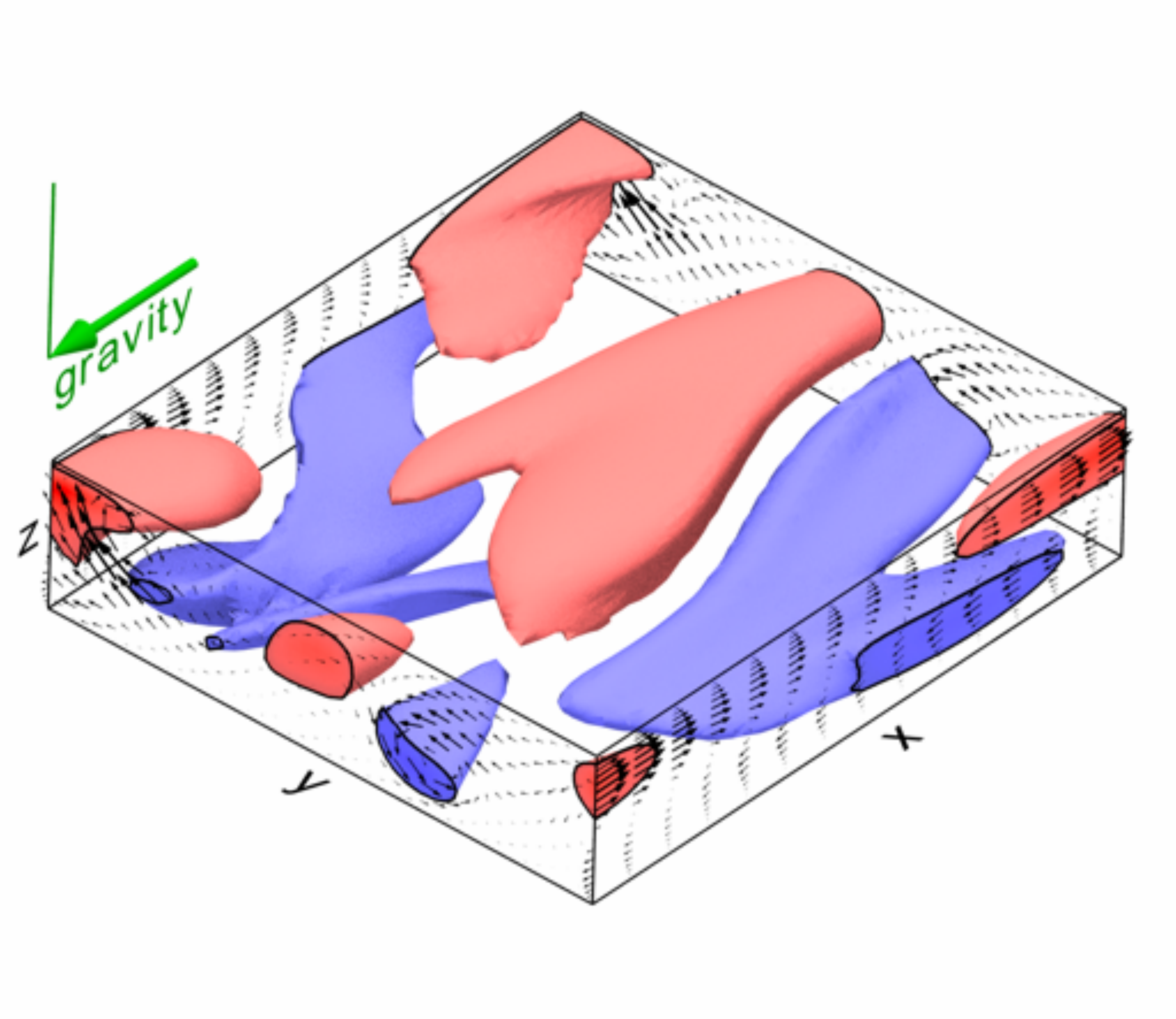}};
    	\draw (-1.4,-1.2) node {\textbf{(l)}};
			\end{tikzpicture}
        \end{subfigure}
       	\caption{\label{fig:app:vertical} Continuation yields eight invariant states in vertical layer convection ($\gamma=90^{\circ}$) at $\epsilon=1.5$ ($\mathrm{Ra}=21\,266$). \textbf{(a)} $TR$, \textbf{(b)} $KN$, \textbf{(c/d)} upper/lower branch of $WR$, \textbf{(e/f)} upper/lower branch of $SSW$, \textbf{(g/h)} upper/lower branch of $STW$. The distinction between upper/lower branches refers to $||\theta ||_2$. The upper branches of $WR$ and $SSW$ are closer to bifurcations from $LR$ than the lower branches, for $STW$ \emph{vice versa}. \textbf{(i)-(l)} Snapshots from a DNS at $\gamma=90^{\circ}$ and $\epsilon=1.5$ show transiently emerging structures resembling the invariant states above.}
\end{figure}

ILC at $\gamma=90^{\circ}$ is a particular case because vertical layer convection has the largest laminar shear forces of all inclinations and can be considered a pure shear flow. Here, buoyancy provides a body force along the channel domain, acting like a pressure gradient in pressure-driven channel flow. Despite the absence of a wall-normal buoyancy force, all main invariant states considered in the present study were numerically continued to $\gamma=90^{\circ}$, with the exception of $LR$ existing only at $\mathrm{Ra}=\infty$ due to (\ref{eq:scinvu}-\ref{eq:scinvc}) and $SV$. The flow structures of the invariant states at $\gamma=90^{\circ}$ and $\epsilon=1.5$ show sharper interfaces and more pointed convective plumes than at the parameters where these states were initially found (Figure \ref{fig:app:vertical}).

\end{document}